\documentclass[a4,11pt]{article}

\makeatletter
\normalsize
\divide\@tempdima\baselineskip
\@tempcnta=\@tempdima
\skip\@mpfootins = \skip\footins
\renewcommand\section{\@startsection{section}{1}{\z@}%
                                   {-3.5ex \@plus -1.3ex \@minus -.7ex}%
                                   {2.3ex \@plus.4ex \@minus .4ex}%
                                   {\normalfont\large\bfseries}}
\renewcommand\subsection{\@startsection{subsection}{2}{\z@}%
                                   {-2.3ex\@plus -1ex \@minus -.5ex}%
                                   {1.2ex \@plus .3ex \@minus .3ex}%
                                   {\normalfont\normalsize\bfseries}}
\renewcommand\subsubsection{\@startsection{subsubsection}{3}{\z@}%
                                   {-2.3ex\@plus -1ex \@minus -.5ex}%
                                   {1ex \@plus .2ex \@minus .2ex}%
                                   {\normalfont\normalsize\bfseries}}
\renewcommand\paragraph{\@startsection{paragraph}{4}{\z@}%
                                   {1.75ex \@plus1ex \@minus.2ex}%
                                   {-1em}%
                                   {\normalfont\normalsize\bfseries}}
\renewcommand\subparagraph{\@startsection{subparagraph}{5}{\z@}%
                                   {1.75ex \@plus1ex \@minus .2ex}%
                                   {-1em}%
                                   {\normalfont\normalsize\itshape}}

\renewcommand{\@dotsep}{10000}

\def\fnum@figure{\textbf{\figurename\nobreakspace\thefigure}}
\def\fnum@table{\textbf{\tablename\nobreakspace\thetable}}

\long\def\@makecaption#1#2{%
  \vskip\abovecaptionskip
  \sbox\@tempboxa{\small #1. #2}%
  \ifdim \wd\@tempboxa >\hsize
    \small #1. #2\par
  \else
    \global \@minipagefalse
    \hb@xt@\hsize{\hfil\box\@tempboxa\hfil}%
  \fi
  \vskip\belowcaptionskip}

\renewenvironment{thebibliography}[1]{%
\begin{oldthebibliography}{#1}%
\small%
\raggedright%
\setlength{\itemsep}{5pt plus 0.2ex minus 0.05ex}%
}%
{%
\end{oldthebibliography}%
}
\makeatother

\usepackage[colorlinks=true, linkcolor=blue, filecolor=blue, urlcolor=blue, citecolor=blue, anchorcolor=blue, menucolor=blue , linktocpage=true]{hyperref}
\usepackage{scalerel}
\usepackage{graphicx}
\usepackage{amsmath,amssymb}
\usepackage{amsfonts}
\usepackage{physics}
\usepackage{bbold}
\usepackage{mathtools}
\usepackage{verbatim}
\usepackage{cite} 
\usepackage{mdframed}
\usepackage{enumitem}
\usepackage[normalem]{ulem}

\allowdisplaybreaks
\numberwithin{equation}{section}

\makeatletter
\newcommand{\subalign}[1]{%
  \vcenter{%
    \Let@ \restore@math@cr \default@tag
    \baselineskip\fontdimen10 \scriptfont\tw@
    \advance\baselineskip\fontdimen12 \scriptfont\tw@
    \lineskip\thr@@\fontdimen8 \scriptfont\thr@@
    \lineskiplimit\lineskip
    \ialign{\hfil$\m@th\scriptstyle##$&$\m@th\scriptstyle{}##$\hfil\crcr
      #1\crcr
    }%
  }%
}
\makeatother

\usepackage[textsize=scriptsize,textwidth=2.45cm]{todonotes}

\definecolor{ForestGreen}{RGB}{34,139,34}

\newcommand{\eqa}[1]{\begin{align}#1\end{align}}
\newcommand{\eq}[1]{\begin{equation}#1\end{equation}}

\newcommand{\eqsp}[1]{\begin{equation}\begin{split}#1\end{split}\end{equation}}

\newcommand{\tauh}{\hat{\tau}}
\newcommand{\sh}{\hat{\sigma}}
\newcommand{\zh}{\hat{z}}

\renewcommand{\title}[1]{\vbox{\center\LARGE{#1}}\vspace{5mm}}
\renewcommand{\author}[1]{\vbox{\center#1}\vspace{5mm}}

\newcommand{\email}[1]{\vspace{5mm}\vbox{\center\footnotesize\tt#1}\vspace{5mm}}

\newcommand{\polar}{\text{\tiny H}|\text{\tiny L}}
\newcommand{\npolar}{\text{\tiny H}|\text{\tiny H}}

\newcommand{\e}{e}
\newcommand{\knl}[3]{\mathbb{#1}_{\{#2\};\{#3\}}}

\newcommand{\dmin}{\Delta_0}
\newcommand{\nn}{\textsf{n}}
\newcommand{\jj}{\,\textsf{j}}

\newcommand{\light}{\text{\tiny L}}
\newcommand{\hj}{\mathsf{a}}

\newcommand{\radc}{c}

\newcommand{\svar}{s}
\newcommand{\dmina}{|\Delta_0|}

\begin{document}

\thispagestyle{empty}

\phantom{a}
\vskip20mm

\begin{center} 

\title{
The light we can see:\\
Extracting black holes from weak Jacobi forms
}


\author{Luis Apolo$^{\,a}$,
Suzanne Bintanja$^{\,b,\,c}$,
Alejandra Castro$^{\,d}$ and
Diego Liska$^{\,b}$}


\vskip4.1mm

\begin{minipage}[c]{0.89\textwidth}\centering \footnotesize \em {\it $^{a}$Beijing Institute of Mathematical Sciences and Applications, Beijing 101408, China}
\\ \vspace{0.5em} 
{\it $^{b}$Institute for Theoretical Physics and $\Delta$-Institute for Theoretical Physics, University of Amsterdam, PO Box 94485, 1090GL Amsterdam, The Netherlands}\\ \vspace{0.5em}
{\it $^c$Kavli Institute for Theoretical Physics, University of California, Santa Barbara, CA 93106, USA}\\ \vspace{0.5em}
{\it $^{d}$Department of Applied Mathematics and Theoretical Physics, University of Cambridge,
Cambridge CB3 0WA, United Kingdom} \end{minipage}

\email{apolo@bimsa.cn, s.bintanja@uva.nl, ac2553@cam.ac.uk, d.liska@uva.nl}
\end{center}

\vskip6mm

\begin{abstract}
\noindent We quantify how constraints on light states affect the asymptotic growth of heavy states in weak Jacobi forms. The constraints we consider are sparseness conditions on the Fourier coefficients of these forms, which are necessary to interpret them as gravitational path integrals. Using crossing kernels, we extract the leading and subleading behavior of these coefficients and show that the leading Cardy-like growth is robust in a wide regime of validity. On the other hand, we find that subleading corrections are sensitive to the constraints placed on the light states, and we quantify their imprint on the asymptotic growth of states. Our approach is tested against the generating function of symmetric product orbifolds, where we provide new insights into the factors contributing to the asymptotic growth of their Fourier coefficients. Finally, we use our methods to revisit the UV/IR connection that relates black hole microstate counting to modular forms. We provide a microscopic interpretation of the logarithmic corrections to the entropy of BPS black holes in ${\cal N}=2,4$ ungauged supergravity in four and five dimensions, and tie it to consistency conditions in AdS$_3$/CFT$_2$. 

\end{abstract}

\eject

\phantom{a}
\vspace{-4em}
{
\tableofcontents
}

\bigskip

\bigskip

\newpage

\section{Introduction}

Modular forms play a pivotal role in the counting of black hole microstates. The underlying modular symmetry of counting formulae was key in the precise match between the Bekenstein-Hawking entropy of five-dimensional supersymmetric black holes and Cardy's formula for the asymptotic growth of states \cite{Strominger:1996sh}. The same principle is used in the derivation of the entropy of these black holes via the AdS$_3$/CFT$_2$ correspondence \cite{Strominger:1997eq}. These developments have led to an understanding of the statistical nature of several black holes in string theory, where the properties of modular forms can be used to cast specific counting formulae as gravitational path integrals. Reviews that highlight the prominence of modular forms on this subject include \cite{Sen:2007qy,Murthy:2023mbc}.

Our goal is to revisit the connection between modular forms and black hole entropy, and tie it with other consistency conditions of AdS/CFT. We will focus our attention on weak Jacobi forms (wJf), which can be thought of as generating functions of the form
\begin{equation}\label{eq:wjf1}
 \varphi(\tau,z) = 
    \sum_{n,\ell} c(n,\ell)  e^{2\pi i \tau n} e^{2\pi i z \ell}\,.
\end{equation}
As we will review in the coming sections, the Fourier coefficients $c(n,\ell)$ are highly constrained by the modular and elliptic properties of these forms. In particular, the $c(n,\ell)$ coefficients are organized in terms of the discriminant $\Delta\coloneqq n- \ell^2/4t$. 
Here $t$ is the index of the wJf which, together with the weight of the form, completely characterize the modular and elliptic properties of $\varphi(\tau,z)$. When $\Delta \geq 0$, one refers to the corresponding state as \textit{non-polar} or heavy, while $\Delta<0$ corresponds to a \textit{polar} or light state. The discriminant of the most polar state is denoted by $\dmin$. The appearance of wJfs is common in the context of ${\cal N}=4$ supersymmetric CFT$_2$, where 
\begin{equation}\label{eq:eg}
    \varphi(\tau,z)=  \Tr_{\text{RR}} (-1)^{F} e^{2\pi i \tau(L_0-\frac{c}{24})} e^{2\pi i zJ_0}   e^{2\pi i \bar \tau (\bar L_0 -\frac{c}{24})}\,,
\end{equation}
is the elliptic genus of the theory --- an index that captures 1/4-BPS states. In these cases, the central charge of the CFT$_2$ is $c=6t$ and the most polar state satisfies $\Delta_0 = -c/24$. However, our analysis will not rely upon this specific interpretation.\footnote{A relation similar to \eqref{eq:eg} exists for theories with ${\cal N}=2$ supersymmetry, for which one needs to account for fractional values of $J_0$. Other instances where wJfs appear include holomorphic theories with a $U(1)$ current and warped CFTs. One can also generalize the notion of wJfs to theories with an enlarged symmetry algebra, leading to wJfs with multiple charges.}

The comparison between black hole entropy and counting formulae like \eqref{eq:wjf1} is usually valid for large values of the parameters involved. We will revisit this comparison by implementing further conditions on $\varphi(\tau,z)$ that are motivated by the AdS$_3$/CFT$_2$ correspondence. These conditions are:
\begin{itemize}[leftmargin=0.3cm]
    \item[] {\bf Large $\dmina$ limit.} The index $t$ and maximal polarity $\dmina$ determine the central charge of the CFT$_2$ (their precise relation will be given when we discuss examples). Taking $|\dmin|\gg1$ and $t\gg1$ is equivalent to taking the large central charge limit characteristic of  AdS$_3$/CFT$_2$.
    \item[] {\bf Sparseness of polar states.} The distribution of polar states is an important criterion to diagnose the class of gravitational theories in AdS$_3$ we are working with. We will quantify the imprint of the sparseness of polar states on the $c(n, \ell)$ coefficients of non-polar states, the latter of which are naturally linked to black hole configurations. 
\end{itemize}
These conditions were first introduced by Hartman, Keller and Stoica (HKS) \cite{Hartman:2014oaa} in the context of the torus partition function of a CFT$_2$. The authors showed that these conditions are necessary for the free energy to be universal and for the asymptotic growth of states to be given by the Cardy formula for all $\Delta\geq \dmina\gg1$. The same universality is present in AdS$_3$ quantum gravity in the semiclassical regime. Hence, the aforementioned conditions are necessary for any CFT$_2$ to be holographic, i.e.~to capture universal features of three-dimensional gravity in the semiclassical approximation. One outcome of this analysis is that the light spectrum does not leave an explicit imprint on the free energy.  

The ideas of HKS were extended to weak Jacobi forms in \cite{Benjamin:2015hsa,Benjamin:2015vkc}, which considered the elliptic genus of a CFT$_2$ and compared its asymptotic growth to the entropy of BPS black holes in AdS$_3$ supergravity. Our analysis will start from those findings. More specifically, we will study the asymptotic growth of states of ``sparse weak Jacobi forms'', which are wJfs that comply with the two conditions mentioned above. Given that polar states do not leave a mark on the asymptotic expansion of the $c(n,\ell)$ coefficients at leading order, the questions we will address are the following. Can we quantify the effects of light states on the subleading corrections to the $c(n, \ell)$ coefficients? And how do these effects translate into aspects of a putative dual theory of gravity in the semiclassical regime?  

One might expect that implementing the large $\dmina$ and sparseness conditions should be somewhat straightforward: the Rademacher expansion gives an exact expression for the $c(n,\ell)$ coefficients of non-polar states in terms of the polar ones, as we will review in the coming sections. However, this expansion is impractical in the large $\dmina$ limit.  In this paper, we will take a different route and use crossing kernels to study the asymptotic behavior of the $c(n,\ell)$ coefficients. In particular, we will quantify the effect of the large $\dmina$ and sparseness conditions have in the analysis, and contrast this approach with the Rademacher expansion, highlighting the differences and similarities. One advantage of using crossing kernels is that they allow us to approximate a discrete spectrum by a continuum, which will be key in our analysis.

In this paper, we give an expression for the $c(n, \ell)$ coefficients of non-polar (heavy) states in sparse wJfs that can be written as
\begin{equation}\label{eq:logs-intro}
    \log c(n,\ell)= 4\pi \sqrt{\Delta\dmina}+ a_{\scaleto{\Delta}{4pt}} \log \Delta + a_{\scaleto{\dmin}{4pt}} \log|\dmin| + \cdots\,,
\end{equation}
where the dots denote terms suppressed in the $\Delta\gg 1$ and $\dmina\gg1$ limits. The leading (first) term indicates a universal exponential growth of states, in accordance with HKS. We will also quantify the subleading or \textit{logarithmic} corrections by providing explicit expressions for $a_{\scaleto{\Delta}{4pt}}$ and $a_{\scaleto{\dmin}{4pt}}$ in terms of the polar spectrum. We will show that \eqref{eq:logs-intro} is valid in the following scenarios:
\begin{itemize}[leftmargin=0.3cm]
\item[]{\bf Universal behavior for $\Delta \gtrsim |\dmin|$.} In this regime, the asymptotic growth of coefficients of a sparse wJf is of the form
\begin{equation}\label{eq:cnl}
     c(n,\ell) \approx  \frac{\rho_0(\dmin)}{\Delta} \sqrt{\frac{|\dmin|}{t}} \e^{4\pi \sqrt{\Delta \dmina}}\,, \qquad \Delta  \gtrsim  |\dmin|\gg 1\,.
\end{equation}
The exponential growth is expected from the HKS bound and it agrees with \cite{Benjamin:2015hsa}. Our analysis also captures the subleading logarithmic corrections. These are controlled by  $\rho_0(\dmin)$, which incorporates the possibility of a degenerate ground state. Comparing to \eqref{eq:logs-intro}, we thus find that $a_{\scaleto{\Delta}{4pt}}$ is universal (independent of the light states), and that $a_{\scaleto{\dmin}{4pt}}$ is only sensitive to the ground state degeneracy. 

It is instructive to compare this result with \cite{Mukhametzhanov:2019pzy}, where Tauberian methods were used to obtain subleading corrections to the HKS analysis of modular forms with positive Fourier coefficients. While the explicit dependence on $\Delta$, $\dmina$, and $t$ in \eqref{eq:cnl} agrees with their results, $\rho_0(\dmin)$ was not accounted for there. In many circumstances, such as in partition functions, it is natural to set $\rho_0(\dmin)=1$. However, we will see that in a supersymmetric context, there can be degenerate Ramond ground states that cause $\rho_0(\dmin)$ to grow polynomially with $\dmina$. This feature is present in BPS black holes and is crucial for the agreement between the gravitational and microscopic account of their entropy.

\item[]{\bf Non-universal behavior for $\Delta \lesssim |\dmin|$.} This regime is not natural from the CFT$_2$ point of view. However, it has appeared on the gravitational side and is thus worth exploring. Our methods can quantify the asymptotic behavior of the $c(n,\ell)$ coefficients in the regime $ |\dmin| \gtrsim \Delta  \gg 1$ when the wJf exhibits ``slow growth''. These are wJfs where the distribution of light states is sub-Hagedorn. An important outcome of our analysis is that we can determine when $c(n,\ell)$ is of the form \eqref{eq:cnl} and quantify precisely how  $a_{\scaleto{\Delta}{4pt}}$ and $a_{\scaleto{\Delta_0}{4pt}}$ depend on the spectrum of light states.
\end{itemize}

We test our analysis by considering the generating function of symmetric product orbifolds. These are wJfs that not only comply with the two conditions described above \cite{Keller:2011xi,Benjamin:2015vkc} but also explicitly realize different types of sparse wJfs \cite{Belin:2019rba,Belin:2020nmp,Benjamin:2022jin}. Within this class of generating functions, we have access to a third independent method to extract the $c(n,\ell)$ coefficients, which follows from the methods of \cite{Sen:2007qy,Belin:2016knb,Cardoso:2021gfg}. This portion of the analysis will place various components of the crossing kernels into context, provide nontrivial checks, and illuminate the interpretation of the results. 

Another motivation for revisiting the analysis of symmetric product orbifolds is comparison with the entropy of 1/4-BPS black holes in ${\cal N}=4$ ungauged supergravity in four and five dimensions. The modular form that captures the relevant microstates in 4D is the Igusa cusp form, while a simple modification of this form covers 5D black holes. In both cases, these modular forms contain information about the elliptic genus of the symmetric product orbifold of K3 since there is an uplift to AdS$_3\times S^3\times$ K3. The seminal work of \cite{Banerjee:2011jp,Sen:2012cj} showed that the logarithmic corrections in gravity agree precisely with those extracted from this modular form. From a gravitational perspective, one evaluates the entropy of the black hole via a Euclidean path integral, which takes the form
\begin{equation}\label{eq:entropy-intro}
    S_{\rm BH} = \frac{A_H}{4G} + a_{\scaleto{\rm grav}{4pt}} \log (\frac{A_H}{4G}) + \cdots \,,
\end{equation}
where the dots denote corrections that are subleading in the area of the horizon $A_H$ relative to the Planck length, i.e.~$A_H/G\gg1$. The logarithmic correction is an IR effect that comes from the one-loop determinant of massless fields and zero modes in supergravity, effects that we are encoding as $a_{\scaleto{\rm grav}{4pt}}$. The analysis of \cite{Banerjee:2011jp,Sen:2012cj} match $a_{\scaleto{\rm grav}{4pt}}$ to the microscopic (UV) data encoded in the $a_{\scaleto{\Delta}{4pt}}$ and $a_{\scaleto{\Delta_0}{4pt}}$ variables parametrizing the $c(n, \ell)$ coefficients of a wJf. We will revisit this match from the point of view of HKS: we are interested in determining the data in the modular form that controls the logarithmic corrections, and how this fits within the framework of AdS$_3$/CFT$_2$. We will see that the 4D black hole falls into the universal regime, and hence the logarithmic corrections are only sensitive to the number of ground states. On the other hand, the 5D black hole falls in the non-universal regime, which is sensitive to the distribution of polar states in the modular form.

The final application of our analysis comprises the microscopic nature of the entropy of 1/2-BPS black holes in 4D $\mathcal N = 2$ supergravity. The leading and logarithmic contribution to the entropy of these black holes have been argued to be compatible with the OSV formula in \cite{Sen:2012kpz}, and there are components of this formula that deal with the asymptotic behavior of a wJf \cite{Gaiotto:2006ns, deBoer:2006vg,Denef:2007vg,Gomes:2019vgy}. As we will see, various subtleties make the comparison challenging because these black holes fall into the non-universal regime. In fact, the charges of the black hole are such that $\Delta$ is situated right at the edge of the regime of validity of our analysis, where the leading Cardy behavior receives order one corrections. Additionally, in the non-universal regime, the logarithmic corrections are sensitive to the distribution of polar states, meaning that a match with gravity requires precise knowledge of the light spectrum. Using our methods, we illustrate how to reproduce the logarithmic corrections to the black hole entropy and highlight the delicate components of the match. 

This paper is organized as follows. In Sec.\,\ref{sec:2} we review the properties of wJfs, obtain general expressions for their Fourier coefficients using crossing kernels, and compare these to the expressions obtained from the Rademacher expansion. In Sec.\,\ref{sec:beyondcardy} we consider wJfs with a sparse spectrum of light states and quantify the imprint of the light states on the asymptotic growth of the $c(n, \ell)$ coefficients. Therein we derive expressions for $c(n,\ell)$ that are valid in the the universal and non-universal regimes characterized by different scalings of $\Delta$ with $\dmina$. In Sec.\,\ref{sec:symmN} we study the asymptotic growth of wJfs obtained from symmetric product orbifolds and compare the resulting $c(n, \ell)$ coefficients to the ones obtained from the crossing kernel and the exponential lift. The expressions for the $c(n, \ell)$ coefficients derived in Sec.\,\ref{sec:beyondcardy} and Sec.\,\ref{sec:symmN} are tested in Sec.\,\ref{sec:bhs}, where they are shown to reproduce Bekenstein-Hawking entropy and its logarithmic corrections in 1/4-BPS black holes in $\mathcal N =4$ supergravity in four and five dimensions. Therein we also revisit the logarithmic corrections to 1/2-BPS black holes in four-dimensional $\mathcal{N}=2$ supergravity. We conclude with a discussion in Sec.\,\ref{sec:disc}. Various results are collected in the appendices. App.\,\ref{app:conventions} describes our conventions for the different approximations used in the paper. App.\,\ref{app:kernels} contains more details on the derivation of the crossing kernel of wJfs with nonzero weight. Some important features (including the crossing kernel) of wJfs with nonzero weight and multiple charges are discussed in App.\,\ref{app:multiplecharges}. In App.\,\ref{app:Rademacher} we write down the Rademacher expansion of wJfs. Finally, in App.\,\ref{app:SMF} we provide more details on the relationship between exponential lifts and the generating function of symmetric product orbifolds.

\section{Asymptotic expansions of sparse weak Jacobi forms}\label{sec:2}

In this section we introduce the basic ingredients necessary to extract the Fourier coefficients of weak Jacobi forms (wJf). The method we use to extract these coefficients is the crossing kernel, which allows us to explore different regimes of parameter space. We then compare this method with the more familiar Rademacher expansion. We also review the well-known derivation of the Cardy-like growth of the Fourier coefficients of wJfs, and introduce the notion of a sparse wJf.

\subsection{Weak Jacobi forms}

The basic object we want to consider is a weak Jacobi form. These forms are labeled by two integers: a weight $k$ and an index $t>0$. A Jacobi form is a function $\varphi(\tau,z)$ defined on $\mathbb{H}\times \mathbb{C} \rightarrow \mathbb{C}$ with a Fourier expansion given by 
\begin{equation}\label{eq:defn-wjf}
    \varphi(\tau,z) = 
    \sum_{n,\ell} c(n,\ell)  q^{n}y^{\ell}\,, \qquad q \coloneqq e^{2\pi i \tau}\,,
    \qquad 
    y \coloneqq \e^{2\pi i z}\,.
\end{equation}
Crucially, the adjective ``weak'' in a wJf means that  
\begin{equation}
    c(n,\ell)=0\,, \quad \text{unless} \quad n\geq0\,. \label{ccondition}
\end{equation}
For integral values of $\ell$ and $n$, $\varphi(\tau,z)$ satisfies the transformation rules
\begin{equation}
\label{eq:modInv}
    \varphi
    \left( \frac{a\tau + b}{c\tau +d} , \frac{z}{c\tau +d} \right) 
    = 
    (c\tau+d)^{k}
    \e^{  \frac{2\pi i t c z^2}{c\tau +d}  }  \varphi(\tau,z)\,,
 \qquad
    \begin{pmatrix}
    a & b \\
    c & d
    \end{pmatrix}
    \in \text{SL}(2,\mathbb{Z})\,.
\end{equation}
In particular, the $S$ and $T$ transformations of $\text{SL}(2,\mathbb{Z})$ correspond to $(a,b,c,d)=(0,-1,1,0)$ and $(a,b,c,d)=(1,1,0,1)$, respectively. In addition, $\varphi(\tau, z)$ transforms under shifts of $z$ as\footnote{Our results also apply to cases where $\ell$ is fractional. In these cases there is an ``unwrapping'' of the potential $z$ that makes $\ell$ an integer, see e.g.~\cite{Belin:2019rba}.}  
\begin{equation}
\label{eq:specFlow}
    \varphi(\tau, z + \lambda \tau + \mu) 
    =
   \e^{  -2 \pi i t (\lambda^2\tau+2\lambda z +\mu )  }
    \varphi(\tau,z)\,, \qquad \lambda,\mu \in \mathbb{Z}\,.
\end{equation}
This transformation is known as an elliptic (or spectral flow) transformation. A simple consequence of \eqref{eq:modInv} is that under the modular transformation $(ST)^3$, the $c(n, \ell)$ coefficients satisfy
\eq{
c(n, \ell) = (-1)^k c(n, -\ell)\,. \label{cneq}
}
Additionally, \eqref{eq:specFlow} implies that 
\begin{equation}
\label{eq:speccoef}
    c(n,\ell) =  c (n+\lambda \ell +t \lambda^2,\ell + 2\lambda t )\,, \qquad \lambda \in \mathbb{Z}\,. 
\end{equation}

There are two important properties of wJfs to keep in mind. First, the transformations \eqref{eq:modInv} and \eqref{eq:specFlow} imply that the $c(n,\ell)$ coefficients depend only on $\ell$ (mod $2 t$) and the \emph{discriminant} $\Delta$ defined by
\begin{equation}\label{eq:delta}
   \Delta \coloneqq n - \ell^2/(4t)\,.
\end{equation}
In this context, we will differentiate between two classes of contributions to \eqref{eq:defn-wjf}: for given values of $(n,\ell)$ we will denote the corresponding states as a
\begin{equation}
    \begin{aligned}
        \text{{\it polar}  state if} \quad \Delta < 0\,, \\
               \text{{\it non-polar} state if} \quad  \Delta \geq 0\,. 
    \end{aligned}
\end{equation}
Second, the discriminant of polar states is bounded from below by the index $t$ of $\varphi(\tau,z)$ such that $\dmin \geq -t/4$, where  $\dmin$ is the minimum value of the discriminant, i.e.~the discriminant of the most polar term. We are interested in wJfs where the most polar term is of the form 
\begin{equation}\label{eq:most-polar}
    q^0 y^{\pm b}\,,
\end{equation}
such that $\dmin = -b^2/(4t)$ where $b$ is taken to be a positive integer with $b\leq t$.

The notion of a wJf can be generalized to an arbitrary rank $M$, where we would instead have  $\varphi: \mathbb{H}\times \mathbb{C}^M \rightarrow \mathbb{C}$. Physically, this means that we have a system with multiple charges $\ell^i$, $i=1,\ldots, M$. Many of the properties of these modular forms are straightforward generalizations of the wJfs discussed in this section. There are however some surprising differences, which will prove to be crucial in Sec.\,\ref{sec:n2bhs}. In particular, the relationship between the index and the maximal polarity of the form can be modified. We discuss wJfs with multiple charges in detail in App.\,\ref{app:multiplecharges}.

\subsection{Crossing kernels}
\label{sec:kernel}

In this section we derive a general formula for the $c(n,\ell)$ coefficients of wJfs using crossing kernels. This formula follows naturally from the standard statistical mechanics approach of extracting the Fourier coefficients using an inverse Laplace transform; see for example \cite{Maloney:2007ud,McGough:2013gka,Keller:2014xba,Benjamin:2016fhe} for an application to non-holomorphic modular forms. It is important to note, however, that this approach does not give an exact count for these coefficients because it approximates the numbers $c(n,\ell)$ with a density (or distribution) that we denote by $\rho({\sf n},{\sf j})$, where $\sf n$ and $\sf j$ are continuous parameters.\footnote{Note that $\rho({\sf n},{\sf j})$ is not strictly a density since the $c(n,\ell)$ coefficients are not necessarily positive. A precise definition of $\rho({\sf n},{\sf j})$ and its relation to $c(n,\ell)$ is given below.} Nevertheless, this approach is conceptually much simpler than the exact methods presented in Sec.\,\ref{sec:rademacher} and Sec.\,\ref{sec:symmN}, and it suffices to capture the leading and subleading corrections to $c(n, \ell)$ when $\Delta$ is large. For simplicity, we only consider weight-zero forms of rank one in this section, deferring the generalization of our results to forms with arbitrary weight and arbitrary rank to App.\,\ref{app:kernels} and \ref{app:multiplecharges}, respectively.

We begin by writing the wJf $\varphi(\tau, z)$ as an integral over the density of states, 
\begin{equation}
\label{eq:wJf}
    \varphi(\tau, z) = 
    \int_0^{\infty} \dd {\sf n} 
    \int_{-\infty}^{\infty} \dd {\sf j} \,
    \rho({\sf n},{\sf j})\, 
    \e^{2\pi i ( \tau {\sf n} + z {\sf j} )}\,,
\end{equation}
where $\rho({\sf n}, {\sf j})$ is given by a sum of delta functions centered at integer values of ${\sf n}=n$ and ${\sf j}=\ell$. For a weight-zero form, invariance under modular $S$ transformations implies
\begin{equation}
\label{eq:SwJf}
    \varphi(\tau, z) = 
    \int_0^{\infty} \dd {\sf n} 
    \int_{-\infty}^{\infty} \dd {\sf j} 
    \,
    \rho({\sf n},{\sf j}) \,
    \e^{\frac{2\pi i}{\tau} \left(- {\sf n} + z {\sf j} -  t z^2 \right)}\,.
\end{equation}
Using an inverse Laplace transform then leads to the \emph{crossing equation} 
\begin{align}
\rho({\sf n},{\sf j}) 
\label{eq:crosseq2}
    = \int_0^{\infty} \dd {\sf n'} 
    \int_{-\infty}^{\infty} \dd {\sf j'} 
    \, \rho({\sf n'},{\sf j'})
    \,\knl{P}{{\sf n},\,{\sf j}}{{\sf n'},\,{\sf j'}}\,,
\end{align}
where $\knl{P}{{\sf n},\,{\sf j}}{{\sf n'},\,{\sf j'}}$ is the \emph{crossing kernel} defined by
\begin{equation}
\label{def:cross-ker0}
    \knl{P}{\nn,\,\jj}{\nn',\jj'} \coloneqq \int\dd \tau \, \dd z \,\e^{-2\pi i(\tau \nn+z \jj)}\e^{\frac{2\pi i}{\tau}\left(-\nn'+ z \jj' - t z^2\right)}\,.
\end{equation} 
Since the density of states $\rho({\sf n},{\sf j})$ is a sum of delta functions, we can rewrite the integral in \eqref{eq:crosseq2} as a discrete sum over the spectrum of $\varphi(\tau, z)$, with the result
\begin{equation}
\label{eq:crosseq3}
    \rho({\sf n},{\sf j}) = \sum_{\subalign{
    \ell'&\in\mathbb{Z} \\n'&\geq0 }} c(n',\ell') \knl{P}{{\sf n},\,{\sf j}}{n',\ell'}\,.
\end{equation}
This equation is an exact relation between distributions and it does not converge in the usual sense. Since the left-hand side of \eqref{eq:crosseq3} is a sum of delta functions, the sum on the right-hand side does not have a smooth asymptotic behavior. Rather, this equation only holds when integrated against a valid test function, which requires some smearing. The level of smearing required for \eqref{eq:crosseq3} to be valid depends on the details of the modular form at hand. For wJfs, which have a discrete and integer-spaced spectrum, we need an integration window that is presumably much larger than one. The most conservative statement we can make about \eqref{eq:crosseq3} is that the total number of states below a certain value of $n$ and $\ell$ is asymptotic to the integral of the sum on its right-hand side. In Sec.\,\ref{sec:rademacher}, we will review the exact Rademacher expansion and contrast it against the distribution $\rho({\sf n},{\sf j})$.  

The crossing kernel $\knl{P}{{\sf n},{\sf j}}{{\sf n'},{\sf j'}}$ for a weight-zero wJf is given by (see App.\,\ref{app:kernels}) 
\begin{equation}
\label{eq:kernel0}
    \knl{P}{{\sf n},{\sf j}}{{\sf n'},{\sf j'}} \!\coloneqq \e^{-i\pi\frac{ {\sf j'}{\sf j}}{t}} \knl{P}{\Delta}{\Delta'} \,,
\end{equation}
where $\mathbb{P}_{\{\Delta\};\{\Delta'\}}$ is defined, in terms of $\Delta= {\sf n}-{\sf j}^2/(4t)$ (and similarly for $\Delta'$), by
\begin{equation}\label{eq:kernel}
    \quad\,\, \knl{P}{\Delta}{\Delta'} \!=
    \sqrt{\frac{2\pi^2}{t}} 
    \left(
     -\frac{\Delta'}{\Delta}
    \right)^{\frac{3}{4}} 
    I_{-\frac{3}{2}}
    \big(4\pi \sqrt{-\Delta \Delta'}\big)
    \Theta(\Delta)\,.
\end{equation} 
In the equation above, $\Theta(\Delta)$ is the Heaviside step function and $I_\nu(z)$ is a modified Bessel function of the first kind. Note that up to the ${\sf j'j}$-dependent phase, the crossing kernel depends only on the discriminants $\Delta$ and $\Delta'$. For this reason, it is convenient to rearrange the sum in \eqref{eq:crosseq3} in terms of $\Delta'$ and ${\ell'}$. Furthermore, since $c(n, \ell)$ depends only on $\Delta$ and $\ell\,(\text{mod}\,2t)$, we can use \eqref{eq:speccoef} to write\footnote{The sum over delta functions in the last line of \eqref{eq:finalestimate} explicitly shows that $\rho({\sf n},{\sf j})$ is a distribution.}
\eqsp{
     \rho({\sf n},{\sf j}) &= 
     \sum_{\Delta'} 
     \knl{P}{\Delta}{\Delta'} 
     \sum_{\ell' = -t}^{t-1}
     \sum_{\lambda \in \mathbb{Z}}
     c\left(\Delta' + \frac{(\ell' + 2\lambda t)^2}{4t}, \ell' +2 \lambda t \right)
     \e^{-i\pi\frac{ (\ell' +2 \lambda t){\sf j}}{t}} \\
    \label{eq:finalestimate}
    &= 
    \sum_{\Delta'}  
    \knl{P}{\Delta}{\Delta'} 
    \sum_{\ell' =-t}^{t-1}
    c\left(\Delta'+ \frac{(\ell')^2}{4t} , \ell' \right)
    \e^{-i\pi\frac{ \ell'{\sf j}}{t}} \sum_{\lambda \in \mathbb{Z}} \e^{-2\pi i\lambda {\sf j}}\\
    &= (-1)^{{\sf j}} \sum_{\Delta'}  
    \knl{P}{\Delta}{\Delta'} 
    c_{\sf j} (\Delta' )
    \sum_{m\in\mathbb{Z}} \delta(m-{\sf j})\,,
}
where in the last line we used the Fourier transform of the Dirac comb and introduced the notation\footnote{When $t=b$, the sum in the definition of $c_\ell(\Delta')$ collapses to a single term. The factor of $(-1)^{\ell}$ is included in this definition so that ${c}_t(\dmin) = c\left(0,t\right)$.}
\begin{equation}\label{eq:c1}
{c}_\ell(\Delta')\coloneqq 
    (-1)^\ell \sum_{\ell' =-t}^{t-1}
    c\left(\Delta'+ \frac{(\ell')^2}{4t} , \ell' \right)
    \e^{-i\pi\frac{ \ell'\ell}{t}}\,.
\end{equation}
In general, very few terms tend to contribute to \eqref{eq:c1} since the arguments of $c(n,\ell)$ have to be integral. Note also that $c(\Delta' + \ell'{}^2/(4t), \ell')$ vanishes unless $\Delta' + \ell'{}^2/(4t)$ is a nonnegative integer.

We are ultimately interested in determining the asymptotic density of states for large values of the discriminant $\Delta$. In this case, $\rho({\sf n}, {\sf j})$ receives qualitatively different contributions from different types of states. This can be seen from the large-$\Delta$ behavior of the kernel
\begin{equation} \label{eq:kernelexp}
  \knl{P}{\Delta}{\Delta'} = \frac{1}{2\Delta} \sqrt{- \frac{\Delta'}{t}} e^{4\pi \sqrt{- \Delta \Delta'}}+\cdots\, , \qquad \Re\big(\sqrt{-  \Delta\Delta'}\big)\rightarrow \infty\,.
\end{equation}
For polar (light) states where $\Delta'$ is negative, the kernel is a smooth exponential function of $\sqrt{|\Delta'|}$, while it becomes a rapidly oscillating function for non-polar (heavy) states where $\Delta'$ is positive. Hence, it is natural to write the density of states  as
\begin{equation}
\label{eq:deco}
    \rho({\sf n},{\sf j}) = \big(
    \rho_{\polar}({\sf n},{\sf j})
    +
    \rho_{\npolar}({\sf n},{\sf j})
    \big)\times
\sum_{m\in\mathbb{Z}} \delta(m-{\sf j})
\,,
\end{equation}
where $\rho_{\polar}({\sf n},{\sf j})$ and $\rho_{\npolar}({\sf n},{\sf j})$ denote the contributions of the light and heavy states, respectively, that is
\eqa{
\label{eq:c-asymp}
    \rho_{\polar}({\sf n},{\sf j}) &\coloneqq (-1)^{\sf j} \sum_{\Delta'<0}  
    \knl{P}{\Delta}{\Delta'} 
    c_{\sf j} (\Delta' ) \,, \\
    \rho_{\npolar}({\sf n},{\sf j}) & \coloneqq  (-1)^{\sf j}\sum_{\Delta'\geq 0}  
    \knl{P}{\Delta}{\Delta'} 
    c_{\sf j} (\Delta' ) \,.
}
As described above, $\rho_{\polar}({\sf n},{\sf j})$ is responsible for the exponential growth of the density $\rho({\sf n}, {\sf j})$ at large $\Delta$, while $\rho_{\npolar}({\sf n},{\sf j})$ is responsible for its oscillatory behavior. We have not included the spin quantization condition in these expressions as they will be directly related to the $c(n,\ell)$ coefficients in the next section. Note that the delta functions corresponding to the spin quantization appear explicitly in \eqref{eq:deco} as a consequence of spectral flow invariance. On the other hand, the quantization of ${\sf n}$ is nontrivially encoded in the infinite sum of oscillating terms in $\rho_{\npolar}({\sf n}, {\sf j})$.

\subsection{Rademacher expansions}\label{sec:rademacher}

Let us now relate the light density $\rho_{\polar}({\sf n},{\sf j})$ to the coefficients $c(n,\ell)$ via the Rademacher expansion. The Rademacher expansion is an exact formula for the coefficients $c(n,\ell)$ of a Jacobi form of weight $k \leq 1/2$ and index $t$ that is valid for states with a positive discriminant. For wJfs with zero weight, the Rademacher expansion reads\footnote{We will follow the conventions and definitions of \cite{eichler2013theory} (see also \cite{Gomes:2017bpi,Dijkgraaf:2000fq}). For completeness, the Rademacher expansion of wJfs of weight $k \leq 1/2$ is given in App.\,\ref{app:Rademacher}.}
\begin{equation}
\label{eq:rademacher}
    c(n,\ell) = 
    \sum_{\Delta'<0}
    \sum_{\ell'=-t}^{t-1}
    c(n',\ell')
    \sum_{\radc=1}^{\infty}
    \frac{2\pi
    }{\radc}
    \left(
    -\frac{\Delta'}{\Delta}
    \right)^{\frac{3}{4}}
    I_{\frac{3}{2}}\left(\frac{4\pi}{\radc}\sqrt{-\Delta \Delta'}\right)
    \text{Kl}(\Delta,\ell,\Delta', \ell';\radc)\,,
\end{equation}
where the summation variable ``$\radc$" is traditional and should not be confused with the central charge of a CFT. The function $\text{Kl}(\Delta,\ell,\Delta', \ell';\radc)$ is known as a generalized Kloosterman sum. Kloosterman sums are finite sums over complex phases that are bounded by $\radc$ and are multiplied by an overall factor of  $\sqrt{1/t}$ \cite{Dijkgraaf:2000fq}. The asymptotic behavior of the $c(n, \ell)$ coefficients turns out to be sensitive only to the $\radc = 1$ term in \eqref{eq:rademacher}, which is given by
\begin{equation}
    \text{Kl}(\Delta,\ell,\Delta',\ell', 1) = \sqrt{\frac{1}{2t}}\e^{-i\pi \frac{\ell \ell'}{t}}\,.
\end{equation}
A general definition of $\text{Kl}(\Delta,\ell,\Delta',\ell', \radc)$ and some of its properties are described in App.\,\ref{app:Rademacher}. 

There are a few small but important differences between the expressions for $\rho({\sf n}, {\sf j})$ and $c(n, \ell)$ obtained from the crossing kernel \eqref{eq:finalestimate} and the Rademacher expansion \eqref{eq:rademacher}. First, the Bessel functions featured in these expressions have different weights, $-3/2$ and $3/2$, respectively.\footnote{The discrepancy in the weight of $I_{-3/2}(x)$ and $I_{3/2}(x)$ can be understood as the result of a choice of integration contour (see App.\,\ref{app:kernels} and \ref{app:Rademacher} for details). Nevertheless, these functions are related via the modified Bessel function $K_{3/2}(x)$ and their difference is non-perturbatively small when the argument is large 
\begin{align*}
    I_{3/2}(x)-I_{-3/2}(x) = \frac{2}{\pi} K_{3/2}(x) \sim \sqrt{\frac{2}{\pi x}}  \e^{-x}\Big(1 + \order{x^{-1}}\Big)\,.
\end{align*}
} 
Second, \eqref{eq:finalestimate} is a sum over the entire spectrum of the wJf while \eqref{eq:rademacher} only sums over polar terms with spin restricted to the range $\ell' \in [ -t  , t-1]$. Finally, the crossing equation does not include a sum over the discrete variable $\radc$ featured in the Kloosterman sum and in fact only captures the behavior of the first term with $\radc = 1$.

The differences between the density $\rho({\sf n}, {\sf j})$ and the exact coefficients $c(n, \ell)$ described above are ``small'' for our purposes, as they correspond to non-perturbative, exponentially suppressed corrections. To make this comparison precise, let us define the ``Rademacher" kernel $\knl{R}{n,\ell}{n',\ell'}$ by
\begin{equation} \label{eq:Rkerneldef}
   \knl{R}{n,\ell}{n',\ell'} \coloneqq
   \sum_{\radc=1}^{\infty}
    \frac{2\pi
    }{\radc}
    \left(
    -\frac{\Delta'}{\Delta}
    \right)^{\frac{3}{4}}
    I_{\frac{3}{2}}\left(\frac{4\pi}{\radc}\sqrt{-\Delta \Delta'}\right)
    \text{Kl}(\Delta,\ell,\Delta', \ell';\radc)\,,
\end{equation}
such that the $c(n, \ell)$ coefficients can be written as
\begin{equation}
\label{eq:expofc0}
    c(n,\ell) = 
    \sum_{\Delta'<0}\sum_{\ell' = -t}^{t-1} c(n',\ell' ) \knl{R}{n,\ell}{n',\ell'}\,.
\end{equation}
For large values of $|\Delta \Delta'|$, the Rademacher and crossing kernels agree up to non-perturbative corrections that originate from the $\radc > 1$ terms in the Kloosterman sum, that is
\begin{equation}\label{eq:rkernel}
    \frac{\knl{R}{n,\ell}{n',\ell'}}{
    \knl{P}{n,\ell}{n',\ell'}
    } = 1+\order{ \e^{-2\pi \sqrt{-\Delta \Delta'}}}\,.
\end{equation}
Using the asymptotic expansion of the Bessel function we can write
\begin{equation}
\label{eq:expofc}
    c(n,\ell) =(-1)^{\ell}\sum_{\Delta'<0} c_{\ell}(\Delta') 
    \frac{1}{2\Delta}\sqrt{-\frac{\Delta'}{t}} \e^{4\pi \sqrt{- \Delta \Delta'}}\left[1+\order{\e^{-2\pi \sqrt{-\Delta \Delta'}}}\right]\,,
\end{equation}
where $c_{\ell}(\Delta')$ is defined in \eqref{eq:c1} and we have omitted terms with $\radc > 1$ that are responsible for the leading corrections to this expression. Comparing \eqref{eq:c-asymp} and \eqref{eq:expofc} then yields
\begin{equation}\label{eq:crho1}
 \frac{c(n,\ell)}{\rho_{\polar}(n,\ell)}
 =1+ \order{\e^{-2\pi \sqrt{-\Delta \Delta'}}}\,.
\end{equation}
We conclude that the $c(n,\ell)$ coefficients are well approximated by the density $\rho_{\polar}(\mathsf{n,j})$. In other words, we can use crossing kernels to extract the leading and subleading behavior of the $c(n, \ell)$ coefficients  when the product $|\Delta \Delta'|$ is large.

\subsection{Asymptotics of sparse weak Jacobi forms}
\label{sec:2.3}

We now describe the leading asymptotic behavior of $\rho_{\polar}(n,\ell)$ when the wJf has a sparse spectrum and a large value of the minimum discriminant $\dmin$. These features are motivated by the AdS$_3$/CFT$_2$ correspondence and control the leading universal behavior of the $c(n, \ell)$ coefficients of wJfs.

Let us first consider the asymptotic behavior of generic wJfs in the regime $\Delta \gg |\dmin|$. In this case, the density \eqref{eq:c-asymp}, which receives contributions from states with $\Delta' < 0$, is dominated by the most polar term with $\Delta' = \dmin$. This can be seen from the crossing kernel, which organizes the spectrum into an exponential hierarchy
\begin{equation}
\label{eq:exphier}
    \frac{
    \knl{P}{\Delta}{\Delta'}
    }{\knl{P}{\Delta}{\dmin}} 
    =\sqrt\frac{\Delta'}{\dmin}
    \e^{ -4\pi \sqrt{\Delta} \left(\sqrt{|\dmin|} -\sqrt{|\Delta'|}\right)
    } + \cdots \,.
\end{equation}  
Consequently, for fixed values of $\dmin$ (or $t$), the coefficient $c(n, \ell)$ with discriminant $\Delta \gg |\dmin|$ is universally given by
\begin{equation} \label{eq:dom-polar0}
    c(n,{\ell}) = \rho_{\polar}(n,\ell) +\cdots = (-1)^\ell {c}_\ell(\dmin)\sqrt{
    \frac{2\pi^2}{t}
    }\left(
    \frac{\dmina}{\Delta}
    \right)^{\frac{3}{4}}
    I_{-\frac{3}{2}}\left(4\pi \sqrt{\Delta \dmina }\right)+ \cdots \,.
\end{equation}
Ignoring the subleading (non-exponential) terms, \eqref{eq:dom-polar0} yields the expected leading behavior of $c(n, \ell)$, namely
\begin{equation}
    c(n,{\ell})  \sim  \e^{  4\pi \sqrt{ \Delta \dmina}  }\,, \qquad \Delta \gg |\dmin|\,. \label{eq:dom-polar}
\end{equation}
It is important to note that as $\Delta$ approaches $|\dmin|$, the corrections to \eqref{eq:dom-polar} become relevant. The effect of these corrections to the leading behavior \eqref{eq:dom-polar} is discussed in the next section. 

It is instructive to compare \eqref{eq:dom-polar} with the ordinary Cardy formula in CFT$_2$, $S_{\text{Cardy}} \sim 2\pi\sqrt{\frac{c}{3} E}$, where $c$ is the central charge of the CFT and $E$ is the energy of the state. We see that \eqref{eq:dom-polar} reproduces the Cardy formula provided that we identify
\begin{equation} \label{eq:dict}
    \dmin \rightarrow -\frac{c}{12}\,, \qquad  \Delta \rightarrow E\,.
\end{equation}
Note that the dictionary \eqref{eq:dict} depends on the precise relationship between $\varphi(\tau,z)$ and the torus partition function or index of the CFT. Nevertheless, it is natural to refer to the regime of validity of \eqref{eq:dom-polar} as the \emph{Cardy regime}. 

Following \cite{Hartman:2014oaa,Benjamin:2015hsa}, let us now introduce the notion of a \emph{sparse weak Jacobi form}. We say that a wJf is sparse if the Fourier coefficients of its polar states satisfy the HKS bound, namely \cite{Hartman:2014oaa,Benjamin:2015hsa}
\begin{equation}\label{eq:sparse}
    c(n, \ell) \lesssim \e^{2\pi (\Delta-\dmin)}\,, \qquad \dmin<\Delta<0\,.
\end{equation}
This is usually referred to as the sparseness condition. An immediate consequence of \eqref{eq:sparse} is that the asymptotic growth \eqref{eq:dom-polar} can be extended to the so-called \emph{universal regime} where
\begin{equation}
\label{eq:ext-cardy}
    c(n, \ell) \sim \e^{  4\pi \sqrt{ \Delta \dmina}  }\,, \qquad  \Delta \gtrsim |\dmin|\gg1 \,.
\end{equation}
In the context of AdS$_3$/CFT$_2$, this result is compatible with holographic conditions on the spectrum. In particular, from a gravitational perspective, we interpret \eqref{eq:ext-cardy} as the appearance of BTZ black holes that dominate the spectrum at high energies.\footnote{Colloquially, we identify non-polar states with black holes, and polar states with ``light'' or perturbative states. However, this is just a rough analogy since the precise statements depend on the gravitational theory.} 

To summarize, the asymptotic growth of wJfs is given by a universal Cardy formula that is determined by the most polar state in the spectrum. Depending on the distribution of the polar states, this formula is valid in the Cardy regime \eqref{eq:dom-polar} or the universal regime \eqref{eq:ext-cardy}. So far, we have only discussed the leading order behavior of the $c(n,\ell)$ coefficients. The subleading corrections to these coefficients, and their sensitivity to other polar states in the spectrum, are the subject of the next section.

\section{Light state imprint on the asymptotics}
\label{sec:beyondcardy}

We now consider the subleading corrections to the Fourier coefficients of sparse wJfs. We begin by refining the concept of sparseness and distinguish between wJfs with a fast and slow  growth of polar states. We will show that in the universal regime where $\Delta \gtrsim |\dmin| \gg 1$, the corrections to the Fourier coefficients of any sparse wJf remain universal, being determined by the most polar state in the spectrum. In addition, we will show that slow-growing wJfs have an extended regime of validity of the Cardy formula that includes states with discriminant $|\dmin| \gtrsim \Delta \gg 1$. In this regime, the subleading corrections to the Fourier coefficients become sensitive to the distribution of light states and are no longer universal. 

\subsection{Fast vs slow growth}

Based on the arguments around \eqref{eq:crho1}, the $c(n, \ell)$ coefficients  of a wJf are approximated by the density \eqref{eq:c-asymp} when the discriminant is large. While the leading asymptotic behavior of these coefficients is determined by the most polar state, their subleading corrections are in principle sensitive to the distribution of light states. Let us model the light spectrum with the density $\rho_{\light}(\nn, \jj)$ defined as 
\begin{equation}
\label{eq:discden}
    \rho_{\light} (\nn , \jj) \coloneqq \Theta\left(  \frac{\jj^{\,2}}{4t}-\nn\right)\sum_{\ell = -t}^{t-1} \sum_{n = 0}^{\infty} \delta(n-\nn)\delta(\ell - \jj)\;c(n,\ell)\,. 
\end{equation}
In contrast to the full density of states $\rho(\nn, \jj)$, this density is restricted to the polar states and values of the spin $\jj \in [-t, t-1]$.\footnote{The range on the spin $\jj$ 
follows from \eqref{eq:speccoef} and is a consequence of invariance under spectral flow.} The density \eqref{eq:discden} is defined so that $\rho_{\polar}(\nn,\jj)$ in \eqref{eq:c-asymp} can be written as
\begin{equation}
\label{eq:intapprox2}
    \rho_{\polar}(\nn,\,\jj) 
    = 
    \int \dd \nn'\,  \dd \jj' \,
    \rho_{\light} (\nn',\jj' )\,
    \knl{P}{\nn,\,\jj}{\nn',\jj'}\,.
\end{equation}

We are interested in characterizing the behavior of the $c(n,\ell)$ coefficients of wJfs in a regime where  $\dmina\gg1$. For rank one wJfs, large $\dmina$  implies that $t \sim b \gg 1$. In order to quantify the effects of the light states on the asymptotic behavior of $c(n,\ell)$, it is necessary to refine the sparseness condition introduced in the previous section. Henceforth, we distinguish between sparse wJfs whose polar states grow fast or slow. These types of growth are characterized as follows:

\begin{itemize}[leftmargin=0.3cm]
\item[]{\bf Fast growth.} A sparse wJf is ``fast'' growing when the density of light states $\rho_{\light}(\nn,\jj)$ is Hagedorn-like, namely 
\begin{equation}\label{eq:fast}
    \rho_\light ( \nn,  \jj) \lesssim \e^{2\pi \gamma (\Delta-\dmin)}\,, \qquad 1\ll\Delta<\dmina\,, 
\end{equation}
where  $\gamma >0$. Generally $\gamma$ is a function of $\nn$ and $\jj$ that does not scale in the large $\Delta$, $|\Delta_0|$ limits. Note that we have broadened the sparseness condition \eqref{eq:sparse} by allowing $\gamma$ to be potentially greater than one. We will quantify the repercussions of this choice in the next subsection. A familiar case of fast growth is the Hagedorn growth that occurs when $\gamma = 1$. In the context of AdS$_3$/CFT$_2$, this growth is associated with a stringy spectrum on the bulk side of the correspondence \cite{Keller:2011xi}.

\item[]{\bf Slow growth.} A sparse wJf is “slow” growing when $\rho_{\light}(\nn,\jj)$ is well below the HKS bound. We can parametrize this growth by the behavior of $\rho_{\light}$ under a rescaling of the charges by a parameter $\Lambda$ that we take to be large. More precisely, we consider
\begin{equation}\label{eq:scaling}
    \nn \mapsto \Lambda \nn\,, \qquad
    \hj  \mapsto \Lambda \hj\,,
\end{equation}
where $\hj\coloneqq \jj+b$ is the spin shifted by the spin of the ground state. In terms of this scaling, slow-growing wJfs satisfy
\begin{equation}
\label{eq:light-spec}
    \rho_\light (\nn, \jj) \lesssim \e^{f(\nn,\,\hj)}, \quad \text{where}\quad f\big(\Lambda\, \nn,\Lambda\,\hj\big) \lesssim \Lambda^\alpha f(\nn,\hj).
\end{equation}
Here $\alpha<1$, and we take $\Lambda$ to be parametrically larger than one but less than $|\Delta_0|$. The condition on $\alpha$ is relevant in the context of holography as it makes the growth of light states compatible with that of a weakly interacting quantum field theory in, for example, AdS$_3\times M_{D-3}$.\footnote{The perturbative spectrum of a weakly coupled QFT on AdS$_3\times M_{D-3}$ has $\alpha= \frac{D-1}{D}<1$.}
\end{itemize}

As illustrated in \cite{Benjamin:2015hsa,Benjamin:2015vkc}, it is difficult to construct wJfs with a slow growth of polar states; fast-growing light states seem to be the norm. One arena where it is possible to design wJfs with either slow or fast growth is in the context of symmetric product orbifolds \cite{Belin:2018oza,Belin:2019rba,Belin:2020nmp,Benjamin:2022jin}. We will use this class of wJfs as testing grounds for our analysis in Sec.\,\ref{sec:symmN}. 

\subsection[Universal behavior for \texorpdfstring{$\Delta \gtrsim |\dmin|$}{delta>delta0}]{Universal behavior for $\mathbf{\Delta \gtrsim |\dmin|}$}\label{sec:fast}

Let us now consider the behavior of $\rho_{\polar}(\nn,\,\jj)$ for sparse wJfs in the regime $\Delta \gtrsim |\dmin| \gg 1$, where the following ratios are held fixed
\begin{equation}
\label{eq:limit1}
\frac{\Delta}{|\dmin|}\,, \qquad \frac{\jj}{|\dmin|}\,.
\end{equation} 
Following \eqref{eq:crho1}, the asymptotic behavior of the $c(n, \ell)$ coefficients is determined by the light density of states via \eqref{eq:expofc}. In the regime $\Delta \gtrsim |\dmin|$, the integral in \eqref{eq:intapprox2} is dominated by light states that lie well below the threshold $\Delta ' = 0$ of non-polar states due to HKS. For this reason, we will assume that
\begin{equation}
\label{eq:condition}
    \nn' \ll b\,, \qquad \jj' = \hj' -b \,, \qquad |\hj'| \ll b\,,
\end{equation}
where we recall that $b$ parametrizes the most polar state of $\varphi(\tau, z)$ via $(n, \ell) = (0, \pm b)$ such that $\dmin = - b^2/4t$. The limit $|\dmin|\gg1$ implies that  $1\ll 2\sqrt{t}\ll b \leq t$.

With this hierarchy, the crossing kernel \eqref{eq:kernel0} can be approximated by 
\begin{align}
\label{eq:kerexp}
    \knl{P}{\nn,\,\jj}{\nn',\hj'-b}  = \frac{1}{2\Delta}
    \sqrt{\frac{\dmina}{t}}
    \e^{\pi i \frac{ b\,\jj}{t}}\e^{4\pi \sqrt{\Delta\dmina}}
    \e^{
    2\pi i \tau_\star \nn' 
    + 2\pi i z_{\star} \hj'}+\cdots\,, 
\end{align}
where we used \eqref{eq:kernelexp} with the exponential term expanded to linear order in $\nn'$ and $\hj'$, while $\tau_{\star}$ and $z_{\star}$ are defined as
\begin{equation} \label{eq:potentials}
    \tau_{\star} \coloneqq  i\sqrt{\frac{\Delta}{|\dmin|}} \,, \qquad z_{\star} \coloneqq -\frac{\jj}{2t} + i \frac{b}{2t} \sqrt{\frac{\Delta}{|\dmin|}} = -\frac{\jj}{2t} + \frac{b}{2t}\tau_{\star}\,.
\end{equation}
The reason for this choice of variables becomes evident once we substitute \eqref{eq:kerexp} in \eqref{eq:intapprox2}. Indeed, we find that \eqref{eq:intapprox2} can be written as
\begin{equation}
\label{eq:cardycorrdef}
    \rho_{\polar}(\nn,\,\jj)   =
    \frac{1}{2\Delta}
    \sqrt{\frac{|\dmin|}{t}}
    \e^{\pi i \frac{ b\,\jj}{t}}\e^{4\pi \sqrt{\Delta \dmina}}
    \varphi_0(\tau_{\star},z_{\star})+\cdots\,,
\end{equation}
where $\varphi_0(\tau,z)$ can be interpreted as a partition function for the light states defined by
\begin{equation}
\label{eq:cardycorr}
   \varphi_0(\tau,z)
    \coloneqq \int_0^{\infty} \dd\nn' \int_{b-t}^{b+t-1}\dd \hj' \, \rho_{\light}(\nn',\hj' - b)\;  \e^{
    2\pi i \tau \nn'
    + 2\pi i z \hj'}\,.
\end{equation}
Note that in this expression, $\rho_{\light}(\nn', \hj' - b)$ only has support on states with $\dmin\leq\Delta'<0$. 

Equation \eqref{eq:cardycorrdef} provides insight into the factors that control the corrections to $\rho_{\polar}(\nn,\,\jj)$. In particular, we see that the subleading corrections to $\rho_{\polar}(\nn,\,\jj)$ are controlled entirely by $\varphi_0(\tau_\star,z_\star)$. Importantly, $\nn$ and $\jj$ only enter $\varphi_0(\tau_\star,z_\star)$ via the potentials \eqref{eq:potentials}, which remain fixed in the regime $\Delta \gtrsim |\dmin| \gg 1$ as they only depend on the ratio $\Delta/|\Delta_0|$. Namely, under the rescaling 
\begin{equation}
    \Delta\mapsto \Lambda \Delta\,, \qquad \dmin\mapsto \Lambda \dmin\,,
\end{equation}
with $\Lambda\gg1$, the partition function  $\varphi_0(\tau_{\star},z_{\star})$  simply contributes a numerical factor to \eqref{eq:cardycorrdef}. In order to sharpen this picture, we will now consider \eqref{eq:cardycorr} more carefully. 
 
In the regime where the ratios \eqref{eq:limit1} are held fixed, the parameters $\tau_\star$ and $z_\star$ are fixed numbers of order one. What we would like to argue is that $\varphi_0(\tau_{\star},z_{\star})$ exhibits no pathologies when the potentials are finite. It suffices to show that this partition function is finite when the range of integration is extended to infinity, that is,
\begin{equation}
    |\varphi_0(\tau_\star,z_\star)| \lesssim \int_0^{\infty} \dd \nn' \int_{-\frac{2t}{b}\nn'}^{\infty}\dd \hj' \, \rho_\light(\nn', \hj'-b)  \,
    \e^{2\pi i \tau_\star \nn' 
    +2\pi i z_\star \hj'}\,,
\end{equation}
where the lower limit of integration of the spin variable $\hj'$ comes from the requirement $\Delta' \ge \dmin$ in the limit $|\dmin|\gg1$. The integrand above is bounded by the sparseness condition, such that the real part of the exponent is bounded by
\begin{equation}
    \Re\big(
    \log(\rho_\light) + 
    2\pi i \tau_\star \nn' 
    +2\pi i z_\star \hj'\big) 
    \leq \left\{ \begin{array}{lccl}
    -2\pi ({\rm Im}(\tau_\star) - \gamma )(\nn' + \frac{b}{2t} \hj')\,,  &\phantom{a}& \text{fast growth}\,, \\
    &&& \\[-6pt]
-2\pi {\rm Im}(\tau_\star)(\nn' + \frac{b}{2t} \hj')\,,  &&  \text{slow growth}\,,
    \end{array}\right.
\end{equation}
where we are using the bounds for fast and slow growth given in \eqref{eq:fast} and \eqref{eq:light-spec}. As a result, the partition function \eqref{eq:cardycorr} is finite for fast-growing wJfs provided that ${\rm Im}(\tau_\star) > \gamma$,  or equivalently $\Delta> \gamma^2 |\dmin|$. For slow-growing wJfs, the explicit exponential factor always dominates over $\rho_\light(\nn, \jj)$. Thus, obtaining a finite integral in this case only requires  ${\rm Im}(\tau_\star) > 0$, which is consistent with the regime chosen. 

An important caveat is that the number of light states can scale as a polynomial of $\dmin$ which does affect  \eqref{eq:cardycorrdef}. This is compatible with the HKS sparseness criteria, and it usually arises from an overall normalization of the modular form.\footnote{\label{foot} An explicit example of this is a system where the ground state is highly degenerate. We will see this when considering symmetric products orbifolds in Sec.\,\ref{sec:symmN}, and it is crucial for the coefficients of the wJf to correctly reproduce the logarithmic correction to black hole entropy. In other cases, a polynomial growth in $\dmin$ can originate from an overall normalization of the partition function due to global symmetries, see for example \cite{Cassani:2021fyv}.} For this reason, we write the effect of the scaling as
\begin{equation}\label{eq:scal-var0}
    \varphi_0(\tau_{\star},z_{\star}) \underset{\substack{\Delta\to \Lambda \Delta \\ \dmin\to \Lambda \dmin}}{\approx} \rho_0(\dmin)\,,
\end{equation}
where $\rho_0(\dmin)$ is a polynomial in $\dmin$. 

We can now gather our results and relate \eqref{eq:cardycorrdef} to the Fourier coefficients of sparse wJfs. We find that the subleading corrections to the $c(n, \ell)$ coefficients with $\Delta \gtrsim |\dmin| \gg 1$ are universal and determined entirely by the modular image of the most polar state in the spectrum, that is
 \begin{equation}
\label{eq:subleadingcorr}
    c(n,\ell) \approx  \frac{\rho_0(\dmin)}{\Delta} \sqrt{\frac{|\dmin|}{t}} \e^{4\pi \sqrt{\Delta \dmina}} \qquad  \text{when} \qquad \left\{ \begin{array}{lcl}
      \Delta > \gamma^2 |\dmin|\,, &\phantom{a}& \text{fast growth}\,,\\
    && \\[-6pt]
    \Delta \gtrsim |\dmin|\,, && \text{slow growth}\,.
    \end{array}\right.
\end{equation}
There are two important takeaways of this analysis. First, we have found that fast-growing wJfs feature a minimum value of ${\rm Im}(\tau_\star)$ for which the asymptotic growth of $c(n, \ell)$ is given by the Cardy formula. This means that sparse wJfs can satisfy a weaker version of the HKS bound \eqref{eq:sparse} where $\gamma>1$ and still feature a universal Cardy growth. On the other hand, when $\Delta \le \gamma^2 |\dmin|$, the partition function $\varphi_0(\tau_\star, z_\star)$ is no longer well defined and it is necessary to consider the Rademacher expansion of $c(n, \ell)$. In this case, we expect large corrections to both the leading and subleading behavior of these coefficients. 

We have also found that slow-growing wJfs feature a Cardy growth that is valid for an extended regime. Note that, while ${\rm Im}(\tau_\star)$ can be arbitrarily small, this regime still requires $\Delta$ to scale linearly with $|\dmin|$, as reflected in \eqref{eq:scal-var0}. It is natural to wonder if a linear scaling is necessary for the Cardy growth, or if other scalings for which ${\rm Im}(\tau_\star) > 0$ are also possible. In the next section we will show that slow-growing wJfs can realize the second option, allowing the leading Cardy growth to extend to a \emph{non-universal} regime where $\Delta \lesssim |\dmin|$. This regime is not universal since the subleading corrections to the $c(n, \ell)$ coefficients turn out to depend on the distribution of the light states.

\subsection[Non-universal behavior for \texorpdfstring{$\Delta \lesssim |\dmin|$}{delta<delta0}]{Non-universal behavior for $\mathbf{\Delta \lesssim |\dmin|}$}\label{sec:slowgrowth}

We now consider the asymptotic behavior of the Fourier coefficients of slow-growing wJfs in the non-universal regime $|\dmin| \gtrsim \Delta \gg 1$ where the ratio $\Delta/\dmina$ is not fixed. In this regime, the potentials \eqref{eq:potentials} at which we evaluate $\varphi_0(\tau_\star, z_\star)$ are no longer of order one, since they scale in the large-$|\dmin|$ limit as
\begin{equation}
\tau_\star\sim z_\star \sim i|\dmin|^{-\theta}\,,
\end{equation}
for some $\theta>0$.
As a result, it is not clear that \eqref{eq:condition} holds, and the approach of Sec.\,\ref{sec:fast} may no longer apply. In this case, we will evaluate the integral in \eqref{eq:intapprox2} via a saddle-point approximation. This will allow us to demonstrate that the corrections to the Cardy growth of the $c(n, \ell)$ coefficients in the regime $|\dmin| \gtrsim \Delta \gg 1$ are no longer universal and carry an imprint of the light spectrum.

We start by considering \eqref{eq:intapprox2} in the non-universal regime $|\dmin| \gtrsim \Delta \gg 1$. An important difference with the analysis of the previous section is that we cannot expand the crossing kernel as in \eqref{eq:kerexp}. This is because in the non-universal regime, \eqref{eq:intapprox2} is not necessarily dominated by states close to the state of maximal polarity. Instead, we use the following approximation of the crossing kernel
\eq{
    \knl{P}{\nn,\,\jj}{\nn',\jj'}
    =\frac{1}{2\Delta}\sqrt{\frac{-\Delta'}{t}} \e^{-\pi i \frac{\jj'\jj}{t}} \e^{4\pi \sqrt{\Delta\big( \frac{(\jj')^2}{4t} - \nn' \big)}} + \cdots \,,
}
which only assumes that $\Delta\gg1$. Thus, the asymptotic behavior of the $c(n, \ell)$ coefficients is determined by the following integral
\begin{equation} \label{eq:integral2}
\int_0^\infty\dd \nn' \int_{-t}^{t-1}\dd \jj'\,
\rho_\light(\nn',\jj')\sqrt{-\Delta'}\,
\e^{4\pi\sqrt{\Delta\big(\frac{(\jj')^2}{4t}-\nn'\big)}}\,.
\end{equation}

In the non-universal regime $|\dmin| \gtrsim \Delta \gg 1$, we can evaluate \eqref{eq:integral2} using a saddle-point approximation. The exponential growth of $\rho_\light(\nn, \jj)$ determines the location of the saddle point and the regime of validity of the Cardy growth of $c(n, \ell)$. The non-exponential terms in the distribution of light states, on the other hand, are crucial in determining the subleading corrections to these coefficients. Therefore, it is convenient to parametrize the density of light states as
\begin{equation}\label{eq:rholightfg}
    \rho_\light(\nn,\jj) = \rho_0(\Delta_0) g(\nn,\hj) \e^{f(\nn,\hj)}\,, \qquad \Delta_0 < \nn - \frac{\jj^2}{4t} < 0\,,
\end{equation}
where we have separated the contribution of light states into an exponential part $f(\nn,\hj)$ that affects the position of the saddle point in \eqref{eq:integral2}, a polynomial part $g(\nn,\hj)$, and a possible overall normalization $\rho_0(\Delta_0)$ (which is the factor we introduced in \eqref{eq:scal-var0}; see footnote \ref{foot}). As highlighted in Sec. \ref{sec:fast}, the relevant variable to describe the growth of light states is not $\jj$ but $\textsf{a} = \jj +b$. Following \eqref{eq:light-spec}, the condition that the wJf is slow growing then translates into the property that, under a rescaling of $\nn$ and $\hj$ by a large parameter $\Lambda$,
\begin{equation}\label{eq:scale-fL}
    f(\Lambda \nn,\Lambda \hj) \lesssim \Lambda^\alpha  f( \nn,\hj) \,, \quad 
    \text{ where}\quad \alpha<1\,.
\end{equation}

To illustrate the sensitivity of logarithmic corrections to the details of the light spectrum, we will examine two specific cases that are relevant for Sec.\,\ref{sec:symmN} and Sec.\,\ref{sec:bhs}. In the first case, both variables $\nn$ and $\jj$ appear on equal footing in $\rho_\light(\nn,\jj)$. In the second case, we assume that $f(\nn,\hj)$ and $g(\nn,\hj)$ do not depend on $\hj$  to leading order, making $\nn$ the only relevant variable. The cases under consideration are by no means exhaustive. They are chosen for their relevance to the explicit examples discussed in the following sections. Moreover, they demonstrate the sensitivity of the logarithmic corrections to the details of the light spectrum.

\paragraph{Democracy.} For simplicity, in this case we restrict to $\alpha =\frac{1}{2}$ in \eqref{eq:scale-fL}. We define ``democracy'' as situations where $f(\nn,\hj)$ and $g(\nn,\hj)$ in \eqref{eq:rholightfg} scale uniformly for large values of $\Lambda$, namely situations where
\begin{equation}
\label{eq:assumptions1}
\begin{split}
     f(\Lambda \,\nn, \hj) \approx f(\nn, \Lambda \,\hj)\approx  f(\Lambda \,\nn,\Lambda \,\hj) &\approx \Lambda^\frac{1}{2} f(\nn,\hj)\,, \\
    g(\Lambda \,\nn,\Lambda \,\hj) &\approx \Lambda^\omega g(\nn,\hj)\,,
\end{split}
\qquad \text{for}\qquad 1\ll\Lambda \le |\Delta_0|\,.
\end{equation}
Both $\alpha=\frac{1}{2}$ and democracy are present in wJfs constructed via symmetric product orbifolds, which will be discussed in Sec.\,\ref{sec:symmN}.  

With this parametrization of the light spectrum, the saddle-point equations of \eqref{eq:integral2} are 
\begin{equation}\label{eq:323}
    \partial_{\nn'} f(\nn',\hj') = 2\pi \sqrt{-\frac{\Delta}{\Delta'}}\,, 
    \qquad 
    \partial_{\hj'} f(\nn',\hj') = -\pi \frac{\jj'}{t} \sqrt{-\frac{\Delta}{\Delta'}}\,.
\end{equation}
As we take $\nn'\approx\hj'\approx\Lambda$, it follows that $\Lambda \approx \Delta_0-\Delta'$ and that the derivatives of $f(\nn',\hj')$ scale like $\Lambda^{-1/2}$. This scaling leads to the following solution to the saddle-point equations
\begin{equation}\label{eq:lambdastar}
    \Lambda_\star \approx - \frac{\Delta_0}{\Delta}\,,
\end{equation}
where we used the fact that $t\sim\dmina\gg1$.

From \eqref{eq:assumptions1}, it follows that the Hessian of the saddle-point approximation is of order $\order{\Lambda_\star^{-3}}$. This can be seen from the fact that the Hessian is given by the determinant of the two-by-two matrix whose entries are the second derivatives of $f(\nn,\hj)$. Using $\Delta'_\star-\dmin\approx\Lambda_\star$, we find that \eqref{eq:integral2} evaluates to
\begin{align}
\begin{split}
&\int_0^\infty\dd \nn' \int_{-t}^{t-1}\dd \jj'\;
\rho_\light(\nn',\jj')\;\sqrt{-\Delta'}\;
\e^{4\pi\sqrt{\Delta\big(\frac{(\jj')^2}{4t}-\nn'\big)}} 
\\
&\hspace{180pt}\approx\rho_0(\dmin)\left(\frac{\-\dmin}{\Delta}\right)^{\frac{3}{2}+\omega}
\sqrt{-\Delta_0}\,\e^{4\pi\sqrt{-\Delta\Delta_0}}\,.
\end{split}
\end{align}
Here we used the fact that for large values of the discriminant $\Delta\gg1\,$, the saddle-point value satisfies $\Lambda_\star\ll\dmina$, which follows from \eqref{eq:lambdastar}.  This means that we can approximate $\Delta'_{\star}$ with $\dmin$ at the saddle point. Combining all these ingredients into \eqref{eq:intapprox2} we obtain
\begin{align}\label{eq:democraticresult}
c(n,\ell)
&\approx\frac{\rho_0(\Delta_0)}{\Delta}
\left(\frac{\-\dmin}{\Delta}\right)^{\frac{3}{2}+\omega}
\sqrt{\frac{-\Delta_0}{t}}\,\e^{4\pi\sqrt{-\Delta\Delta_0}}\,,
\end{align}
which is valid for a democratic light spectrum with $\alpha=\frac{1}{2}$ in the regime $|\dmin| \gtrsim \Delta \gg 1$. This expression shows that the logarithmic corrections are now sensitive to properties of the light spectrum via $\rho_0(\Delta_0)$ and $\omega$ (the scaling behavior in \eqref{eq:assumptions1}).

\paragraph{Autocracy.} The second example we consider is one where light states are not sensitive to $\hj\mapsto \Lambda\hj$ to leading order (relative to $\nn\mapsto \Lambda\nn$). This means that we can safely assume $\rho_\light(\nn,\jj)$ as being ``neutral'' in the large-$\Lambda$ limit, that is
\begin{equation}\label{eq:assumptions2}
    \rho_\light(\nn,\jj) \approx \rho_0(\Delta_0)
    \nn^\omega \e^{2\pi\gamma \nn^\alpha}  \delta(\jj +b)\,, \qquad 1\ll\nn<\dmina\,,
\end{equation}
where $\alpha<1$ and $\gamma>0$. This is what we call ``autocracy'':  one variable dominates over the other, and it is the case relevant for Sec. \ref{sec:n2bhs}. This time, the integral over the spin variable in \eqref{eq:integral2} collapses due to the delta function and we are left with a single integral over $\nn$
\begin{equation}
\label{eq:singleint}
    \int_0^{|\dmin|} \dd \nn \; \nn^\omega\sqrt{\dmina-\nn}\,\e^{F(\nn)}\,, \qquad 
     F(\nn) = 2\pi \gamma \nn^{\alpha} + 4 \pi \sqrt{\Delta\big( |\dmin| - \nn\big)}\,.
\end{equation}
We can evaluate the integral via saddle-point approximation once again. In this case, the saddle-point equation for $\nn_\star$ is given by  
\begin{equation}
\label{eq:saddlepoint2}
 \frac{\alpha \gamma}{(\nn_{\star})^{1-\alpha}} - \sqrt{\frac{\Delta}{|\dmin|-\nn_{\star}}} = 0\,.
\end{equation}
This equation cannot be solved analytically for generic values of $\alpha$. Nevertheless, the equation can be solved perturbatively in the following cases:
\begin{itemize}[leftmargin=0.4cm]
    \item $\Delta \gg |\Delta_0|^{2\alpha-1}$: In this regime, the solution to the saddle-point equation is given by 
    \begin{equation}
        \nn_\star = \left(\gamma\alpha \frac{|\Delta_0|}{\Delta}\right)^{\frac{1}{2(1-\alpha)}}\Big[1+\mathcal O\Big(\tfrac{|\dmin|^{2\alpha - 1}}{\Delta}\Big)\Big]\,.
    \end{equation}
    Furthermore, we have $-F''(\nn_\star) \approx \nn_\star^{\alpha-2}$ and
    \begin{equation}
    \label{eq:expexp}
        F(\nn_\star) = 4\pi \sqrt{-\Delta_0\Delta}\Big[1 + \mathcal O\Big(\tfrac{|\dmin|^{2\alpha - 1}}{\Delta}\Big)\Big]\,.
    \end{equation}
    We conclude that the saddle-point equation of this integral, including the one-loop corrections, results in the following scaling for the Fourier coefficients of the wJf
    \begin{align}
    \label{eq:slowres}
    c(n,\ell)
    \approx\frac{\rho_0(\Delta_0)}{\Delta}\sqrt{\frac{|\Delta_0|}{t}}
    \left(\frac{|\Delta_0|}{\Delta}\right)^{\frac{2-\alpha}{4(1-\alpha)}+\omega}
    \e^{4\pi\sqrt{-\Delta\Delta_0}}\,.
    \end{align}
    \item $\Delta \ll |\Delta_0|^{2\alpha-1}$: In this regime, the saddle-point equation \eqref{eq:saddlepoint2} tells us that $\nn_\star$ approaches $\dmina$ in the large-$\dmina$ limit, i.e.~it approaches the threshold of non-polarity. The saddle point is located at
    \begin{equation}
         \nn_\star  =|\Delta_0| - \frac{|\Delta_0|}{(\alpha \gamma)^2} \bigg(\frac{\Delta}{|\Delta_0|^{2\alpha - 1}}\bigg) + \cdots\, .
    \end{equation}
    For this value of $\nn_\star$, the light states dominate over the contribution of the kernel. This implies that the exponential term in the integrand of \eqref{eq:singleint} is no longer given by the Cardy formula. Instead, we find that the exponent is given by 
    \begin{equation}
    \label{eq:expexp2}
     2\pi \gamma (\nn_\star)^\alpha + 4\pi \sqrt{-\Delta \Delta'_\star} =2 \pi \gamma |\Delta_0|^\alpha \Big[1 + \mathcal O\Big(\tfrac{\Delta}{|\dmin|^{2\alpha - 1}}\Big)\Big]\,,
    \end{equation}
    where $\Delta'_\star = \nn_\star - |\Delta_0|$. In this case, both the leading and subleading behavior of the $c(n, \ell)$ coefficients depend on the distribution of the light states as evidenced by  
    \begin{equation}
    \label{eq:slowres2}
    c(n, \ell) \approx \frac{\rho(\dmin)}{
    \sqrt{\Delta t}} \dmina^{\frac{4-3\alpha}{2}+\omega} e^{2\pi \gamma\dmina^\alpha}\,.
    \end{equation}
    \item $\Delta \sim |\Delta_0|^{2\alpha-1}$: This is an intermediate regime where $\nn_\star$ is of order $|\Delta_0|$, but it is not quite yet at the threshold of non-polarity. As a result, the saddle-point equation can only be satisfied when
    \begin{equation}
    \label{eq:saddle3}
    \Delta'_\star \approx \Delta_0\,.
    \end{equation}
    In this case, the contribution of the light states becomes comparable to that of the kernel, as can be seen from
    \begin{equation} 
    2\pi \gamma \nn_\star^\alpha \approx  \dmina^\alpha \approx 4\pi \sqrt{-\Delta \Delta'_\star}\,.
    \end{equation}
    In this scaling regime, evaluating the integral \eqref{eq:integral2} at the saddle point \eqref{eq:saddle3} yields
    \begin{equation}
    \label{eq:slowres3}
    c(n, \ell) \approx 
    \frac{\rho_0(\dmin)}{\sqrt{t}} 
    |\dmin|^{\frac{5(1-\alpha)}{2}+\omega}
    \e^{4\pi\zeta \dmina^\alpha}\,, \qquad \zeta = \frac{\gamma}{2}\frac{\nn_\star}{\dmina} + \frac{\sqrt{-\Delta \Delta_\star'}}{\dmina^\alpha}\,,
    \end{equation}
    where $\zeta$ is a number that is independent of $\dmin$ in the large-$\dmina$ limit but instead depends on the ratios $\Delta/\dmina^{2\alpha - 1}$ and $\Delta_\star'/\dmina$. As a consistency check, we note that extending the $c(n, \ell)$ coefficients in \eqref{eq:slowres} and \eqref{eq:slowres2} to the regime $\Delta \approx \dmina^{2\alpha - 1}$ reproduces the subleading (non-exponential) terms in \eqref{eq:slowres3}. However, these expressions cannot reproduce the leading behavior in \eqref{eq:slowres3}. This follows from the fact that the exponential growth in \eqref{eq:expexp} and \eqref{eq:expexp2} receives $\mathcal O (1)$ corrections when $\Delta \approx \dmina^{2\alpha - 1}$.
\end{itemize}

Let us conclude this section by summarizing the main insights derived from the two cases considered above. A general feature we observe is that the leading exponential behavior of the $c(n, \ell)$ coefficients, i.e.~the Cardy behavior, persists as long as the condition $\Delta \gg |\dmin|^{2\alpha -1}$ is fulfilled. When $\alpha = 1/2$ this regime only requires that $\Delta \gg 1$, but for generic values of $\alpha$, there exists a threshold beyond which the Cardy term undergoes significant corrections. On the other hand, the subleading (non-exponential) corrections to the leading Cardy behavior are not universal and depend sensitively on the specific details of the wJf. This sensitivity has multiple origins: the possible scaling of $\rho_0(\dmin)$, the polynomial growth of the density of light states, and the one-loop corrections involved in the integral \eqref{eq:integral2}.

\section{Symmetric product orbifolds}\label{sec:symmN}

In this section we consider the asymptotic behavior of weak Jacobi forms constructed via a symmetric product orbifold. A symmetric product orbifold of a two-dimensional CFT is built by tensoring and orbifolding a CFT$_2$ as follows. Consider a compact and unitary \textit{seed} CFT$_2$ $\mathcal{C}$ with central charge $\mathtt{c}_o$. The symmetric product orbifold $\text{Sym}^N(\mathcal{C})$ is obtained by taking the $N$-th tensor power of $\mathcal{C}$ and orbifolding with respect to the discrete $S_N$ symmetry that permutes the different copies of the tensor product, that is
\eq{\label{eq:defn-symmn}
\text{Sym}^N(\mathcal{C})\coloneqq \frac{\mathcal{C}^{\otimes N}}{S_N}\,.
}
The central charge of the resulting theory is $\mathtt{c}=N\mathtt{c}_o$. Moreover, the spectrum consists of the $S_N$-invariant states of the tensor product $\mathcal{C}^{\otimes N}$, which form the so-called untwisted sector, supplemented by twisted sectors labeled by the conjugacy classes of $S_N$ \cite{Dixon:1985jw,Dixon:1986jc,Dijkgraaf:1996xw}. 

We will use \eqref{eq:defn-symmn} as an operation acting on a wJf.\footnote{The analysis in this section does not require a Hilbert space interpretation in terms of a two-dimensional CFT. However, in many cases, it is possible to give a very precise CFT$_2$ interpretation to the wJf. We refer to \cite{Belin:2019rba,Belin:2020nmp,Benjamin:2022jin} for a recent discussion on this interpretation in the context of ${\cal N}=2$ CFTs.} More precisely, let $\varphi(\tau,z)$ be a wJf with weight zero and index $t_o$ that we identify with $\mathcal{C}$. In order to make this dependence explicit, we write
\eq{
\varphi(\tau,z;\mathcal{C})=\sum_{n,\ell}c_{o}(n,\ell)q^ny^\ell\,,
}
where $c_o(n,\ell)$ is the data of the seed wJf. We assume that $\varphi(\tau,z;\mathcal{C})$ is of the form \eqref{eq:most-polar}, and hence the state with maximal polarity is $q^0 y^{\pm b_o}$. Based on this seed data, we construct a new wJf of weight zero, index $t=t_oN$, and maximal polarity $b=b_oN$, by following the rules in \eqref{eq:defn-symmn}. This results in a new wJf denoted by $\varphi\left(\tau,z;\text{Sym}^N(\mathcal{C})\right)$. A simple way to obtain $\varphi\left(\tau,z;\text{Sym}^N(\mathcal{C})\right)$ is to consider the generating function of partition functions (or grand canonical ensemble) known as the DMVV formula \cite{Dijkgraaf:1996xw}. This generating function is given by
\eq{\label{eq:dmvv}
{\cal Z}(\tau,z,\sigma)=\sum_{N=0}^\infty  p^{t_oN}\varphi\left(\tau,z;\text{Sym}^N(\mathcal{C})\right)=\prod_{\substack{m>0\\n,\ell}}\frac{1}{\left(1-q^ny^\ell p^{t_om}\right)^{c_o(nm,\ell)}}\,,
}
where $p\coloneqq e^{2\pi i\sigma}$. Note the explicit dependence on the seed data due to the appearance of the coefficients $c_o(n,\ell)$ on the right-hand side of \eqref{eq:dmvv}.  This formula shows that we can easily access wJfs with arbitrarily large values of $|\dmin|=Nb_o^2/(4t_o)$, i.e.~large $N$, by expanding ${\cal Z}(\tau,z,\sigma)$ to large powers of $p$.

In the following, we will extract the Fourier coefficients of 
\eq{\label{eq:Zdnml}
\mathcal{Z}(\tau,z,\sigma)=\sum_{t,n,\ell}d(n,\ell;t)\,p^{t}q^ny^\ell\,,
}
by performing a suitable contour integral. This method is conceptually different from the Rademacher and crossing kernel expansions considered in Sec.\,\ref{sec:2}. Depending on the seed data, $\varphi\left(\tau,z;\text{Sym}^N(\mathcal{C})\right)$ can be either a fast or slow growing wJf \cite{Belin:2019jqz,Belin:2019rba}, providing explicit examples of modular forms that comply with \eqref{eq:fast} and \eqref{eq:light-spec}.  The bottom line of this analysis is that for $\Delta |\dmin|\gg1$ we find that $d(n,\ell;t)$ agrees with $c(n,\ell)$, giving us a nontrivial check of the results of Sec.\,\ref{sec:beyondcardy}.

\subsection{Asymptotic expansions}\label{sec:symexp}

The procedure to extract the $d(n, \ell; t)$ coefficients starts by casting $d(n,\ell;t)$ as a contour integral that is simply given by
\eq{ \label{eq:d-integral}
d(n,\ell;t)=\oint \frac{\dd p}{2\pi i p} \oint \frac{\dd q}{ 2\pi i q} \oint \frac{\dd y}{ 2\pi i y} \, \mathcal{Z}(\tau,z,\sigma)\, p^{-t} q^{-n} y^{-\ell} \,,
}
where each of the contours encloses the origin and we recall that $p=e^{2\pi i\sigma}$, $q=e^{2\pi i\tau}$, and $y=e^{2\pi iz}$. The integral \eqref{eq:d-integral} is defined in the Siegel upper half-plane ($\mathbb{H}_2$). On $\mathbb{H}_2$, a contour $C$ that renders the expansion of $\mathcal{Z}(\tau,z,\sigma)$ convergent and encloses $p=q=y=0$ is 
\begin{equation}
\text{Im}\  \tau \gg 1 \,, \qquad \text{Im}\ \sigma \gg 1 \,, \qquad \text{Im}\ z \gg 1 \,, \qquad \text{Im}\ \tau\ \text{Im}\ \sigma - (\text{Im} \ z)^2 \gg 1 \,,
\end{equation}
while the real parts lie in
\begin{equation}
0 \leq \text{Re}\ \tau,\ \text{Re}\ \sigma,\ \text{Re}\ z <1 \,.
\end{equation}
Note that the contour is not closed in terms of $\tau$, $\sigma$, and $z$, but it is closed in terms of $p$, $q$, and $y$.

In solving the integral \eqref{eq:d-integral}, one has two resources at hand, namely the modular and the meromorphic properties of $\mathcal{Z}$. In the latter, the strategy is to locate the poles of $\mathcal{Z}$ on $\mathbb{H}_2$ and deform the contour such that it encloses the poles \cite{Sen:2007qy}. The key to obtaining asymptotic expansions for $d(n,\ell;t)$ is to argue that for a range of $t$, $n$, and $\ell$, the integral is well approximated by a single residue, i.e., that one pole gives the most dominant contribution to the asymptotic expansion.

Following \cite{Sen:2007qy}, we add new segments to the contour $C$ defined above that closes the contour in a clockwise fashion in the Siegel upper half-plane. The addition of these segments can be viewed as deforming the contour $C$ to a new contour by moving the imaginary parts of $\tau$, $\sigma$, and $z$ downward to the following values
\eq{
\text{Im}\ \tau \sim 1 \,, \qquad \text{Im} \ \sigma \sim 1 \,, \qquad \text{Im}\ z \sim 1\,,
}
while their real parts are kept unchanged. Ref.~\cite{Sen:2007qy} argued that this deformation only leads to exponentially suppressed contributions such that the leading and the perturbative corrections to the leading result of the integral \eqref{eq:d-integral} remain unchanged. 

Now that we have closed the contour in the Siegel upper half-plane, we can evaluate \eqref{eq:d-integral} using Cauchy's integral theorem. Therefore, one of the integrals reduces to a sum over residues of poles that are enclosed by the contour. In fact, we will argue that the leading result of this integral, as well as its perturbative corrections, are captured by the contribution of a single pole. In order to make this argument we need one additional assumption: we assume that for $\Delta\abs{\dmin}\gg1$,  the integrand in \eqref{eq:d-integral} is dominated by the explicit exponential factor $p^{-t}q^{-n}y^{-\ell}$. In particular, this assumption implies that the residue of $\mathcal{Z}(\tau,z,\sigma)$ does not compete with the explicit exponential. 

Let us now examine the residues of $\mathcal{Z}$ at each of its poles. This can be done systematically by relating $\mathcal{Z}$ to an exponential lift, the details of which are described in App.\,\ref{app:SMF}. In short, the procedure is as follows. The poles of $\mathcal{Z}$ are simple to detect from \eqref{eq:dmvv}: when $c_o(nm,\ell)>0$, we will have a pole at $q^ny^\ell p^{t_om} =1$ and modular images of this equation. In the regime $4nt-\ell^2\gg1$, the integrand of \eqref{eq:d-integral} is exponentially dominated by the residue of one pole, which is located at
\eq{\label{eq:zero}
t_o(\tau\sigma-z^2)+b_oz=0\,,
}
where $t_o$ and $b_o$ are the index and maximal polarity of the seed wJf.  To manipulate $\mathcal{Z}(\tau,z,\sigma)$ near this point, it is convenient to note that \eqref{eq:zero} has a simple relation to the pole 
\eq{\label{eq:zerohat}
b_o\zh-t_o\sh=0\,,\quad\text{or equivalently}\quad \hat y^{b_o}=\hat{p}^{t_o}\,,
}
where $(\hat \tau, \hat z, \hat \sigma)$ correspond to the $S$ transformation of $(\tau, z, \sigma)$, that is,
\eq{\label{eq:stransf}
\tauh=-\frac{1}{\tau}\,,\qquad\zh=-\frac{z}{\tau}\,,\qquad \sh=\frac{\sigma\tau-z^2}{\tau}\,.
}
Note that $\hat{p}^{t_o}=\hat{y}^{b_o}$ has $t_o$ solutions when solving for $\hat{p}$ since
\eq{
\hat{p}=\hat{y}^{b_o/t_o}\zeta_{t_o}\,,
}
is a solution for all $t_o$-th roots of unity $\zeta_{t_o}$. Therefore, there are $t_o$ poles of the form \eqref{eq:zerohat} that lie in the interval $0\leq\text{Re}\ \sigma<1 $.

We can then expand $\mathcal{Z}$ around \eqref{eq:zerohat} (see App.\,\ref{sec:exppole} for details), which yields
\eq{\label{eq:Zexp}
\mathcal{Z}(\tauh,\zh,\sh)= \frac{\varphi_\infty(\tauh,\zh)}{(2\pi i)^{c_o(0,-b_o)}}\frac{1}{(b_o\zh-t_o\sh)^{c_o(0,-b_o)}} + \cdots\,,
}
where we have only shown the leading behavior around the pole. The residue is controlled by
\eq{
\varphi_\infty(\tauh,\zh)=\prod_{\substack{n\ge0\\\ell\in\mathbb{Z}\\\left(n, \ell\right)\neq(0,0)}}\left(1-\hat{q}^n\hat{y}^{\ell}\right)^{-f\left(n, \ell\right)}\eqqcolon\sum_{n',\ell'}d_{\infty}\left(n',\ell'\right)\hat q^{n'}\hat y^{\ell'}\,,\label{eq:chiinf}
}
where the function $f$ is defined as
\eq{\label{eq:f}
f(n,\ell)\coloneqq\sum_{m=1}^\infty c_o(nm,\ell-b_om)\,.
}
The function $\varphi_\infty(\tauh,\zh)$ was first introduced in \cite{Belin:2019jqz,Belin:2019rba}, where it was shown that $d_{\infty}\left(n',\ell'\right)$ captures the degeneracies of polar states in the limit $N\to \infty$, appropriately regularized. We will discuss the interpretation of $d_\infty$ in more detail below. Using the ingredients described above, we can now manipulate the integrals in \eqref{eq:d-integral}. First, we use modular invariance to write 
\eqsp{
d(n,\ell;t)&=\int_{C} \dd\tau\ \dd z\  \dd\sigma\ \mathcal{Z}(\tau,z,\sigma)e^{-2\pi i\left(t\sigma+\ell z+n \tau\right)}\\
&\hspace{0pt}=\int_{{C}}\dd\tau \ \dd z\ \dd\sigma\ \mathcal{Z}(\tauh,\zh,\sh)e^{-2\pi i\left(t\sigma+\ell z+n \tau\right)}\,.\\
}
We can then use $\eqref{eq:Zexp}$, and rewrite the expansion in the unhatted variables, which yields
\eqsp{\label{eq:nohatpole}
d(n,\ell;t)&=\int_{C} \dd\tau\ \dd z\  \dd\sigma\ \frac{1}{(-2\pi i t_o)^{c_o(0,-b_o)}}\frac{1}{\left(\sigma-\frac{z^2}{\tau}+\frac{b_o}{t_o}\frac{z}{\tau}\right)^{c_o(0,-b_o)}}\\
&\quad\times\ \sum_{n',\ell'}d_{\infty}\left(n',\ell'\right)\ \text{exp}\left[-2\pi i\left(n\tau+t\sigma+\ell z+\frac{n'}{\tau}+\frac{\ell'z}{\tau}\right)\right]+\cdots\,.
}
The dots in the equation above correspond to higher order terms in the expansion that lead to exponentially suppressed contributions in $4nt-\ell^2$. In other words, \eqref{eq:nohatpole} is the leading order approximation to  $d(n,\ell;t)$ when $4nt-\ell^2\gg1$. 

We now approximate the integral over $\sigma$ by the residue of the integrand at $\sigma=\frac{z^2}{\tau}-\frac{b_o}{t_o}\frac{z}{\tau}$, taking into account the fact that there are $t_o$ of these residues, such that\footnote{Note that there is a minus sign because of the clockwise orientation of the contour.} 
\eqsp{\label{eq:987}
d(n,\ell;t)&= \sum_{n',\ell'}d_\infty\left(n',\ell'+b_oN\right)\frac{N^{c_o(0,-b_o)-1}}{(c_o(0,-b_o)-1)!}e^{\pi i\frac{\ell\ell'}{t}}\\
&\hspace{40pt}\times\int\ \dd\tau\ \text{exp}\left[-2\pi i\left(\frac{\Delta'}{\tau}+\Delta\tau\right)\right]\\
&\hspace{40pt}\times\int \dd z\ \text{exp}\left[-\frac{2\pi it}{\tau }\left(z+\frac{\ell \tau+\ell'}{2t}\right)^2\right]+\cdots\,,
}
where we shifted $\ell'\mapsto \ell'+b_oN$ and defined $\Delta'\coloneqq n'-\frac{(\ell')^2}{4t}\,$.

Next, we turn to the remaining two integrals over $z$ and $\tau$. The integral over $z$ is a Gaussian. Recall that the contour of integration is given by $0<\text{Re}(z)<1$ for a fixed positive value of Im$(z)\eqqcolon\epsilon$. To evaluate the integral, we change variables to
\eq{
u\coloneqq\sqrt{\frac{2\pi i t}{\tau}}\left(z+\frac{\ell \tau+\ell'}{2t}\right)\,,
}
so that for $t\gg1$
\eq{
\int_{0+i\epsilon}^{1+i\epsilon} \dd z\ \text{exp}\left[-\frac{2\pi it}{\tau }\left(z+\frac{\ell \tau+\ell'}{2t}\right)^2\right]=\sqrt{-\frac{i\tau}{2\pi t}}\int_{-\infty}^\infty \dd u\ e^{-u^2}=\sqrt{-\frac{i\tau}{2t}}\,.
}

We are then left with the integral over $\tau$ in \eqref{eq:987}, which can be written as
\eq{\label{eq:symbessel}
\mathcal{I}(\Delta,\Delta')\coloneqq\int_{0+i\delta}^{1+i\delta}\dd\tau \sqrt{\tau}\ \text{exp}\left[-2\pi i\left(\frac{\Delta'}{\tau}+\Delta\tau\right)\right]\,,
}
where $\delta>0$ denotes the constant imaginary value of the $\tau$ contour. The integrand has an essential singularity at $\tau=0$ and we take the branch cut to lie along the negative imaginary  $\tau$-axis. Using Rademacher's prescription, we split the contour into smaller pieces, and we deform the contour such that its different parts follow Ford circles in the interval $(0,1)$ (see e.g.~\cite{Alday:2019vdr,Cardoso:2021gfg}). The integral over (parts of) the different Ford circles then sources the different terms in the Kloosterman sum. Since we are only concerned with the leading term in the Kloosterman sum, we can take the contour to be given by the leading Ford circle, which is a circle of radius $1/2$ and is centered around $i/2$ with the branch point $\tau=0$ removed. Next, we change variables to
\eq{
w=\frac{i}{\Delta\tau}\,,
}
such that
\eq{
\mathcal{I}(\Delta,\Delta')=-\Delta^{-3/2}\int_{\nu-i\infty}^{\nu+i\infty} \frac{dw}{(iw)^{5/2}}\text{exp}\left[2\pi\left(\frac{1}{w}-\Delta\Delta'w\right)\right]+\cdots\,,
}
for some $\nu>0$. Above, the dots denote the exponentially suppressed contributions from the subleading Ford circles. After this change of variables, the branch cut now lies along the negative real $w$-axis. The evaluation of this integral depends on the sign of $\Delta'$. When $\Delta'\ge0$, we trivially have $\Delta\Delta'\ge0$, and we can close the contour in the half-plane Re$(w)>0$. There are no poles or branch cuts in this region, and hence the integral evaluates to zero. On the other hand, when $\Delta'<0$ we need to deform the contour to the left half-plane Re$(w)<0$ to get a convergent answer. The contour is deformed such that it surrounds the branch cut on the negative real axis and the integral converges to a Bessel function. We thus obtain 
\eq{
\mathcal{I}(\Delta,\Delta')=2\pi i^{-3/2}
\left(\frac{\abs{\Delta'}}{\Delta}\right)^{3/4}I_{3/2}(4\pi\sqrt{\abs{\Delta'}\Delta})\Theta(\Delta)+\cdots\,.
}
Combining the results above, we arrive at the final result
\eq{
d(n,\ell;t)=\sum_{\substack{n',\ell'\\ \Delta'<0}}c_\infty(n',\ell')e^{\pi i\frac{\ell\ell'}{t}}\sqrt{\frac{2\pi^2}{t}}\left(\frac{\abs{\Delta'}}{\Delta}\right)^{3/4}I_{3/2}(4\pi\sqrt{\abs{\Delta'}\Delta})+\cdots\,,\label{eq:symprodres}
}
where
\eq{
 c_{\infty}(n',\ell')= \frac{N^{c_o(0,-b_o)-1}}{(c_o(0,-b_o)-1)!}  d_\infty(n',\ell'-b_oN)\,,\label{eq:light-symmn}
}
and $d_\infty(n,\ell)$ is defined in \eqref{eq:chiinf}. 

\bigskip
At this stage, it is worth making three important remarks:

\paragraph{Light spectrum of symmetric product orbifolds.}
Let us elaborate further on the interpretation of $d_\infty(n,\ell)$ and   $c_\infty(n,\ell)$ in \eqref{eq:light-symmn}. These coefficients appear as the residue of $\mathcal{Z}(\tauh,\zh,\sh)$ in \eqref{eq:Zexp}. However, these coefficients also count a class of light states in $\text{Sym}^N(\mathcal{C})$ that is important to highlight.  

As mentioned above,  $d_\infty(n,\ell)$ controls the degeneracy of polar states in $\text{Sym}^N(\mathcal{C})$ at $N\to \infty$. In particular, the analysis of  \cite{Belin:2019jqz,Belin:2019rba}  shows that they corresponds to the normalized degeneracies of states in $\text{Sym}^N(\mathcal{C})$ that satisfy
\begin{equation}\label{eq:lightsymm}
n\ll b_0 N \,, \qquad \ell = \hj - b_o N \,, \qquad |\hj|\ll b_o N \,,
\end{equation}
in the strict $N\to\infty$ limit. Therefore, the states captured by $d_\infty(n,\ell)$ are polar states for which $n$ and $\hj$ are kept fixed in the $\dmina\to \infty$ limit.

The degeneracy of these states is actually given by $c_\infty(n,\ell)$, since there is also an overall normalization that is important to take into account. If the seed ground state degeneracy is $c_0(0,-b_o)$, then the number of ground states in $\text{Sym}^N(\mathcal{C})$ at finite $N$ is given by 
\eq{\label{deg_sym_gs}
\sum_{k=1}^{c_o(0,-b_o)}\binom{c_o(0,-b_o)}{k}\binom{N-1}{k-1} = \binom{N + c_o(0,-b_o) - 1}{c_o(0,-b_o) -1}\,.
}
This expression can be obtained by counting all the possible distributions of the seed ground states over the $N$ tensor factors, a result that also follows from the DMVV formula \eqref{eq:dmvv}. As a result, the ground state degeneracy of the $N$-th symmetric product orbifold scales like
\eq{ \label{deg_sym_gs-1}
\binom{N+c_o(0,-b_o)-1}{ c_o(0,-b_o)-1}\underset{N\rightarrow\infty}{=}\frac{N^{c_o(0,-b_o)-1}}{(c_o(0,-b_o)-1)!}+\cdots\,.
}
This is the prefactor in \eqref{eq:light-symmn}, which accounts for the high degeneracy of states due to the intrinsic properties of the symmetric product orbifold. Therefore, $c_\infty(n,\ell)$ counts light states of the form \eqref{eq:lightsymm} in the limit $N\to\infty$.

It is interesting to note that although the asymptotic formula \eqref{eq:symprodres} does not require $N\to \infty$, the relevant coefficients that enter there, even at finite $N$, can be identified as the light states for the wJf for $\text{Sym}^N(\mathcal{C})$ at $N\to \infty$.

\paragraph{Comparison with the Rademacher expansion.}
Using the fact that the $c_\infty(n',\ell')$ are Fourier coefficients of a wJf, we can use spectral flow to restrict the sum over $\ell'$ to $\ell'$ mod $2t$. In analogy with Sec.\,\ref{sec:kernel} and Sec.\,\ref{sec:rademacher}, we can define a ``symmetric product" kernel by
\eq{
\mathbb{S}_{\{n,\ell\},\{n',\ell'\}}\coloneqq e^{-i\pi \frac{\ell\ell'}{t}}\sqrt{\frac{2\pi^2}{t}}\left(\frac{\abs{\Delta'}}{\Delta}\right)^{3/4}I_{3/2}\left(4\pi\sqrt{\Delta\abs{\Delta'}}\right)\,.
}
In terms of this kernel, \eqref{eq:symprodres} becomes\footnote{Note that compared to \eqref{eq:symprodres}, we have added a minus sign to the phase appearing in the symmetric product kernel. This is allowed because weight zero wJfs are symmetric under $z\mapsto -z$, which implies that $c(n,\ell)=c(n,-\ell)$. The sole reason to add the minus sign is to compare more readily to the crossing and Rademacher kernels, which have the minus sign in the phase.}
\eq{\label{eq:symprodkernel2}
d(n,\ell;t)=\sum_{\substack{\Delta'<0\\\abs{\ell'}\text{ mod }2t}} c_\infty(n',\ell')\ \mathbb{S}_{\{n,\ell\},\{n',\ell'\}}+\cdots\,.
}
The dots above contain exponentially suppressed corrections that arise from the fact that (1) we have deformed the contour and we have taken only the leading contribution to the integrals, and (2) we have approximated the integrals by a single residue.

The symmetric product kernel agrees with the Rademacher kernel up to nonperturbative corrections that originate from the aforementioned approximations
\eq{\label{eq:srcomp}
\frac{\mathbb{S}_{\{n,\ell\},\{n',\ell'\}}}{\mathbb{R}_{\{n,\ell\},\{n',\ell'\}}}= 1+\mathcal{O}\left(e^{-2\pi\sqrt{\Delta\abs{\Delta'}}}\right)\,.
}
Properly taking into account the corrections to the leading contour gives rise to the Kloosterman sums, while taking into account subleading residues will generate the $c>1$ terms in the Rademacher expansion (see App.\,\ref{sec:dominantpole}). We believe it is possible to carefully track these corrections and reproduce the full Rademacher expansion for symmetric product orbifolds. For the special case of the Igusa cusp form, which can be cast as an exponential lift, this has been done in \cite{Cardoso:2021gfg,LopesCardoso:2022hvc}.

\paragraph{Comparison with crossing kernels.} In Sec.\,\ref{sec:beyondcardy}, we use crossing kernels to get a reliable approximation to the coefficients of a sparse wJf. The results of this section are compatible with that approximation. In order to make this explicit, we write \eqref{eq:987} as
\eqsp{\label{eq:3.41}
d(n,\ell;t)&=\int_{C} \dd\tau\ \dd z\ \frac{N^{c_o(0,-b_o)-1}}{(c_o(0,-b_o)-1)!}\varphi_{\infty}(\tauh,\zh)\ \\
&\hspace{80pt} \times\text{exp}\left[-2\pi i\left(n\tau+\frac{tz^2}{\tau}-\frac{bz}{\tau}+\ell z\right)\right]+\cdots\,,
}
where $\varphi_{\infty}(\tau,z)$ is given in \eqref{eq:chiinf}.
In the derivation of \eqref{eq:symprodres}, we evaluated the $\tau$ and $z$ contour integrals explicitly. Here we will approximate these integrals via saddle point. We are assuming that the residue of $\mathcal{Z}$, or more explicitly $\varphi_{\infty}(\tau,z)$, does not compete with the explicit exponential dependence of \eqref{eq:3.41}. It follows that the location of the saddle point is controlled, to a good approximation, by the explicit exponential. Under this assumption, the location of the saddle point is given by
\eq{\label{eq:symprodsaddle}
\tau_\star=\frac{b}{\sqrt{\ell^2-4nt}}\,,\qquad z_\star=\frac{b}{2t}-\frac{b\ell}{2t\sqrt{\ell^2-4nt}}\,.
}
A simple check shows that the exponential evaluated at the saddle point scales as expected
\eq{
\Re\left[-2\pi i\left(n\tau_\star+\frac{tz_\star^2}{\tau_\star}-\frac{bz_\star}{\tau_\star}+\ell z_\star\right)\right]=4\pi\sqrt{\abs{\dmin}\Delta}\,.
}
In the saddle-point approximation we thus have
\begin{equation}\label{eq:sym-saddle}
   d(n,\ell;t) = \frac{1}{2\Delta}
    \sqrt{\frac{|\dmin|}{t}}
    \e^{\pi i \frac{ b\ell}{t}}\e^{4\pi \sqrt{\Delta |\dmin|}}
    \frac{N^{c_o(0,-b_o)-1}}{(c_o(0,-b_o)-1)!}\varphi_\infty(\hat\tau_{\star},\hat z_{\star})+\cdots\,.
\end{equation}
This expression is exactly of the form \eqref{eq:cardycorrdef}. In particular, the location of the saddle point \eqref{eq:symprodsaddle} is equivalent to the effective potentials in \eqref{eq:potentials}. We can also recognize 
\begin{equation}
   \varphi_0(\tau,z)=\frac{N^{c_o(0,-b_o)-1}}{(c_o(0,-b_o)-1)!}\varphi_\infty(\hat\tau,\hat z)\,,
\end{equation}
as the light states that contribute to the approximation. Furthermore, it is clear from this comparison, and the discussion around \eqref{deg_sym_gs} and \eqref{deg_sym_gs-1}, that 
\begin{equation}\label{eq:rho0-symm}
    \rho_0(\Delta_0)= \frac{N^{c_o(0,-b_o)-1}}{(c_o(0,-b_o)-1)!}\approx |\Delta_0|^{c_o(0,-b_o)-1} \,.
\end{equation}
Recall that  $\rho_0(\Delta_0)$ was introduced in \eqref{eq:scal-var0} as the possibility of having a ground state degeneracy controlled by $\Delta_0$.

One important difference between \eqref{eq:cardycorrdef} and \eqref{eq:sym-saddle} is the regime of validity of the approximation. The analysis of Sec.\,\ref{sec:fast} only applies to states of sparse wJfs with $\Delta\gtrsim|\Delta_0|\gg1$. In contrast, the analysis here applies when $\Delta|\Delta_0|\gg1$ and the residue does not compete against the exponential factors. For slow-growing forms, this implies that we can also use \eqref{eq:sym-saddle} in the regime $|\Delta_0|\gtrsim\Delta\gg1$, which is special to symmetric product orbifolds (and the exponential lifts discussed below).\footnote{It is worth stressing that \eqref{eq:sym-saddle} is also valid when we hold $\Delta_0$ fixed and $\Delta$ is large, i.e.~in the universal Cardy regime.} 

\subsection{Logarithmic corrections}\label{sec:symprodlog}

Now that we have \eqref{eq:sym-saddle}, it is straightforward to extract the subleading corrections to the exponential growth of $d(n,\ell;t)$. The nontrivial information needed is contained in $\varphi_\infty(\tau,z)$ which we will now discuss. 

An important consequence of the symmetric product orbifold construction is that the gauging by the symmetric group leads to a maximal possible growth of light states. In fact, the growth of wJfs obtained from a symmetric product orbifold satisfy \cite{Keller:2011xi}
\eq{\label{eq:dHKS}
d(n,\ell;t)\lesssim e^{2\pi(\Delta-\dmin)}\,, \qquad \dmin<\Delta<0\,.
}
This means that any symmetric product orbifold is sparse and complies with either \eqref{eq:fast} or \eqref{eq:light-spec}. We can, however, obtain a more refined version of \eqref{eq:dHKS}. The degeneracies of light states as $N\to\infty$ are determined by the seed coefficients, or more precisely, by the values of $f(n,\ell)$ defined in \eqref{eq:f}. It was proven in \cite{Belin:2019jqz,Belin:2019rba} that there are only two possible kinds of growth for wJfs obtained from symmetric product orbifolds:
\begin{itemize}[leftmargin=0.3cm]
    \item[]{\bf Fast growth.} The $f(n,\ell)$ functions grow exponentially as functions of the charges. This corresponds to the fast growth defined in \eqref{eq:fast} with $\gamma\le1$.
    \item[]{\bf Slow growth.} The $f(n,\ell)$ functions are constants, i.e.~the dependence on $n$ and $\ell$ is extremely weak. Therefore, $\varphi_\infty$ is basically a ratio of theta and Dedekind-eta functions. We are in the democratic case discussed in Sec.\,\ref{sec:slowgrowth}. 
\end{itemize}

What determines fast versus slow growth are simple conditions on the polar states of the seed theory, which are presented in  \cite{Belin:2019jqz,Belin:2019rba,Keller:2020rwi}. These conditions can be used to classify large families of symmetric product orbifolds. We can now characterise the subleading corrections to $d(n,\ell;t)$ for slow and fast growing forms in the two regimes discussed in Sec.\,\ref{sec:beyondcardy}, namely $\Delta \gtrsim |\dmin|$ and $\Delta \lesssim |\dmin|$.

\paragraph{Universal behavior for $\Delta \gtrsim |\dmin|$.} In this regime $\tau_\star$ and $z_\star$ are order one parameters such that $\varphi_\infty(\tau_\star,z_\star)$ does not contribute to the scaling behavior of $d(n,\ell;t)$. From \eqref{eq:sym-saddle} we find 
 \begin{equation}
\label{eq:subleadingcorr-sym}
    d(n,\ell;t) \approx  \frac{|\dmin|^{c_o(0,-b_o)}}{\Delta|\dmin|} \e^{4\pi \sqrt{\Delta \dmina}} \qquad  \text{when} \qquad \left\{ \begin{array}{lcl}
      \Delta > \gamma^2 |\dmin|\,, &\phantom{a}& \text{fast growth}\,,\\
    && \\[-6pt]
    \Delta \gtrsim |\dmin|\,, && \text{slow growth}\,.
    \end{array}\right.
\end{equation}
This expression agrees with \eqref{eq:subleadingcorr} where $\rho_0(\dmin)$ is given by \eqref{eq:rho0-symm}. The result also reinforces the fact that, in the regime $\Delta \gtrsim |\dmin|\gg1$, the asymptotic growth of states is universal and determined entirely by the degeneracy of ground states.

\paragraph{Non-universal behavior for $\Delta \lesssim |\dmin|$.} In this case, we will only focus on forms that exhibit slow growth. The subleading corrections to $d(n, \ell; t)$ can be obtained from \eqref{eq:sym-saddle}. However, in contrast with the previous case, we now need information about $\varphi_{\infty}$. 

Since $|\dmin| \gtrsim \Delta\gg1$, the effective potentials \eqref{eq:symprodsaddle} are small, which implies that \eqref{eq:sym-saddle} is sensitive to the high-temperature behavior of $\varphi_{\infty}(\tau,z)$. The latter is given by a ratio of theta and Dedekind-eta functions, whose high-temperature behavior can be extracted from its low-temperature regime. In App.\,\ref{sec:omega}, we show that the relevant contribution to $\varphi_\infty(\tau,z)$ is schematically given by
\eq{\label{eq:symeta}
\varphi_{\infty}(\tau,z)\sim\eta(\tau)^{c_o(0,0)}\eta(\tau-z)^{-c_o(0,0)-2c_o(0,-b_o)}\eta(z)^{-c_o(0,0)-2c_o(0,-b_o)}\,,
}
where $\eta(\tau)$ is the Dedekind-eta function, and that the weight of $\varphi_\infty(\tau,z)$ is
\eq{\label{eq:weight-sym}
\text{weight of } \varphi_\infty = 2c_o(0,-b_o)+c_o(0,0)/2\,.
}
Using these ingredients in the regime $|\dmin| \gtrsim \Delta\gg1$, we find that \eqref{eq:sym-saddle} gives
\eq{\label{eq:sym-non-uni}
d(n,\ell;t) \approx \left(\frac{\Delta}{|\dmin|}\right)^{c_o(0,0)/4}\frac{\Delta^{c_o(0,-b_o)}}{\Delta|\dmin|} e^{4\pi\sqrt{\Delta\abs{\dmin}}}\,,\qquad |\dmin|\gtrsim\Delta \gg 1 
\,.
}
In contrast to \eqref{eq:subleadingcorr-sym}, information about the density of light states encoded in the properties of $\varphi_\infty$ was crucial to determine the subleading corrections to the asymptotic growth of states.

The result in \eqref{eq:sym-non-uni} is in perfect agreement with \eqref{eq:democraticresult}, a democracy, once we take into account the scaling of the saddle-point \eqref{eq:lambdastar}. This provides a nontrivial verification of the methods used in Sec.\,\ref{sec:slowgrowth}. In particular, it is not difficult to check that due to the weight of $\varphi_\infty$ in \eqref{eq:weight-sym}, applying a Laplace transform on $\varphi_\infty$ yields a density of light states of the form \eqref{eq:rholightfg} with $f$ and $g$ as in \eqref{eq:assumptions1}, $\rho_0$ given in \eqref{eq:rho0-symm}, and
\eq{ \label{se3:omegaval}
\omega=-c_o(0,-b_o)-c_o(0,0)/4-3/2\,.
}

\subsection{Exponential lifts}\label{sec:exp-lift-log}

In this section we will use the exponential lift as a tool to construct wJfs with weight $k$ and a large index. The exponential lift is closely related to the symmetric product orbifold construction of wJfs. Given a seed wJf $\varphi(\tau, z)$, the exponential lift of $\varphi(\tau, z)$ is constructed as follows
\eq{
\text{Exp-Lift}(\varphi)(\tau,z,\sigma)\coloneqq q^Ay^Bp^{t_oA}\prod_{(n,\ell,m)>0}\left(1-q^ny^\ell p^{t_om}\right)^{c_o(nm,\ell)}\,.\label{eq:explift3}
}
Here $(n,\ell,m)>0$ means $n,m\in\mathbb{Z}_{\ge0}$ and $\ell\in\mathbb{Z}$ such that $m>0\vee (m=0\wedge n>0)\vee(n=m=0\wedge \ell<0)$, while $p$, $A$, and $B$ are defined by
\eq{
p= e^{2\pi i \sigma}\,,\qquad A\coloneqq \frac{1}{24}\sum_{\ell\in\mathbb{Z}}c_o(0,\ell)\,,\qquad B\coloneqq\frac{1}{2}\sum_{\ell\in\mathbb{Z}_{>0}}\ell c_o(0,\ell)\,.} 
The exponential lift is a meromorphic modular form of weight $k\coloneqq\frac{1}{2}c_o(0,0)$ with respect to the paramodular group $\Gamma^+_{t_o}$ \cite{Gritsenko:1996tm}, see App.\,\ref{app:SMF} for details. In particular, the inverse of the generating function of the symmetric product $\mathcal{Z}$ can be recognized as a factor of \eqref{eq:explift3}. In what follows we will focus on slow-growing forms, obtained from the inverse of the exponential lift, as defined in \cite{Belin:2019jqz}. 

One can find the asymptotic expansion of the Fourier coefficients of the inverse of the exponential lift (which has weight $-k$) along the lines of Sec.\,\ref{sec:symexp},
\eq{
\Phi(\tau,z,\sigma)\coloneqq \frac{1}{\text{Exp-Lift}(\varphi)(\tau,z,\sigma)}=\sum_{t,n,\ell}d_{\Phi}(n,\ell;t)p^tq^ny^\ell\,.
} 
In evaluating the inverse Laplace transform of $\Phi$, we follow the same steps as in Sec.\,\ref{sec:symexp}. In this case, the dominant pole is located at \eqref{eq:zero}, and its residue is controlled by
\eq{
\varphi^\Phi_\infty(\tauh,\zh)=\hat{q}^{-A}\hat{y}^{-B-b_oA}\prod_{\ell<0}\left(1-\hat{y}^\ell\right)^{-c_o(0,\ell)}\prod_{\substack{n\ge0\\\ell\in\mathbb{Z}\\\left(n, \ell\right)\neq(0,0)}}\left(1-\hat{q}^n\hat{y}^{\ell}\right)^{-f^\Phi\left(n, \ell\right)}\,,\label{eq:chiinfEL}
}
with
\eq{
f^\Phi\left(n, \ell\right)\coloneqq\sum_{m=0}^\infty c_o(nm,\ell-b_om)\,.
}
This can be seen by comparing to \eqref{eq:chiinf} and using the relation between symmetric products and exponential lifts \eqref{eq:symel}. 
Compared to the analysis of Sec.\,\ref{sec:symexp}, the residue that for symmetric products is characterized by $\varphi_\infty$, is now multiplied by a form of weight $-k$. $\varphi_\infty$ counts polar states in the limit $t\to \infty$ \cite{Belin:2019jqz}, and it ties elegantly to wall crossing for $\Phi_{10}$ and CHL models \cite{Cheng:2007ch,Sen:2011mh}.   By carefully tracking the effects of the weight through the computation, we can then reproduce the Rademacher kernels for weighted wJfs up to exponentially suppressed corrections (see eqs. \eqref{eq:cross-we} and \eqref{eq:rademacherap} for the explicit form of the kernels). For the special case of the Igusa cusp form, the Rademacher expansion has been reproduced exactly in \cite{Cardoso:2021gfg,LopesCardoso:2022hvc}.

As in the case for wJfs constructed via a symmetric product orbifold, we can connect to the analysis of Sec.\,\ref{sec:beyondcardy}. We can find a reliable approximation to the coefficients of sparse wJfs constructed via an exponential lift by evaluating the inverse Laplace transform by saddle point, leading to
\begin{equation}\label{eq:EL-saddle}
   d_\Phi(n,\ell;t) = \frac{1}{2\Delta}
    \sqrt{\frac{|\dmin|}{t}}\left(\frac{\dmina}{\Delta}\right)^{\frac{k}{2}}
    \e^{\pi i \frac{ b\ell}{t}}\e^{4\pi \sqrt{\Delta |\dmin|}}
    \frac{N^{c_o(0,-b_o)-1}}{(c_o(0,-b_o)-1)!}\varphi^\Phi_\infty(\hat\tau_{\star},\hat z_{\star})+\cdots\,,
\end{equation}
where the location of the saddle is once again given by \eqref{eq:symprodsaddle}. This equation should be compared to \eqref{eq:sym-saddle}.

Using steps similar as those of Sec.\,\ref{sec:symprodlog}, we arrive at the following asymptotic expansion for the $d_\Phi(n, \ell; t)$ coefficients in the universal regime
\eq{\label{eq:dphi-uni}
d_{\Phi}(n,\ell;t)\approx   \left(\frac{|\dmin|}{\Delta}\right)^{c_o(0,0)/4} \frac{|\Delta_0|^{c_o(0,-b_o)}}{\Delta|\dmin|} \e^{4\pi \sqrt{\Delta \dmina}}\,, \qquad \Delta \gtrsim\abs{\dmin}\,,
}
and the following one for the non-universal regime
\eq{\label{eq:dphi-non-uni}
d_{\Phi}(n,\ell;t)\approx\left(\frac{\Delta}{|\dmin|}\right)^{c_o(0,0)/4}\frac{\Delta^{c_o(0,-b_o)}}{\Delta|\dmin|} e^{4\pi\sqrt{\Delta\abs{\dmin}}}\,, \qquad |\dmin|\gtrsim\Delta \gg 1
\,.
}
An important difference with the analysis of wJfs constructed via a symmetric product orbifold is that, due to the weight of the exponential lift, the parameter $\omega$ given in \eqref{se3:omegaval} is shifted by the weight of $\Phi$,
\eq{
\omega_{\Phi}=\omega-k\,.
}
Note that the expressions \eqref{eq:dphi-uni} and \eqref{eq:dphi-non-uni} apply only when $\Phi(\tau, z ,\sigma)$ is slow growing \cite{Belin:2019jqz}.
Corresponding expressions can be obtained for fast-growing forms, which we omit for brevity. 

Interestingly, the logarithmic corrections to the leading Cardy exponent only differ from those obtained from the symmetric product orbifold in the Cardy regime $\Delta\gg\dmin$. In the non-universal regime, the modifications due to the weighted kernels and the shift in $\omega$ cancel against each other such that \eqref{eq:dphi-non-uni} has the same form as \eqref{eq:sym-non-uni}. Finally, we note that this result agrees perfectly with the results of \cite{Belin:2016knb}.

\section{Logarithmic corrections to the entropy of BPS  black holes} \label{sec:bhs}

We now turn to the gravitational side of the analysis and revisit the microscopic nature of the logarithmic corrections to the entropy of 4D and 5D BPS black holes in ungauged supergravity. These corrections can be computed using the quantum entropy function, which is the Euclidean path integral in the near-horizon AdS$_2$ region of the black hole.\footnote{For a review of the original work see \cite{Sen:2014aja}.}  In this setup, the logarithmic corrections appear as a scaling effect. To illustrate this, consider a 4D BPS black hole where the area of the horizon  is a function of charges $q_i$ such that  rescaling $q_i\mapsto \Lambda q_i$ induces 
\begin{equation}
    A_H(q_i) \mapsto \Lambda^2 A_H(q_i)\,. 
\end{equation}
When this rescaling is implemented in the quantum entropy function, it shows that there is a correction to the black hole entropy of the form
\begin{equation}
    S_{BH}(q_i) \mapsto \cdots + a_{\scaleto{\rm grav}{4pt}} \log \Lambda^2 + \cdots \,.
\end{equation}
The logarithmic term arises as a one-loop correction to the leading saddle point in the path integral that originates from massless fields. This correction is determined entirely by the low energy effective action and is therefore independent of the UV completion. Thus, the logarithmic corrections to the entropy provide an infrared window into the microstates of the black hole.  

There are several cases where $a_{\scaleto{\rm grav}{4pt}}$ is matched with a microscopic description, where the microscopic counting formula is a weak Jacobi form. In these cases, the near-horizon AdS$_2$ geometry can be uplifted to AdS$_3$, and we would like to see what the match says about AdS$_3$/CFT$_2$.  In the following, we revisit the ingredients that control the logarithmic corrections on both the macroscopic and microscopic sides. In particular, we highlight the cases where $a_{\scaleto{\rm grav}{4pt}}$ is in the universal or non-universal regimes, and demonstrate how the logarithmic corrections could have been predicted from an AdS$_3$ perspective. In that process, we also highlight what data of the CFT$_2$ determines the logarithmic corrections to the black hole entropy.

Before we proceed, it is important to describe a feature of $a_{\scaleto{\rm grav}{4pt}}$ that will contrast the analysis for 4D and 5D black holes. Generically, this coefficient receives two contributions 
\begin{equation}
 a_{\scaleto{\rm grav}{4pt}}= a_{\rm local} + a_{\rm zm}\,,
\end{equation}
where $a_{\rm local}$ is the contribution from non-zero eigenvalues of the one-loop determinant while $a_{\rm zm}$ accounts for zero modes in the path integral. For an even-dimensional background, both of these terms contribute. As a result, in four dimensions the logarithmic corrections are sensitive to the supergravity spectrum. Following the conventions in \cite{Charles:2015eha}, we write
\begin{equation}\label{eq:all-4d}
     a_{\rm local,4D} = \frac{1}{12}\left(11(3-{\cal N})-n_V+n_H\right)\,, \qquad a_{\rm zm,4D} = \frac{1}{2}(-6 +8)\,,
\end{equation}
where ${\cal N}$ denotes the number of supercharges while $n_V$ and $n_H$ are the number of ${\cal N}=2$ vector and hyper multiplets, respectively. The $a_{\rm local,4D}$ term is evaluated using the heat kernel method where the contribution of modes with zero eigenvalue has been explicitly removed. For this reason, the zero modes in $a_{\rm zm,4D}$ come solely from the gravity multiplet, with the negative contribution arising from the six isometries of the metric while the positive one comes from the gravitinos. 

For an odd-dimensional background, the local contribution to $a_{\scaleto{\rm grav}{4pt}}$ always vanishes. Therefore, the supergravity spectrum plays no role. In particular,  for the five-dimensional BMPV black hole, we have \cite{Sen:2012cj} 
\begin{equation}\label{eq:all-5d}
    a_{\rm local, 5D}=0\,,\qquad a_{\rm zm, 5D}= -\frac{1}{6}\left(\mathfrak{n}_{V}+21 -24\right)\,.
\end{equation}
In this case, the vector fields, metric, and gravitinos have zero modes that contribute to the $a_{\scaleto{\rm grav}{4pt}}$, with each contribution respectively highlighted. In the language of ${\cal N}=2$ 5D supergravity, the total number of vector fields is $\mathfrak{n}_V=n_V^{5d}+2n_S^{5d}+1$, where one contribution comes from the graviton multiplet, while $n_V^{5d}$ and $n_S^{5d}$ are respectively the number of vector and gravitino multiplets. We note that $a_{\rm zm, 5D}$ is independent of any other aspect of the supergravity theory.

\subsection{1/4-BPS black holes in 4D \texorpdfstring{$\mathcal{N}=4$}{N = 4} supergravity}\label{sec:4D-N4}

Let us begin by considering the logarithmic corrections to 1/4-BPS black holes in 4D $\mathcal N = 4$ supergravity \cite{Banerjee:2011jp}. This theory can be obtained by a compactification of type IIB supergravity on K3$\,\times\,T^2$. One way to characterize these black holes is as dyonic solutions which carry electric charges $\vec Q$ and magnetic charges $\vec P$. Following the conventions of \cite{Sen:2007qy}, the area law for these black holes reads
\begin{equation}
    \frac{A_H}{4G_4} = \pi \sqrt{\vec Q^2 \vec P^2 - (\vec Q\cdot \vec P)^2}\,.
\end{equation}
The logarithmic corrections in this case are parametrized by
\begin{equation}
    a_{\rm local, 4D} = -1 \,,\qquad  a_{\rm zm, 4D}=1\,.
\end{equation}
where we used the fact that $n_V=n_H+1$ for ${\cal N}=4$ supergravity. Famously, it follows that the logarithmic correction vanishes
\eq{\label{eq:sbh-4}
S_{\rm BH} =  \frac{A_H}{4G_4} + 0 \times \log(\frac{A_H}{G_4}) + \cdots \,.
}

On the microscopic side, the relevant counting formula is the reciprocal of the Igusa-cusp form $\Phi_{10}$, which is the exponential lift of the wJf  $2\varphi_{0,1}$ \cite{Dijkgraaf:1996it,Sen:2007qy}.  Hence, this counting formula falls into the cases discussed in Sec.\,\ref{sec:exp-lift-log}, and it is instructive to see how the parameters are related, and the logarithmic corrections matched, to the macroscopic result \eqref{eq:sbh-4}. For the seed theory we have $t_o=b_o=1$, $c_o(0,-b_o)=2$, and $c_0(0,0)=20$. The relation between the parameters used in Sec.\,\ref{sec:exp-lift-log} and the gravitational charges is \cite{Dijkgraaf:1996it,Sen:2007qy}
\begin{equation}
    \dmin= -\frac{t}{4}=- \frac{\vec Q^2}{8}\,, \qquad n = \frac{\vec P^2}{2} \,, \qquad \ell= \vec Q\cdot \vec P\,.
\end{equation}
 The rescaling of parameters relevant to match the gravitational computation is 
\begin{equation}
    (\vec Q, \vec P) \mapsto  (\Lambda\vec Q, \Lambda\vec P) \quad \Rightarrow \quad   (\dmina, \Delta) \mapsto  (\Lambda^2\dmina, \Lambda^2\Delta)\,,
\end{equation}
where $\Lambda$ is large. This indicates that we are in the universal regime such that  $\Delta\sim \dmina$. Using \eqref{eq:dphi-uni} we then see that
\begin{equation}\label{eq:dphi}
   \log d_{\Phi_{10}}(n,\ell;t)= 4\pi \sqrt{\Delta \dmina} + (c_o(0,-1)-1)\log(\dmina) - \log \Delta + \cdots\,,
\end{equation}
which perfectly reproduces \eqref{eq:sbh-4} since $c_o(0,-1)=2$. This is the result originally reported in \cite{Banerjee:2011jp}.

Let us now revisit this success from an AdS$_3$/CFT$_2$ perspective. The near-horizon AdS$_2 \times S^2$ region of the black hole can be uplifted to an AdS$_3 \times M$ spacetime that is expected to describe the near-horizon region of a black string. In this case, the AdS$_2 \times S^2$ geometry is uplifted to the near horizon region of the 6D BPS black string with charges $Q_1$, $Q_5$, and momentum $n$ considered in  \cite{Strominger:1996sh}. The near-horizon region of this black string is described by
\eq{\label{se4:ads3s3}
\text{AdS}_3 \times S^3 \times \text{K3}\,. 
}
The spectrum of IIB supergravity on an $\text{AdS}_3 \times S^3$ spacetime is organized into short multiplets of $SU(1,1|2)_L \times SU(1,1|2)_R$, i.e.~into chiral primary representations of the $\mathcal N = (4,4)$ superconformal algebra. The elliptic genus provides a signed count of these states which are all 1/4-BPS from the point of view of the $\mathcal N = (4,4)$ algebra. The contributions of the light states to the elliptic genus of type IIB supergravity on $\text{AdS}_3 \times S^3 \times  \text{K3}$ were computed in \cite{deBoer:1998kjm} and shown to match those of the symmetric product orbifold of the nonlinear sigma model on K3,
\eq{\label{eq:Nq1q5}
\text{Sym}^N \text{K3}\,, \qquad N = Q_1 Q_5+1\,,
}
where $Q_1$ and $Q_5$ are large. In fact, the supergravity spectrum matches  the regularized degeneracies of polar states in the $N\rightarrow\infty$ limit denoted by $d_\infty$ in \eqref{eq:chiinf}. The elliptic genus of the nonlinear sigma model on K3 is given by $2\varphi_{0,1}(\tau, z)$. Therefore, the wJf associated to $\text{Sym}^N \text{K3}$ has vanishing weight and index 
\eq{\label{eq:tk3}
t = N t_o = N\,.
}
In the notation of Sec.\,\ref{sec:4D-N4} we have $N=\vec Q^2/2$. The relationship between the generating function of the symmetric product and $\Phi_{10}$ is
\eq{ \label{eq:ZK3}
\mathcal{Z}(\tau,z,\sigma)=\sum_{N\in\mathbb{Z}_{\ge0}} p^{N}\varphi\left(\tau,z;\text{Sym}^N(\text{K3})\right)=\frac{p\,\phi_{10,1}(\tau,z)}{\Phi_{10}(\tau,z,\sigma)}\,,
}
where $\phi_{10,1}(\tau,z)=\eta(\tau)^{18}\vartheta_{1,1}^2(\tau,z)$ is a wJf of weight 10 and index 1. 

The logarithmic correction is controlled by $c_o(0,-1)=2$, which is the degeneracy of ground states in the elliptic genus of K3 (as opposed to that of Sym$^N$ K3) and is determined entirely by the symmetries of the theory. This follows from the fact that the ground states in the elliptic genus of $\mathcal N = (4,4)$ CFTs take the form
\eq{
|0 \rangle_L \times |\text{chiral primary}\rangle_R, \label{ramondGS}
}
where $|0 \rangle_L$ and $|\text{chiral primary}\rangle_R$ are the trivial and chiral primary representations of the $SU(1,1|2)_L \times SU(1,1|2)_R$ supergroup. The only states of the form \eqref{ramondGS} are either the vacuum (empty AdS$_3 \times $ S$^3$) or the generators of the right-moving part of the $R$-symmetry group. For the $\mathcal N = (4,4)$ superconformal algebra, only one of the generators of the $SU(2)_R$ R-symmetry belongs to a chiral primary representation. As a result, there are only 2 ground states in the elliptic genus of K3 such that $c_o(0,-1) = 2$.\footnote{By the same reasoning, the elliptic genus of a theory with only an $\mathcal N = (2,2)$ symmetry algebra admits only one ground state, namely the vacuum.} 

The degeneracy of ground states in the elliptic genus of K3 translates into the following ground state degeneracy in the elliptic genus of Sym$^N$ K3 (see \eqref{eq:rho0-symm})
\eq{ \label{eq:rhovac}
\rho_0(\dmin) \sim \dmina\,.
}
As described above, this result is independent of K3 and a consequence of supersymmetry. In fact, \eqref{eq:rhovac} holds even if the CFT is not a symmetric product orbifold --- it is an outcome of the $SU(2)_R$ R-symmetry group. In order to see this, we note that under spectral flow, the NS vacuum flows to a lowest $SU(2)_R$ weight state in the Ramond sector with $\bar J_0^3 = -c_R/12$, where $c_R$ is the right-moving central charge. Consequently, the degeneracy of the ground state is $2|\bar J_0^3| + 1 \sim \dmina$, from which \eqref{eq:rhovac} follows. 

Although the logarithmic corrections to the entropy of 4D $\mathcal N = 4$ black holes vanish, it is interesting to contrast the origins of this result from the AdS$_3$ and AdS$_2$ perspectives.  An important aspect to keep in mind is that we are working in the \textit{universal regime} $\Delta \sim \dmina$: from our findings in Sec.\,\ref{sec:beyondcardy}, we expect the logarithmic corrections to depend only on the ground state degeneracy $\rho_0(\dmin)$ and to be insensitive to the distribution of the polar states.  From the point of view of the near-horizon AdS$_2 \times S^2$ geometry, the vanishing logarithmic correction is the result of a careful cancellation between the local contribution of massless fields and the contribution of zero modes. This cancellation is due to the constrained spectrum of ${\cal N}=4$ theories. From the point of view of AdS$_3$ (or the dual CFT$_2$), the degeneracy of Ramond ground states is universally determined by the $SU(2)$ R-symmetry group, and leads to vanishing logarithmic corrections. In particular, the information about the massless spectrum does not enter in the AdS$_3$ calculation. In the next section we will see an example where this situation is reversed, and it is the near-horizon AdS$_2$ calculation that is universal, while the AdS$_3$ analysis depends on the details of the theory.

\subsection{1/4-BPS black holes in 5D \texorpdfstring{$\mathcal{N}=4$}{N = 4} supergravity}\label{sec:bh-5d-n4}

Let us now turn to the logarithmic corrections of 1/4-BPS black holes in 5D $\mathcal N = 4$ supergravity \cite{Sen:2012cj}. These are the black holes originally conceived  in \cite{Strominger:1996sh,Breckenridge:1996is}. We will focus on the BMPV solution which carries electric charges $(Q_1,Q_5,n)$ and angular momentum $(J)$, following the conventions in \cite{Castro:2008ys,Sen:2012cj}. The near-horizon region of these black holes is described locally by an AdS$_2\times S^3$ geometry. The area law in this case is 
\begin{equation}
    \frac{A_H}{4G_5}=2\pi \sqrt{Q_1 Q_5 n - J^2}\,.
\end{equation}
When the ${\cal N}=4$ theory arises as a compactification of type IIB string theory on $K3\times S^1$, the number of vector fields is $ \mathfrak{n}_V=22+2\times 2+1=27$, and therefore $a_{\rm zm,5D}=-4$ in \eqref{eq:all-5d}.\footnote{This compactification is dual to Heterotic strings on $T^5$ or M-theory on $K3\times T^2$.} As a result, the leading expression for the entropy is given by
\begin{equation}\label{eq:ent5d}
S_{\rm BH}=  \frac{A_H}{4G_5} - 4  \log(\frac{A_H}{G_5}) + \cdots\,.
\end{equation}

In analogy with the previous section, we can uplift the near-horizon AdS$_2 \times S^3$ region of the black hole to an AdS$_3 \times S^3$ geometry. We expect the entropy of the black hole to be captured by the elliptic genus of $\text{Sym}^N K3$ such that the relevant counting formula is given in \eqref{eq:ZK3}. Using the fact that $\ell = J$, together with \eqref{eq:Nq1q5}, \eqref{eq:tk3}, and the results of Sec.~\ref{sec:symexp}, we find that in the large charge limit
\begin{equation}
\Delta = n- \frac{J^2}{Q_1 Q_5}\,, \qquad \dmina = \frac{Q_1Q_5}{4}\,.
\end{equation}
As before, the supergravity spectrum on AdS$_3 \times S^3$ captures the regularized density of polar states in the large-$N$ limit \cite{deBoer:1998kjm}; for symmetric product orbifolds this is $d_\infty$  given by \eqref{eq:chiinf}.

The crucial difference between the analyses of this and the previous section lies in the different scaling limits. The relevant rescaling needed to obtain \eqref{eq:ent5d} from a five-dimensional gravitational path integral is
\begin{equation}
   Q_1 \mapsto \Lambda Q_1\,,\qquad Q_5 \mapsto \Lambda Q_5\,,\qquad n \mapsto \Lambda n\,,\qquad J \mapsto \Lambda^{3/2}J\,,
\end{equation}
where $\Lambda$ is large. In terms of the discriminant, this implies that
\begin{equation}\label{se4:nonuniversal1}
      \dmina \mapsto \Lambda^2\dmina \,,\qquad  \Delta \mapsto \Lambda \Delta\,.
\end{equation}
This scaling corresponds to the non-universal regime where $\dmina\gg\Delta$. 
The asymptotic density of states of a wJf in the non-universal regime \eqref{se4:nonuniversal1} can be obtained from the microcanonical or canonical ensembles as described in Sec.\,\ref{sec:beyondcardy} and Sec.\,\ref{sec:symmN}, respectively. In either case, the asymptotic density of states is sensitive to the distribution of light states and is given by
\begin{equation}\label{se4:cnl2}
    c(n,\ell) \approx   \frac{\rho_0(\dmin)}{\Delta}\sqrt{\frac{|\dmin|}{t}} \left(\frac{|\dmin|}{\Delta}\right)^{\omega+\frac{3}{2}} e^{4\pi\sqrt{\dmina\Delta}}\,,
\end{equation}
where $\rho_0(\dmin)$ and $\omega$ are determined from the density of states $c_o(n, \ell)$ of the K3 seed by \eqref{eq:rho0-symm} and \eqref{se3:omegaval}, which we reproduce here for convenience
\eq{
\omega = -c_o(0,-1) - \frac{c_o(0,0)}{4} - \frac{3}{2}\,, \qquad \rho_0(\Delta_0)\approx |\Delta_0|^{c_o(0,-1)-1} \,.
}
For the elliptic genus of K3, the number of Ramond ground states  is given by $c_o(0,-1)=2$, while 
\eq{
c_o(0,0) = h^{1,1} = 20\,, 
}
with $h^{1,1}$ the Hodge number of K3.
As a result, the asymptotic density of states of $\text{Sym}^N \text{K3}$ in the non-universal regime \eqref{se4:nonuniversal1} is given by
\eq{\label{eq:logK35d}
\log c(n,\ell) =4\pi\sqrt{\dmina\Delta} - 6 \log\dmina + 6 \log \Delta + \dots\,.
}
The first term corresponds to the area of the black hole in Plank units while the second and third terms encode the logarithmic corrections to the entropy. In terms of the area $A_{H} \sim \Lambda^{3/2}$, $\dmina\sim \Lambda^2$, and $\Delta\sim \Lambda$, we find that \eqref{eq:ent5d} and \eqref{eq:logK35d} perfectly agree.

It is important to reinforce that, in contrast to the 4D case, heat kernels in odd dimensions do not contribute to the logarithmic corrections of the black hole entropy. From the five-dimensional AdS$_2 \times S^3$ perspective, these logarithmic corrections arise instead from the contributions of zero modes associated with the symmetries of the background. In this sense, the 5D result is universal. Conversely, the AdS$_3$/CFT$_2$ calculation is not universal, as it is sensitive to the distribution of the light states $\rho_\light(\nn, \jj)$, as can be explicitly seen from its dependence on the light state data parametrized by $\omega$ in \eqref{se4:cnl2}. From an AdS$_3\times S^3$ perspective, we need input from all the matter fields in supergravity that contribute to the elliptic genus, not just zero modes. From a CFT$_2$ perspective, \eqref{se4:cnl2} is sensitive to the exponential and polynomial growth of the light spectrum, in addition to its dependence on the degeneracy of the Ramond ground state. The fact that the macroscopic computation in 5D matches the microscopic counting in the CFT$_2$ is remarkable and not a coincidence.

\subsection{1/2-BPS black holes in 4D \texorpdfstring{$\mathcal{N}=2$}{N=2} supergravity}\label{sec:n2bhs}
 
Let us now turn to a more delicate case: the logarithmic corrections to the entropy of 1/2-BPS black holes in 4D $\mathcal N =2$ supergravity \cite{Sen:2012kpz}. Following the conventions in \cite{Pioline:2006ni}, the black hole carries magnetic charges $p^I$ with $I=1,\ldots,n_V$, and one electric charge $q_0$. Viewing this as a solution of Type IIA supergravity on $CY_3$, the charges $p^I$ and $q_0$ correspond to the charge of D4-branes and a D0-brane, respectively. On the other hand, from the point of view of M-theory on $CY_3\times S^1$, the $p^I$ are M5-brane charges wrapping four-cycles of $CY_3$, while $q_0$ is the momentum along the M-theory direction. The area law is then given by
\begin{equation}
    \frac{A_H}{4G_4}=2\pi\sqrt{q_0 P^3}\,,
\end{equation}
where $P^3\coloneqq c_{IJK} p^I p^J p^K$ and $c_{IJK}$ are the intersection numbers of the $CY_3$. It is possible to add additional electric charges $q_I$ associated to D2-branes in Type IIA (M2-branes in M-theory). The entropy remains of the same form where now $q_0\mapsto \hat q_0=q_0 + \frac{1}{12} c^{IJ}q_I q_J$ and $c^{IJ}$ is the inverse of $c_{IJ}=c_{IJK}p^K$. The leading expression for the entropy including the logarithmic correction follows from \eqref{eq:all-4d} and is given by \cite{Sen:2012kpz}
\eq{ 
S_{\rm BH} = \frac{A_H}{4G_4}+ \frac{1}{12} \big( 23 - n_V + n_H \big) \log \bigg(\frac{A_H}{G_4} \bigg)+\cdots\,. \label{se4:CY_log}
}
The entropy was obtained from the quantum entropy function applied to the two-derivative ${\cal N}=2$ supergravity in 4D. In particular, this result applies in the limit where all the charges of the theory scale uniformly with a large parameter $\Lambda$ such that
\eq{\label{eq:scaling-n2bhs}
(q_0, q_I, p^I) \mapsto (\Lambda q_0, \Lambda q_I, \Lambda p^I)\,,
} 
and hence $A_H\to \Lambda^2 A_H$. It was argued in \cite{Sen:2012kpz} that the logarithmic corrections are compatible with the OSV formula \cite{Ooguri:2004zv} refined in \cite{Denef:2007vg}. Here we would like to revisit some of the ingredients that control this agreement.

The logarithmic correction to the entropy of the black hole is sensitive to the massless degrees of freedom in the near-horizon AdS$_2 \times S^2$ region. This geometry can be uplifted to an AdS$_3 \times S^2$ spacetime with $\mathcal N = 2$ supersymmetry. The corresponding five-dimensional theory can be obtained from compactification of $M$-theory on a $CY_3$, with the desired AdS$_3 \times S^2$ background arising from the near-horizon limit of the wrapped $M5$-branes described above. The resulting supergravity theory is expected to be dual to an ${\cal N}=(0,4)$ CFT$_2$ whose left and right-moving central charges are given by \cite{Maldacena:1997de}
\eq{ \label{eq:centralcharge}
c_L = 6 P^3 + c_2\cdot p\,,\qquad c_R = 6 P^3 +\frac{1}{2} c_2\cdot p\,, 
}
where $c_{2I}$ is the second Chern class of $CY_3$. It is important to note that the symmetry algebra of the dual CFT consists of a small $\mathcal N = 4$ superconformal algebra that is enhanced by the presence of an additional ``center of mass'' multiplet \cite{Minasian:1999qn}. This multiplet includes four right-moving bosons and four right-moving Goldstinos, and it plays a significant role in our analysis.

Following \cite{deBoer:2006vg}, the relevant index we should consider is 
\begin{equation}\label{eq:EG(0,4)}
  Z\left(\tau,\bar\tau,z^I\right) = {\rm Tr}_{R}\Big(F^2 (-1)^F e^{2\pi i \tau (L_0 -c_L/24)}e^{-2\pi i \bar \tau (\bar L_0 -c_R/24)} e^{2\pi i z^I \ell_I}\Big)\,,
\end{equation}
where we identify
\eq{
q_0 = L_0 -\frac{c_L}{24} - (\bar L_0 -\frac{c_R}{24}) \,, \qquad \ell_I = q_I\,.
}
$Z\left(\tau,\bar\tau,z^I\right)$ is a modular form of rank $n_V$ that is equipped with a lattice specifying the quantization of the charges, with $c_{IJ}:=c_{IJK}p^K$ controlling the lattice spacing \cite{Maldacena:1997de,deBoer:2006vg,Gomes:2019vgy}.
There are three subtle aspects of $Z(\tau,\bar\tau,z^I)$ that are important to mention and are explained in more detail in \cite{deBoer:2006vg}. First, the term $F^2$ inside the trace is necessary to absorb the contribution of the four fermion zero modes from the center of mass multiplet, the latter of which make the elliptic genus (without the $F^2$ insertion) vanish. Second, there is a continuous part of the spectrum associated with three of the bosonic zero modes in the center of mass multiplet that describe the motion of the M5 branes in $\mathbb R^3$. The removal of these states makes $ Z(\tau,\bar\tau,z^I)$ a weight $(-3/2,1/2)$ form with respect to $\tau$ and $\bar\tau$. Third, the $\bar\tau$ dependence is controlled by $z^I$ and is determined by a heat equation. For our purposes, this means that we can set $\bar\tau=\tau$ so that we are working with a wJf that we denote by
\eq{\label{eq:EG(0,4)2}
Z(\tau, z^I) \coloneqq Z(\tau, \bar\tau = \tau, z^I) = \sum_{n, \ell_I} c(n, \ell_I) e^{2\pi i\tau n + 2\pi i z^I \ell_I}\,,
}
which has rank $n_V$, weight $-1$, and matrix index $t_{IJ}\sim c_{IJ}$. In addition,  we have $n=q_0$, $\Delta = \hat q_0$, and $\Delta_0 =- c_L/24$.

We would like to establish whether the $c(n,\ell_I)$ coefficients of \eqref{eq:EG(0,4)2} can reproduce \eqref{se4:CY_log}, given what we know about this instance of AdS$_3$/CFT$_2$. From the analysis of crossing kernels in Sec.\,\ref{sec:kernel}, we know that a convenient way to cast the asymptotic expansion of the $c(n,\ell_I)$ coefficients is   
\begin{equation}\label{eq:cnl_I2}
    c(n,\ell_I)=   \int \dd \nn'\,  \int\prod_{I}\dd \textsf{r}_I\!\,' \,
    \rho_{\light} (\nn',\textsf{r}_I\!\,' )\,
    \mathbb{P}_{\{n,\ell_I\};\{\nn',\textsf{r}_I\!\,'\}}^{(-1,n_V)}+ \cdots \,,
\end{equation}
where $\rho_{\light} (\nn,\textsf{r}^i )$ is the density of light states of $Z(\tau, z^I)$ and $ \mathbb{P}_{\{n,\ell_i\};\{\nn',\textsf{r}_i\!\,'\}}^{(k,M)}$ is the crossing kernel of a wJf of rank $M$ and weight $k$ given in \eqref{eq:crossing-rankM}. In particular, when $|\Delta \Delta'| \gg 1$, the crossing kernel is approximately given by
\eq{\label{eq:crossingkernelnV}
\mathbb{P}_{\{n,\ell_I\};\{\nn',\textsf{r}_I\!\,'\}}^{(-1,n_V)} = \sqrt{\frac{1}{2} \det\bigg(\frac{t^{IJ}}{2}\bigg)} \bigg(\sqrt{-\frac{\Delta}{\Delta'}}\bigg)^{-2-\frac{n_V}{2}} e^{-i\pi(t^{IJ}\ell_I \textsf{r}_J')} \frac{e^{4\pi\sqrt{-\Delta \Delta'}}}{(-\Delta \Delta')^{\frac{1}{4}}} + \dots\,,
}
where $t^{IJ}$ is the inverse of the matrix index $t_{IJ}$.

The first step in evaluating \eqref{eq:cnl_I2} is to establish what the scaling limit \eqref{eq:scaling-n2bhs} does to the parameters defining $c(n,\ell_I)$ and $Z(\tau, z^I)$. In this scaling limit, the central charge \eqref{eq:centralcharge} scales as $c_L \mapsto \Lambda^3 c_L$ such that, in terms of the discriminant, we find
\eq{\label{eq:regimeN2}
\Delta \mapsto \Lambda \Delta\,, \qquad |\Delta_0|  \mapsto \Lambda^3 |\Delta_0|\,.
}
Since $\Delta \ll |\Delta_0|$ in the large-$\Lambda$ limit, we are working in the non-universal regime described in  Sec.\,\ref{sec:slowgrowth}. This means that the subleading behavior of the $c(n,\ell_I)$ coefficients is sensitive to the distribution of  light states. 

Our next task is to characterize the distribution of light states in the modified elliptic genus \eqref{eq:EG(0,4)2}. The light part of \eqref{eq:EG(0,4)2} receives two types of contributions at large $c_{L/R}$:  perturbative contributions from the supergravity spectrum and non-perturbative ones from M2 and anti-M2 branes wrapping two-cycles of $CY_3$ \cite{Gaiotto:2006ns}. The polar part of the modified elliptic genus which determines the distribution of light states can therefore be written as
\eq{\label{eq:polarMEG}
Z_{\text{light}}(\tau, z^I) \coloneqq  Z_{\text{sugra}}(\tau, z^I) Z_{M2}(\tau, z^I )\,.
}
In the dilute gas approximation where the M2 and anti-M2 contributions decouple, the general form of $Z_{M2}(\tau, z^I)$ is given in \cite{Gaiotto:2006ns}. An explicit expression for $Z_{M2}(\tau, z^I)$ is not known for arbitrary $CY_3$, a fact that makes obtaining a general expression for the density of light states difficult. Nevertheless, in the limit $c_L,\dmina\to \infty$, the distribution of light states should be captured by the spectrum of supergravity fluctuations on AdS$_3 \times S^2$. We expect this to be the case to leading order in the large-$\Lambda$ limit, since the logarithmic corrections to the entropy of the corresponding 4D black holes depend only on the supergravity spectrum. In what follows, we will extract the density of light states from the supergravity part $Z_{\text{sugra}}(\tau, z^I)$ of the modified elliptic genus.

The perturbative spectrum of supergravity on AdS$_3 \times S^2$ is organized into chiral primary representations of the corresponding symmetry group, namely $SL(2,R) \times SU(1, 1|2)$. The contributions of supergravity modes to the modified elliptic genus take the form \cite{Gaiotto:2006ns,Kraus:2006nb} 
\eq{\label{Z_sugra_def}
Z_{\text{sugra}}(\tau, z^I) = 
\e^{2\pi i z^I b_I} Z_{\text{sugra}}(\tau)\,, \qquad Z_{\text{sugra}}(\tau) = \sum_n c_{\text{sugra}}(n) q^n\,, 
}
where the $b_I$ are the minimum values of the $\ell_I$ charges such that $\dmin = - c_L/24$ (one could also cast the $e^{2\pi i z^I b_I}$ term as $q^{-c_L/24}$ using spectral flow). Note that up to spectral flow transformations, supergravity states are not charged since the theory has $(0,4)$ supersymmetry. The partition function \eqref{Z_sugra_def} has been explicitly computed in \cite{Gaiotto:2006ns,Kraus:2006nb} and is given by
\eq{\label{Z_sugra}
Z_{\text{sugra}}(\tau) = q^{-\frac{\svar}{24}} \eta(q)^{\svar} M(q)^{-\chi}\,.
}
In this expression, $M(q)$ is the MacMahon function
\eq{
M(q) \coloneqq \displaystyle\prod_{n=1}^\infty (1 - q^n)^n\,,
}
while $\chi$ is the Euler characteristic of the Calabi-Yau threefold\footnote{The Hodge numbers of $CY_3$ are given in terms of 5D vector ($n_V^{5d}$) and hyper ($n_H^{5d}$) multiplets by
\begin{equation}
h^{1,1} = n_V^{5d} + 1, \qquad \text{and}\qquad h^{1,2} = n_H^{5d} -1\,. \notag
\end{equation}
The 5D multiplets are related to the 4D ones by ($n_V$,$n_H$) = ($n_V^{5d} + 1$, $n_H^{5d}$).
} 
\eq{\label{Euler}
\chi = 2(h^{1,1} - h^{1,2}) = 2(n_V  - n_H + 1)\,,
}
where $n_V$ and $n_H$ denote the number of vector and hyper multiplets in 4D. 

The variable $\svar$ in \eqref{Z_sugra} parametrizes a disagreement in the elliptic genera reported in \cite{Gaiotto:2006ns,Kraus:2006nb}. The value found in  \cite{Gaiotto:2006ns} is $\svar =1$, which was then set to $\svar=0$, while the analysis of \cite{Kraus:2006nb} found  $\svar = 3$. Both \cite{Gaiotto:2006ns,Kraus:2006nb} compute the spectrum of $\mathcal N = 2$ supergravity fluctuations on AdS$_3 \times S^2$, which is compatible with the $\mathcal N = (0,4)$ supersymmetry of the dual CFT$_2$. However, as discussed earlier, the algebra of the dual CFT is enhanced by the presence of the center of mass multiplet, a fact that was not taken into account in \cite{Gaiotto:2006ns,Kraus:2006nb}. We expect the modifications to the symmetry algebra to cancel the $s$-dependent terms in \eqref{Z_sugra} effectively setting $s = 0$. One piece of evidence in favor of this value, besides the supergravity derivation of the OSV formula \cite{Gaiotto:2006ns}, comes from the computation of the central charge \eqref{eq:centralcharge} from supergravity. As shown in \cite{ArabiArdehali:2018mil}, a value of $s= 3$ in \eqref{Z_sugra} leads to a finite correction to the central charge that is missing from the originally proposed value \cite{Maldacena:1997de}. For these reasons, we will assume that $s = 0$ such that 
\eq{\label{Z_sugra2}
Z_{\text{sugra}}(\tau) = M(q)^{-\chi}\,.
}
We will show that this partition function leads to $c(n, \ell_I)$ coefficients that precisely match the logarithmic corrections to the entropy of the corresponding 4D black holes. 

The distribution of light states corresponding to the perturbative supergravity spectrum in \eqref{Z_sugra2} can be obtained by approximating the MacMahon function for large values of $n$, see e.g.~\cite{Dabholkar:2005dt,Castro:2010ce} for a derivation. Using the approach described there, we find that up to a numerical constant,
\eq{
c_{\text{sugra}}(n) \approx   n^{- \frac{\chi + 24 }{36}} \exp\bigg( 3\bigg(\frac{\chi \,\zeta(3)}{4} \bigg)^{\frac{1}{3}} n^{\frac{2}{3}}  \bigg)\,. \label{c_sugra_assymp}
}
We see that the growth of light states is slow, i.e.~sub-Hagedorn, as expected for a theory of supergravity.

In order to determine the density of light states, we also need to take into account the contribution from the ground state degeneracy in \eqref{eq:polarMEG}. The latter is given, in the $c_L, \dmina \to \infty$ limit, by the universal formula \eqref{eq:rhovac}, which we reproduce here for convenience
\eq{  \label{eq:gsdegen}
\rho_0(\dmin) = 2|\bar J^3_0| + 1\sim \dmina\,.
}
This degeneracy is a consequence of the $SU(2)_R$ R-symmetry group, as discussed in more detail in Sec.\,\ref{sec:4D-N4}. Note that the degeneracy is actually controlled by $c_R$ and subleading corrections from the center of mass multiplet as well as M2 and anti-M2 branes \cite{Gaiotto:2006wm,Gomes:2017bpi}. However, in the limit $c_L, \dmina \to \infty$, we have $c_R = c_L + \cdots$, which justifies \eqref{eq:gsdegen}. From a 4D perspective, the ground state degeneracy can be understood as originating from bound states of multicentered solutions \cite{Denef:2007vg}.

Altogether, we find that the density of light states of the modified elliptic genus \eqref{eq:polarMEG} takes the form
\eq{
\rho_\light(\nn,\jj_I)  \approx \dmina \nn^{\omega} e^{2\pi \gamma \nn^\alpha} \prod_{I = 1}^{n_V} \delta(\jj_I-b_I)\,,\label{eq:rho-L-N=2}
}
where $\omega$, $\gamma$, $\alpha$ are respectively given by 
\eq{\label{eq:parametersN2}
\omega = - \frac{\chi + 24}{36}\,, \qquad \gamma = \frac{3}{2\pi} \bigg(\frac{\chi \,\zeta(3)}{4} \bigg)^{\frac{1}{3}}\,, \qquad \alpha = \frac{2}{3}\,.
}

We now have all the ingredients necessary for the evaluation of \eqref{eq:cnl_I2}. Several of the steps follow the autocracy example presented in Sec\,\ref{sec:slowgrowth}, with the most important difference being that the kernel in \eqref{eq:cnl_I2} now includes a weight and a rank. In terms of the parameter $\alpha$, the scaling limit \eqref{eq:regimeN2} can be written as
\eq{\label{eq:regimeN2detail}
\Delta \sim \dmina^{2\alpha - 1}\,.
}
This shares the features of the third autocratic case in \eqref{eq:saddle3} -- \eqref{eq:slowres3}. One important feature of this case is that the leading asymptotic behavior of the $c(n, \ell_I)$ coefficients is not $e^{4\pi\sqrt{\dmina\Delta}}$; instead, the exponential growth receives corrections that depend on $\gamma$ and the ratio $\Delta/\dmina^{1/3}$. 

One simplification in this case is that the density of light states depends only on one variable. Therefore, the integrals over the spin variables in \eqref{eq:cnl_I2} collapse and the saddle-point equation is given by \eqref{eq:saddlepoint2}. This follows from the fact that the exponential behavior of the rank-$n_V$ crossing kernel \eqref{eq:crossingkernelnV} is independent of the number of $U(1)$ charges in the scaling limit \eqref{eq:regimeN2}. In the regime \eqref{eq:regimeN2detail}, the location of the saddle-point is given in \eqref{eq:saddle3}, which we reproduce here for convenience
\eq{
\Delta'_\star \approx \Delta_0\,.
}
Using the crossing kernel \eqref{eq:crossingkernelnV} together with the results of Sec.\,\ref{sec:slowgrowth}, we find that the asymptotic value of the $c(n, \ell_I)$ coefficients is approximately given by
\eq{ \label{eq:cnl_N2_1}
c(n, \ell_I) \approx \dmina \sqrt{ \det\big(t^{IJ}\big)} \Bigg(\sqrt{\frac{\Delta}{\dmina}}\Bigg)^{-2-\frac{n_V}{2}}  \frac{\dmina^{\omega + \frac{2}{3}} e^{4\pi\zeta \dmina^{\frac{2}{3}}}}{(\dmina\Delta)^{\frac{1}{4}}}\,,
}
where $\zeta$ is an order one number given in \eqref{eq:slowres3}. Note that the location of the saddle and the leading asymptotic behavior of $c(n, \ell_J)$ follow from the analysis of the autocratic regime in Sec.\,\ref{sec:slowgrowth}, and  are unchanged by the weight and the rank of the crossing kernel.  Letting $\Delta \sim \Lambda$ and $\dmina \sim \Lambda^3$ with $\Lambda \gg 1$, we find that 
\eq{
\begin{split}
\log\,c(n, \ell_I) &=  4\pi \zeta \Lambda^2 + \frac{3 (\omega + 2)}{2}\log \Lambda^2 + \dots  \\
& = 4\pi \zeta \Lambda^2 +\frac{1}{12} \left(23 - n_V + n_H\right) \log \Lambda^2 + \dots\,, \end{split}\label{eq:N2result}
}
where we used the value of $\omega$ in \eqref{eq:parametersN2} characterizing the growth of the light states. It is important to remark that the logarithmic corrections are sensitive to the weight of $ \mathbb{P}_{\{n,\ell_i\};\{\nn',\textsf{r}_i\!\,'\}}^{(k,M)}$,  but not to its rank. The independence on the rank follows from \eqref{eq:regimeN2detail} and the fact that $t_{IJ} \sim \dmina^{1/3} \sim \Lambda$, which guarantees that the explicit factor of $n_V$ in \eqref{eq:cnl_N2_1} cancels out and enters \eqref{eq:N2result} only via $\omega$.

Let us conclude this section by comparing \eqref{eq:N2result} with its gravitational counterpart \eqref{se4:CY_log}. Two important features come to light:
\begin{itemize}[leftmargin=0.3cm]
    \item Remarkably, the coefficient of the logarithmic correction in \eqref{eq:N2result}  matches precisely the logarithmic correction to the black hole entropy. The spectrum of supergravity on AdS$_3\times S^2$ together with an appropriate account of the center of mass modes and the ground state degeneracy are responsible for this correction. This indicates that we have correctly captured all the ingredients in the CFT$_2$ that account for the logarithmic corrections. 
    \item Due to the the regime \eqref{eq:regimeN2} and the fact that the MacMahon function controls the growth of light states, the leading order behavior of $c(n, \ell_I)$ is not Cardy-like. Therefore, the leading term in \eqref{eq:N2result} does not reproduce the area law in \eqref{se4:CY_log} (albeit both terms scale as $\Lambda^2$). Interpreting this result from the point of view of a gravitational path integral on AdS$_3\times S^2$, we see that this is a case where quantum effects take over and overcome the classical action. This is also manifest on AdS$_2\times S^2$, where the MacMahon function is a resummation of degree zero Gromov-Witten invariants that contribute to OSV, see e.g.~\cite{Dabholkar:2005dt, Denef:2007vg}. 
\end{itemize}

\section{Discussion} \label{sec:disc}

In this paper, we studied the asymptotic behavior of the Fourier coefficients of wJfs using crossing kernels, focusing on the features relevant for a holographic interpretation in terms of the black hole entropy. For large values of the discriminant, we recovered the universal Cardy behavior at leading order. Importantly, we quantified how the large $\dmina$-limit and sparseness conditions on the polar spectrum extend the regime of validity of the Cardy growth. In this extended regime, we computed the logarithmic corrections to the Fourier coefficients, and quantified how these corrections depend on the distribution of polar states in the wJf. We found two qualitatively different behaviors. There is a \textit{universal regime} characterized by $\Delta\gtrsim\dmina$ where the logarithmic corrections are universal and depend only on the ground state degeneracy. Additionally, there is a \textit{non-universal regime} with $\Delta\lesssim\dmina$, accessible only for slow-growing wJfs, where the logarithmic corrections depend on the distribution of polar states. 

As a way to verify the results obtained via crossing kernels, we considered wJfs constructed via a symmetric product orbifold and exponential lift. The pole structure of these modular forms gives an independent way to extract the  $c(n,\ell)$ coefficients of certain wJfs, and we found perfect agreement with the result obtained from crossing kernels. More importantly, these forms illustrate the stark difference between the universal and non-universal regimes, and the key role played by the ground state degeneracy in matching the $c(n, \ell)$ coefficients to the entropy of black holes.   

A key motivation for decoding these corrections is to see their imprint in the gravitational path integral. On the gravitational side, these corrections show up as one-loop corrections to the entropy of black holes. Although both the microscopic and macroscopic sides of these corrections have been shown to agree in specific examples, it was not clear what is the microscopic data that controls the logarithmic corrections. In other words, it was not clear what was learned about the microscopic system through the logarithmic corrections computed in gravity. In this context, we revisited the matching of the logarithmic corrections of $1/4$-BPS black holes in $\mathcal{N}=4$ ungauged supergravity in four and five dimensions, and of $1/2$-BPS black holes in $\mathcal{N}=2$ ungauged supergravity in four dimensions. It is interesting to note that the agreement, at least in these cases, is never controlled solely by symmetries. For $1/4$-BPS black holes in 4D, the logarithmic correction on the microscopic side is universal, but the gravitational answer is (in intermediate steps) sensitive to the supergravity spectrum.  For $1/4$-BPS black holes in 5D, the logarithmic correction of the $c(n, \ell)$ coefficients is not universal, but the gravitational answer is dictated only by zero modes, showing that very different pieces of data contribute to the corrections.  For $1/2$-BPS black holes in 4D, both the microscopic and macroscopic sides are controlled by detailed aspects of their spectrum, and their match requires a careful understanding of the appropriate scaling regimes. 

We conclude with some comments and open questions.

\subsection*{Applications of crossing kernels to holographic CFTs}

In this paper, we use crossing kernels as the main tool to study the coefficients of wJfs. Crossing kernels and asymptotic expansions, analogous to the Cardy formula, are known for other CFT$_2$ observables. There, the leading order expressions are obtained by taking the limit where the ``vacuum" kernel dominates, see e.g.~\cite{Collier:2018exn,Collier:2019weq,Belin:2021ryy,Anous:2021caj,deBoer:2024kat}. It would be interesting to study under what conditions the regime of validity of these formulae can be extended in the large central charge limit, and how the light data controls the regime of validity and subleading corrections. We believe the main difficulty is to come up with a reasonable parametrization of the light data, analogous to \eqref{eq:rholightfg}, \eqref{eq:assumptions1}, and \eqref{eq:assumptions2}, argued for by holographic considerations. With such a parametrization at hand, one could in principle generalize the analysis of Sec.\,\ref{sec:beyondcardy}. One instance where an analysis like this has been done (without crossing kernels) is in the context of thermal correlation functions of holographic CFTs \cite{Kraus:2017kyl}.

For the coefficients of wJfs, we have argued for the validity of the crossing kernel analysis through comparison with the Rademacher expansion in Sec.\,\ref{sec:rademacher}. The Rademacher expansion provides an exact relation between light and heavy states that is possible because wJfs are holomorphic modular objects.\footnote{See, for example, \cite{Alday:2019vdr,Kaidi:2020ecu,Benjamin:2021ygh} for a detailed discussion on the limitations of the Rademacher method when objects are not holomorphic.} To ensure the validity of a crossing kernel analysis of more general objects one cannot rely on an exact expansion \`{a} la Rademacher. In some cases, it has been shown that Tauberian theorems can be used to investigate the robustness of these formulae, see e.g. \cite{Mukhametzhanov:2019pzy,Pal:2019zzr,Pal:2019yhz,Das:2017vej}. An essential ingredient in the proof of Tauberian theorems is the positivity of the coefficients of the corresponding object \cite{Qiao:2017xif}. For many observables of interest, including wJfs and CFT correlation functions, these coefficients are not necessarily positive. It would be interesting to investigate if and how Tauberian theorems can be used as inspiration for theorems for observables that are not necessarily positive.

\subsection*{Rademacher expansions for sparse modular forms}

For wJfs constructed via symmetric product orbifolds or exponential lifts, it is clear how one can improve on the approximations made in the derivation of the symmetric product kernel in \eqref{eq:symprodkernel2}. First, one can take into account the subleading poles, in addition to the single dominant one. From the analysis of App.\,\ref{app:SMF}, we see that the contributions of these poles resemble the $c>1$ terms of the Rademacher expansion. Furthermore, there are subleading contributions from the $\tau$-contour that resemble the different terms in the Kloosterman sums. Finally, one can improve on the approximations in the expansion around the poles and the range of integration of the Gaussian integral over $z$. By considering these subleading corrections, one could presumably reproduce the exact Rademacher expansion, as done for the Igusa cusp form (which can be cast as an exponential lift) \cite{Cardoso:2021gfg}. It would be interesting to generalize this result to general wJfs constructed via symmetric products or exponential lifts.

As we mentioned in Sec.\,\ref{sec:2}, it is not obvious how to implement the $t\to\infty$ limit and the HKS condition on the  Rademacher expansion.  The analysis of Fourier coefficients for symmetric product orbifolds and exponential lifts provides insights to overcome this obstacle. The residue at each pole corresponds to polar state data of a wJf at $t\to \infty$. This is not obvious coming from a more traditional approach to the Rademacher expansion: why would polar data at $t\to \infty$ be needed to describe $d(n,\ell;t)$ for finite $t$? From a gravitational perspective, one always starts at $t\to \infty$ ($G_N\to 0$), and the interpretation of this residue is natural in the gravitational path integral: $c_\infty(n,\ell)$ in \eqref{eq:symprodkernel2} is the input from the perspective of the HKS bound. It would be interesting to make this approach mathematically rigorous. For instance, instead of thinking about the Rademacher expansion for a single wJf, can we think of the implementation of the expansion as a collective? The specification of $c_\infty(n,\ell)$ as polar data potentially defines a family of wJfs with varying values of $t$. The question is then, are there constraints on $c_\infty(n,\ell)$ such that each member of the family leads to a modular form? 

\subsection*{Logarithmic corrections to black hole entropy}

In recent years, interest in exploring quantum properties of black holes in AdS/CFT has been revived, with developments on both the microscopic and macroscopic sides. The logarithmic corrections discussed in this paper are nontrivial indicators of these quantum properties. In this work, we revisited the role of these logarithmic corrections from a microscopic point of view when the object counting the black hole microstates is a wJf. 

One class of supersymmetric black holes we did not consider here are 1/8-BPS  black holes in ${\cal N}=8$ ungauged supergravity in four and five dimensions. The appropriate index is known explicitly \cite{Maldacena:1999bp}, and the logarithmic corrections are known to match with the gravitational counterpart \cite{Banerjee:2011jp,Sen:2012cj}. In fact, in four dimensions the analysis can be pushed even further: it is possible to show that the Rademacher expansion can be exactly reproduced by a gravitational path integral \cite{Dabholkar:2011ec,Dabholkar:2014ema,Iliesiu:2022kny}. For this reason, and because of its similarities with the black holes in ${\cal N}=4$, this case was omitted here although it might be interesting to analyze it explicitly.  

For the classes of black holes in 4D discussed in Sec.\,\ref{sec:bhs} there is an implicit coincidence. The local contribution to the logarithmic correction $a_{\rm local,4D}$ is topological, meaning that it is independent of the parameters of the black hole (e.g., electric or magnetic charges, angular momentum). This coincidence was first noticed in \cite{Charles:2015eha}, and it was attributed to the properties of the theory: it is a result of the black hole being embedded in ungauged supergravity.\footnote{For ungauged theories, the AdS$_2 \times S^2$ background solution has vanishing Weyl tensor. This explains why $a_{\rm local,4D}$ is independent of the parameters of the solution. However, the analysis of \cite{Charles:2015eha} applies to non-extremal black holes that are continuously connected to the BPS black hole. For black holes in the non-BPS branch of ungauged supergravity, we expect $a_{\rm local,4D}$ to be non-topological \cite{Castro:2018hsc}. It is also non-topological for several 4D EFTs that are not supersymmetric \cite{Sen:2012dw}.} In contrast, if we consider supersymmetric black holes in gauged supergravity, which are asympotically AdS$_4$ rather than Mink$_4$, one finds that $a_{\rm local,4D}$ is generically non-topological \cite{David:2021eoq,Karan:2022dfy}. Only when the background is lifted to M-theory, the additional fields coming for that uplift make $a_{\scaleto{\rm grav}{4pt}}$ topological since any local contribution vanishes in an odd number of dimensions \cite{Bhattacharyya:2012ye,Liu:2017vbl,Bobev:2023dwx}. 

In fact, whenever $a_{\scaleto{\rm grav}{4pt}}$ is matched with a microscopic counting formula, the logarithmic correction is topological. What explains this coincidence in CFT? Is it possible to engineer a counting formula where the logarithmic correction is non-topological? Or is it possible to justify from the CFT point of view that this correction must be topological? Our analysis addresses this question for the index of a CFT$_2$ that meets the HKS sparseness condition: the logarithmic correction is always topological, both in the universal and non-universal regime. There is also evidence of this for SCFT$_4$ at large $N$ \cite{Cassani:2021fyv}. It would be interesting to extend this reasoning to CFT$_3$, and some arguments are given in \cite{Bobev:2023dwx}. A clear understanding of the microscopic origin of the logarithmic corrections will have an important impact on the EFT that controls the gravitational path integral as one connects the UV to the IR. In particular, ruling out non-topological corrections in the UV would place strong constraints on how matter couples to gravity in the IR.  

\bigskip
\section*{Acknowledgements}
We thank Mert Besken, Alejandro Cabo-Bizet, Frederik Denef, Victor Godet, Gabriel Lopes-Cardoso, Sameer Murthy, Marti Rosello, Wei Song, and Fengjun Xu for helpful discussions. We also thank the participants and organizers of the Physics Sessions Initiative, Crete, 2023, and the participants of the Isaac Newton Institute for Mathematical Sciences programme ``Black holes: bridges between number theory and holographic quantum information." We also thank the organizers of the Amsterdam Summer String Workshop 2024 for their hospitality during the last stages of this work.  

The work of LA was supported in part by the Dutch Research Council (NWO) through the Scanning New Horizons programme (16SNH02). The work of SB is supported by the Delta ITP consortium, a program of the Netherlands Organisation for Scientific Research (NWO) that is funded by the Dutch Ministry of Education, Culture and Science (OCW), and in part by the National Science Foundation under Grant No. NSF PHY-1748958, the Heising-Simons Foundation, and the Simons Foundation (216179, LB). DL is supported by the European Research Council under the European Unions Seventh
Framework Programme (FP7/2007-2013), ERC Grant agreement ADG 834878. SB and AC would like to thank the Isaac Newton Institute for Mathematical Sciences, Cambridge, for support and hospitality during the programme ``Black holes: bridges between number theory and holographic quantum information" where work on this paper was undertaken. This work was supported by EPSRC grant no EP/R014604/1. This work has been partially supported by STFC consolidated grants ST/T000694/1 and ST/X000664/1.
 This research was supported in part by grant NSF PHY-2309135 to the Kavli Institute for Theoretical Physics (KITP). 

\appendix

\numberwithin{equation}{section}
\section{Conventions}\label{app:conventions}

In this appendix we summarize our conventions for the various approximations made when extracting the leading and subleading behavior of the Fourier coefficients of wJfs. We use 
\begin{itemize}
    \item ``$+\cdots$" to denote terms that are suppressed compared to the leading term;
    \item ``$\approx$" to denote equality up to an order one factor;
    \item ``$\sim$" when we only keep track of the leading order scaling.
\end{itemize}
These conventions are best explained through an example. Let us consider the function
\eq{
f(x)=\pi (x^2+x) e^{2\pi x}+1+23 e^{-x}\,.
}
We are interested in approximating the behavior of $f(x)$ at large values of $x$. In our conventions, this means that
\begin{equation}
\begin{split}
    f(x) & =\pi x^2 e^{2\pi x}+\cdots\,,\\
    f(x) &\approx x^2 e^{2\pi x}\,,\\
    f(x) &\sim e^{2\pi x}.
\end{split}
\end{equation}

\section{Crossing kernels for weak Jacobi forms}\label{app:kernels}

\begin{figure}[t]
    \centering
    \includegraphics[width=65mm]{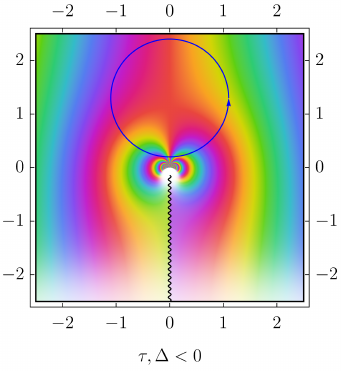}
    \includegraphics[width=65mm]{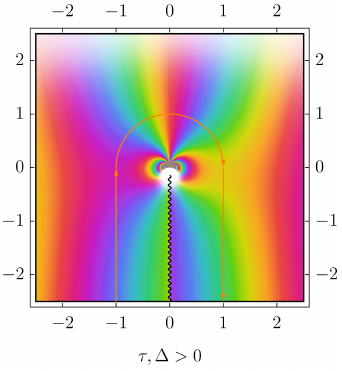}
    \caption{Plot of the $\tau$ integrand in the definition of the crossing kernel. In this diagram, lighter shading corresponds to a larger absolute value. The black line indicates the position of the branch cut while the orange and blue lines are the contours of integration. If $\Delta < 0$, the blue contour closes, and the integral evaluates to zero; if $\Delta > 0$, the orange contour surrounds the branch cut.}
    \label{fig:contours}
\end{figure}

In this appendix we derive the crossing kernel $ \knl{P}{\nn,\,\jj}{\nn',\jj'}^{(k)}$ associated with wJfs of weight $k$ and index $t$. This kernel encodes the transformation rule associated to the modular transformation
\begin{equation}
       \varphi\left(-\frac{1}{\tau}, \frac{z}{\tau}\right) = \tau^{k} \e^{\frac{2\pi i t  z^2}{\tau}}\varphi(\tau,z)\,,
\end{equation}
and it is defined by 
\begin{equation}
\label{def:cross-ker}
    \knl{P}{\nn,\,\jj}{\nn',\jj'}^{(k)} \coloneqq \int\dd \tau \, \dd z \, \tau^{-k}\e^{-2\pi i(\tau \nn+z \jj)}\e^{\frac{2\pi i}{\tau}\left(-\nn'+ \jj' z - t z^2\right)}\,.
\end{equation}
We have introduced the label $(k)$ to indicate the weight of the kernel (we will omit this label whenever $k=0$). One can check that this kernel is the expression that appears in the crossing equation, namely
\begin{equation}\label{eq:a3}
\begin{split}
    \rho(\nn,\,\jj) &= \int \dd \tau  \, \dd z  \,\e^{-2\pi i (\tau \nn + z\jj)} \varphi(\tau,z)\\
    &=  
    \int \dd \tau  \, \dd z \;\tau^{-k} e^{-2\pi i (\tau {\nn} + z {\jj})} \e^{-\frac{2\pi i t z^2}{\tau}}\varphi\left(-\frac{1}{\tau},\frac{z}{\tau}\right)\\
    &= \sum_{n',\ell'} c(n',\ell') \int \dd \tau  \, \dd z \, \tau^{-k} e^{-2\pi i (\tau \nn + z\jj)} \e^{\frac{2\pi i}{\tau}\left(-n'+\ell' z -  t z^2 \right)}\,.
\end{split}
\end{equation}
As it is written, the integrals over $\tau$ and $z$ in \eqref{def:cross-ker} need regularization. The integral over the $z$ variable is Gaussian, and to guarantee its convergence we deform the contour to $z \rightarrow (1-i\epsilon)z$ for $\epsilon>0$. The resulting expression is  
\begin{equation}
    \int_{(-1-i \epsilon)\infty}^{(1+i \epsilon)\infty} \dd z \,
    \e^{2\pi i \left(-z \jj +\frac{ \jj' z}{\tau} - \frac{ t z^2}{\tau}\right)}
\underset{\epsilon\rightarrow0}{=} \sqrt{\frac{-i\tau }{2t}} \e^{\frac{i \pi (\jj'-\jj \tau)^2}{2t\tau}}\,.
\end{equation}
Plugging this result into \eqref{def:cross-ker} leads to an integral over the variable $\tau$ that naturally rearranges itself in terms of the discriminants $\Delta$, $\Delta'$  and a $\jj'\jj$-dependent phase
\begin{equation}
 \knl{P}{\nn,\,\jj}{\nn',\jj'}^{(k)}  = \e^{-\frac{i \pi \jj' \jj}{t}}\int \dd \tau\;
 \tau^{-k} \sqrt{\frac{-i \tau}{2t}} \e^{-2\pi i \left(\frac{\Delta'}{\tau} + \Delta \tau\right)}\,,
\end{equation}
where 
\begin{equation}
    \Delta' = \nn'-\frac{(\,\jj'\,)^2}{4t}\,,\qquad \Delta = \nn-\frac{\jj^2}{4t}\,.
\end{equation}
The contour that regularizes this integral is continuously connected to the real line but depends on the sign of $\Delta$.\footnote{Since we are regularizing a distribution, it is important that the deformed contour is continuously connected to the original contour.}  When $\Delta>0$, we must close the contour in the lower half of the complex plane, such that it surrounds the branch cut produced by the square root. When $\Delta<0$, we must close the contour in the upper half of the complex plane. There is no branch cut in this case and the integral evaluates to zero. See Fig.\,\ref{fig:contours} for details. 

To solve the $\tau$ integral when $\Delta>0$, we consider the change of variables $\tau = i\sigma/(2\pi \Delta)$. The integral in $\sigma$ reads
\begin{multline}
    \frac{-1}{\sqrt{4\pi \Delta t}}\left(\frac{i}{2\pi \Delta}\right)^{-k+1}
    \e^{-\frac{i\pi  \jj'\jj}{t}}
    \int_\gamma \frac{\dd\sigma}{\sigma^{k-\frac{3}{2}+1}} \; \e^{\frac{16 \pi ^2 (-\Delta'\Delta)}{4\sigma} + \sigma} \\ = \sqrt{\frac{2\pi^2}{t}} \bigg(\sqrt{-\frac{\Delta}{\Delta'}}\,\bigg)^{k-\frac{3}{2}}\e^{-\pi i\left(\frac{\jj'\jj}{t}+\frac{k}{2}\right)} I_{k-\frac{3}{2}}\big(4\pi \sqrt{-\Delta'\Delta}\,\big)\,.
\end{multline}
Here, we have used the following representation of the Bessel function
\begin{equation}
\label{def:bess-app}
    I_{\nu}(z) = 
    \frac{1}{2\pi i} \left(\frac{z}{2}\right)^{\nu}\int_\gamma \frac{\dd \sigma}{\sigma^{\nu+1}}\; \e^{\sigma+\frac{z^2}{4\sigma}}\,,
\end{equation}
where the contour $\gamma$ encloses the negative real axis and surrounds the branch cut in a counterclockwise direction. The final result is the following expression for the crossing kernel,
\begin{equation}
\label{eq:cross-we}
        \knl{P}{\nn,\,\jj}{\nn',\jj'}^{(k)}  = \sqrt{\frac{2\pi^2}{t}} \bigg(\sqrt{-\frac{\Delta}{\Delta'}}\,\bigg)^{k-\frac{3}{2}}\e^{-\pi i\left(\frac{\jj'\jj}{t}+\frac{k}{2}\right)} I_{k-\frac{3}{2}}\big(4\pi \sqrt{-\Delta'\Delta}\,\big)\Theta(\Delta)\,.
\end{equation}
To check that this is the correct result, we verify that the kernel satisfies the property
\begin{equation}
    \tau^{-k} \exp{-\frac{2\pi i\, \nn'}{\tau}+\frac{2\pi i \jj' z}{\tau}-\frac{2\pi i t z^2}{\tau}} = \int_{-\infty}^{\infty} \dd \jj\int_0^\infty \dd \nn \;
    \knl{P}{\nn,\,\jj}{\nn',\jj'}^{(k)}\e^{2\pi i (\nn\tau + \jj z)}\,.
\end{equation}
This integral can be solved analytically and validates \eqref{eq:cross-we}. 

\section{Weak Jacobi forms of matrix rank \textit{M}}\label{app:multiplecharges}

The notion of a weak Jacobi form can be extended to a situation where the form $\varphi(\tau,z^i)$ is defined on $\mathbb{H}\times \mathbb{C}^M \rightarrow \mathbb{C}$ (see e.g. \cite{G-Jacobi,bringmann2019rank}). Such forms occur when the Kac-Moody algebra of the CFT$_2$ is generalized from just a single $U(1)$ to having multiple currents whose algebra is of rank $M$. Such a wJf of matrix rank $M$ has a Fourier expansion of the form
\eq{
\varphi\left(\tau,z^i\right)=\sum_{n,\ell^i}c\left(n,\ell_i\right)e^{2\pi i(n\tau+z^i\ell_i)}\,,
}
where
\eq{
c(n,\ell_i)=0\qquad\text{unless}\qquad n\ge0\,.
}

To describe the modular and elliptic properties of $\varphi(\tau,z^i)$ we introduce a matrix $t_{ij}$, which corresponds to the killing form of the Kac-Moody algebra. 
As in Sec.\,\ref{sec:2}, wJfs with matrix rank $M$ satisfy stringent transformation rules. Modular transformations are given by 
\eq{
    \varphi\left(\frac{a \tau + b}{c\tau+d}, \frac{z^i}{c\tau+d}\right) = \tau^{-k} \e^{2\pi i \frac{c z^2}{c\tau + d}} \varphi(\tau,z^i)\,,\qquad  \begin{pmatrix}
    a & b \\
    c & d
    \end{pmatrix}
    \in \text{SL}(2,\mathbb{Z})\,,
}
where $k$ is the weight of the wJf and we use the shorthand notation $z^2 = z^iz^jt_{ij}\,$. Moreover, the spectral flow transformation generalizes to
\eq{
\varphi(\tau,z^i+\lambda^i\tau+\mu^i)=e^{-2\pi i(\lambda^2\tau+2\lambda\cdot z)}\varphi (\tau,z^i )\,,
}
where $\lambda^2=\lambda^i\lambda^jt_{ij}$ and $\lambda\cdot z=\lambda^iz^jt_{ij}\,$. Here 
the spectral flow vectors $\lambda$ and $\mu$ lie on lattices $\Gamma$ and $\Gamma^\mu$, respectively, which are contained in $\mathbb{Z}^{M\times M}$; in addition, we have $\mu^i\ell_i\in\mathbb{Z}$. We will describe the lattice $\Gamma$ by a set of basis vectors $\hat v^i\in\mathbb{Z}^M$ 
\eq{
\lambda\in\Gamma=\left\{\sum_{i=1}^M \alpha^i \hat v^i\ \Big\vert \ \alpha^i\in\mathbb{Z}\ \forall \ i\right\}\,,
}
and similarly for $\Gamma^\mu$.  In physically relevant cases, the lattices associated with spectral flow symmetry are nontrivial. This is the case considered in Sec.\,\ref{sec:n2bhs}, which is a specific instance of string theory/M-theory \cite{Maldacena:1997de,deBoer:2006vg}. 

The spectral flow symmetry implies that the coefficients of the wJf satisfy
\eq{\label{eq:mcsf}
c (n,\ell_i )=c (n+\lambda^i\ell_i+\lambda^i\lambda^jt_{ij},\ell_i+2\lambda^jt_{ij} )\,.
}
It follows from \eqref{eq:mcsf} that the coefficients $c\left(n,\ell_i\right)$ only depend on 
\begin{equation}
    \Delta:=n-\frac{1}{4}\ell^2 \,,\qquad \ell_i\ \text{mod}\ 2\hat v^jt_{ij} \,,
\end{equation}
where $\Delta$ is the discriminant. Another way to state the implications of \eqref{eq:mcsf} is that
\begin{equation}
c (n,\ell_i )=c (m,r_i )\,, 
\end{equation}
whenever $ n-\frac{1}{4}\ell^2=m-\frac{1}{4}r^2$ and  there exists $\lambda\in\Gamma$ such that 
\begin{equation}
\lambda^j=\frac{1}{2}t^{ij} (r_i-\ell_i )\,.
\end{equation}
Here $t^{ij}$ is the matrix inverse of $t_{ij}$, i.e.~$t_{ij}t^{jl} = \delta_{i}^{\ l}$.

In practice, spectral flow symmetry tells us that for any $c\left(n,\ell_i\right)$  coefficient of a wJf we can find a spectral flow transformation to $c\left(m,r_i\right)$ with
\eq{\label{eq:bound}
\abs{r^2}\le\det\left[ T \right]^{1/M}\,,
}
where $T\in\mathbb{Z}^{M\times M}$ is given by
\eq{
T=
\begin{pmatrix}
t_{11}\langle\hat v^1,\hat v^1\rangle & \cdots & t_{1M}\langle\hat v^1,\hat v^M\rangle\\
\vdots&\ddots&\vdots\\
t_{M1}\langle\hat v^M,\hat v^1\rangle&\cdots &t_{MM}\langle\hat v^M,\hat v^M\rangle
\end{pmatrix}\,.
}
This bound can be understood as the maximal radius within the fundamental region of the lattice $\Gamma$ scaled by the matrix index $t_{ij}$, where we used the fact that the $t_{ij}$-scaled volume of the fundamental region of $\Gamma$ is given by 
\eq{
\text{Vol}(\Gamma)=\sqrt{\det[T]}\,.
}
Crucially,  \eqref{eq:bound} is the strongest bound on the length of $r$ we can obtain using the symmetries of wJfs.

An important consequence of the generalization to the spectral flow symmetry is that the bound on the maximal polarity $\Delta_0$ is modified by the lattice spacings $\hat v$. This can be shown as follows. Consider a state $c\left(n,\ell_i\right)$ with discriminant $\Delta$, and choose $r_i$ such that
\eq{
\abs{r^2}\le\det[T]^{1/M}\,, \qquad r_i=\ell^i+2\lambda^jt_{ij} \,,
}
where $\lambda\in\Gamma$. Performing a spectral flow transformation with this $\lambda$ then leads to
\eq{
c\left(n,\ell_i\right)=c\left(\Delta+\frac{1}{4}r^2,r_i\right)=0\qquad\forall\qquad\Delta<-\frac{r^2}{4}\le-\frac{1}{4}\det[T]^{1/M}\,.
}
Hence, we conclude that the maximal polarity $\Delta_0$ is bounded by
\eq{
\Delta_0\ge-\frac{1}{4}\det[T]^{1/M}\,.
}
Note that for $M=1$ the matrix $t_{ij}$ reduces to a single integer $t$, and the lattices $\Gamma$ and $\Gamma^\mu$ are taken to be trivial, i.e.~$\hat v^1=1$. Then the bound on $\Delta_0$ reduces to the familiar one given in Sec.\,\ref{sec:2}, that is
\eq{
\Delta_0\ge-t/4\,.
}

In analogy with the previous appendix, we can define crossing kernels for wJfs of matrix rank $M$. The crossing kernel is defined as
\eq{
    \mathbb{P}_{\{\nn,\textsf{r}_i\};\{\nn',\textsf{r}_i\!\,'\}}^{(k,M)} \coloneqq \int \dd \tau \;\tau^{-k}\int\prod_{i=1}^{M} \dd z^i\;
    \e^{-2\pi i(\tau \nn + z^i\textsf{r}_i)}\e^{\frac{2\pi i}{\tau}\left(-\nn' + z^i\textsf{r}_i\!\,'  - z^2\right)}\,,
}
where we have modified  the label $(k) \to (k,M)$ to include the number of charges in the crossing kernel. We can solve this integral by first evaluating the Gaussian integral over the chemical potentials $z_i$, and then evaluating the $\tau$ integral. The resulting expression is also given in terms of a Bessel function such that
\eq{
    \mathbb{P}_{\{\nn,\textsf{r}_i\};\{\nn',\textsf{r}_i\!\,'\}}^{(k,M)} = \sqrt{4\pi^2\det(\frac{t^{ij}}{2})}
    \bigg(\sqrt{-\frac{\Delta}{\Delta'}}\,\bigg)^{k-\frac{M}{2}-1}\e^{-i \pi \left(\textsf{r}'\cdot\textsf{r}+\frac{k}{2}\right)}
    I_{k-\frac{M}{2}-1}\big(4\pi \sqrt{-\Delta\Delta'}\,\big)\,,\label{eq:crossing-rankM}
}
where $\textsf{r}' \cdot \textsf{r} = t^{ij}\textsf{r}_i \textsf{r}_j $.

\section{Rademacher expansion}
\label{app:Rademacher}

We now describe the Rademacher expansion for wJfs of weight $k$ and index $t$. The following expansion is valid for all wJf of weight less than or equal to $\frac{1}{2}$,
\begin{equation}
\label{eq:rademacherap}
    \!\! c(n,\ell) = 
    \sum_{\Delta'<0} \,
    \sum_{\ell'=-t+1}^{t}
    c(n',\ell')
    \sum_{\radc=1}^{\infty}
    \frac{2\pi
    }{\radc}
    \left(
    \frac{|\Delta'|}{\Delta}
    \right)^{\frac{3-2k}{4}}
    \! I_{\frac{3}{2}-k}\bigg(\frac{4\pi}{\radc}\sqrt{\Delta |\Delta'|}\bigg)
    \text{Kl}(\Delta,\ell,\Delta', \ell';\radc)\,,
\end{equation}
where $I_{\frac{3}{2}-k}\big(\frac{4\pi}{\radc}\sqrt{\Delta |\Delta'|}\big)$ is the Bessel function and the Kloosterman sum $\text{Kl}(\Delta,\ell,\Delta', \ell';\radc)$ is defined by
\begin{equation}
   \text{Kl}(\Delta,\ell,\Delta',\ell';\radc) \coloneqq \;\;
   \e^{i\frac{\pi}{4}}\!\!\!\!\sum_{\substack{
    0\leq-d<\radc;\;(d,\radc)=1\\
    ad = 1 \text{ mod }\radc
    }}
    \e^{2\pi i \Delta \frac{d}{\radc}}
    \left[M^{-1}(\gamma)\right]_{\ell'\ell}\;
    \e^{2\pi i \Delta'\frac{a}{\radc}},\quad 
    \gamma =
    \bigg(\begin{array}{cc}
        a & b \\
        \radc & d
    \end{array} \bigg)
    \in 
    \text{SL}\left(2,\mathbb{Z}\right)\,.
\end{equation}
The matrix $M$ is a $2t$-dimensional representation of $\text{SL}\!\left(2,\mathbb{Z}\right)$ that describes the transformation properties of the standard theta functions of weight $\frac{1}{2}$ and index $t$. In this paper, we follow the conventions of \cite{Gomes:2017bpi} and use the explicit representation for $M$ derived in \cite{kloosterman} and given by
\begin{equation}
    \left[M^{-1}(\gamma)\right]_{\ell'\ell}
    =
    \frac{1}{(2it\radc)^{\frac{1}{2}}} 
    \sum_{m=0}^{\radc-1}
    \exp[
    2\pi i\left(
    \frac{a}{\radc}\frac{(\ell' + 2tm)^2}{4t}
    - \frac{\ell(\ell' +  2tm)}{2t\radc}
    +\frac{d}{\radc}\frac{\ell^2}{4t}
    \right)
    ]\,.
\end{equation}

\section{Properties of exponential lifts and the generating function for \texorpdfstring{$\text{Sym}^N(\mathcal{C})$}{SymN} }\label{app:SMF}

In this appendix we provide the technical tools needed to  understand the residues of $\mathcal{Z}(\tau,z,\sigma)$. The main tool we use is the simple relation between $\mathcal{Z}$ and the exponential lift of the seed wJf $\varphi(\tau,z;\mathcal{C})$. Such an exponential lift is an example of a Siegel paramodular form. Siegel paramodular forms satisfy stringent transformation properties under the Siegel paramodular group. Moreover the zeros and poles of Siegel paramodular forms can be understood using the technology of so-called Humbert surfaces. We employ these properties to study the residues of $\mathcal{Z}$. This technology was used for the same purpose in \cite{Belin:2016knb,Sen:2012cj,Castro:2008ys}, which this work is a refinement and generalization of.

We start by defining Siegel paramodular forms through their transformation properties under the Siegel paramodular group in \ref{sec:paramodulargroups}. Then in \ref{sec:explift} we introduce the exponential lift of a wJf, and relate it to $\mathcal{Z}$. \ref{sec:zeros} summarizes how Humbert surfaces can be used to describe the zeros and poles of Siegel paramodular forms, and we explain the simplifications that arise when studying exponential lifts of wJf. Then in \ref{sec:dominantpole} we apply the technology of Humbert surfaces to find the dominant contribution to \eqref{eq:d-integral}. We show that the dominant contribution is given by the residue of a single pole, and we expand $\mathcal{Z}$ around this pole in \ref{sec:exppole}. Finally, in \ref{sec:omega} we compute the weight of the function $\varphi_\infty$ --- the function multiplying the dominant pole of $\mathcal{Z}$ --- which characterizes the growth of polar states in symmetric product orbifolds, and is related in a simple way to $\omega$ in \eqref{eq:assumptions1}. 

\subsection*{\refstepcounter{subsection}\label{sec:paramodulargroups}\thesubsection\quad Fun with paramodular forms}

In this section we define and give some useful properties of Siegel (para)modular forms. A Siegel modular form is defined with respect to the Siegel modular group $Sp(4,\mathbb{Z})$. We use the shorthand notation
\eq{
\Omega=\begin{pmatrix} \tau & z\\z& \sigma\end{pmatrix}\,.
}
Using this notation, the Siegel upper half-plane $\mathbb{H}_2$ is characterized by
\eq{
\det[\text{Im}(\Omega)]>0\,, \qquad\text{Im}(\tau)\,,\,\text{Im}(\sigma)>0\,.
}
Moreover, we will denote a matrix $\gamma\in Sp(4,\mathbb{Z})$ as
\eq{
\gamma=\begin{pmatrix} A & B\\C& D\end{pmatrix}\,,
}
where the $2\times 2$ block matrices $A$, $B$, $C$, and $D$ satisfy
\eq{
AB^T=BA^T\,,\quad CD^T=DC^T\,, \qquad AD^T-BC^T=\mathbb{1}_{2}\,.
}
The coordinates on the upper half-plane transform under the Siegel modular group as
\eq{
\gamma(\Omega)=(A\Omega+B)(C\Omega+D)^{-1}\,.
}
Now we are ready to define a \textit{meromorphic} Siegel modular form. A meromorphic function $\Phi$ on the Siegel upper half-plane is a Siegel modular form of weight $k$ if it satisfies the following transformation rule under the Siegel modular group
\eq{
\Phi(\gamma(\Omega))=\det[C\Omega+D]^k\Phi(\Omega)\,.
}
We denote the space of meromorphic Siegel modular forms of weight $k$ by $M_k$. We can Fourier expand $\Phi\in M_k$ in the $\sigma$ variable to obtain
\eq{
\Phi(\Omega)=\sum_t p^t\varphi_{k,t}(\tau,z)\,,
}
where $p=e^{2\pi i \sigma}$. Due to the transformation rules under $Sp(4,\mathbb{Z})$, it follows that the coefficients in the Fourier expansion are weak Jacobi forms of weight $k$ and index $t$. To see this more explicitly we note that the transformation rules for wJfs \eqref{eq:modInv} and \eqref{eq:specFlow} are respectively given by
\eq{\label{eq:sl2zmatrix}
\gamma_1=\begin{pmatrix}
        a&0&b&0\\
        0&1&0&0\\
        c&0&d&0\\
        0&0&0&1    \end{pmatrix}\,, \qquad \gamma_2=\begin{pmatrix}
        1&0&0&\mu\\
        \lambda&1&\mu&0\\
        0&0&1&-\lambda\\
        0&0&0&1    
        \end{pmatrix}\,,
}
where $ ad-bc=1$. It turns out that these elements, together with the element that interchanges $\tau$ and $\sigma$ that is given by
\eq{
\gamma_3=\begin{pmatrix}
        0&1&0&0\\
        1&0&0&0\\
        0&0&0&1\\
        0&0&1&0    
        \end{pmatrix}\,,
}
generate the full Siegel modular group.

We can generalize Siegel modular forms to Siegel paramodular forms by restricting the allowed transformations to a subgroup of $Sp(4,\mathbb{Q})$. In particular, we are interested in the paramodular group of level $t_o$, which is denoted by $\Gamma_{t_o}$, and can be defined as
\eq{
\Gamma_{t_o}\coloneqq\begin{pmatrix}
    \mathbb{Z}&t_o\mathbb{Z}&\mathbb{Z}&\mathbb{Z}\\
    \mathbb{Z}&\mathbb{Z}&\mathbb{Z}&\mathbb{Z}/t_o\\
    \mathbb{Z}&t_o\mathbb{Z}&\mathbb{Z}&\mathbb{Z}\\
    t_o\mathbb{Z}&t_o\mathbb{Z}&t_o\mathbb{Z}&\mathbb{Z}
\end{pmatrix}
\ \cap \ 
Sp(4,\mathbb{Q})\,.
}
This group has the following extension
\eq{
\Gamma_{t_o}^+=\Gamma_{t_o}\cup\Gamma_{t_o}V_{t_o}\,,\quad\text{where}\quad V_{t_o}\coloneqq\frac{1}{\sqrt{t_o}}\begin{pmatrix}
    0&t_o&0&0\\
    1&0&0&0\\
    0&0&0&1\\
    0&0&t_o&0
\end{pmatrix}\,.
}
Notice that $\gamma_1,\gamma_2\in\Gamma_{t_o}$. Combining this with the fact that a paramodular form $\Phi\in M_k(\Gamma_{t_o})$ has to be invariant under
\eq{
\gamma_4=\begin{pmatrix}
    1&0&0&0\\
    0&1&0&1/t_o\\
    0&0&1&0\\
    0&0&0&1
\end{pmatrix}\,,
}
it follows that the Fourier expansion of $\Phi$ that is given by
\eq{
\Phi(\Omega)=\sum_t p^t\varphi_{k,t}\,,
}
contains only multiples of $t_o$ in the power of $p$. Hence $t=t_oN$ and the coefficients correspond to a series of wJfs of weight $k$ and index $t_oN$. This should ring some bells: the Fourier expansion of a paramodular form has many similarities with the generating function of a symmetric product orbifold $\mathcal{Z}(\tau,z,\sigma)$ as defined in \eqref{eq:dmvv}. We make the connection explicit in the next subsection.

\subsection*{\refstepcounter{subsection}\label{sec:explift}\thesubsection\quad Exponential lifts and Siegel paramodular forms}

In analogy with the main text, we consider weak Jacobi forms of weight 0 and index $t_o$ that have integral coefficients $c(n,\ell)$
\eq{
\varphi(\tau,z)=\sum_{\substack{n\in\mathbb{Z}_{\ge0}\\\ell\in\mathbb{Z}}}c_o(n,\ell)q^ny^\ell\,.
}
We can construct a Siegel paramodular form from a wJf by taking the exponential lift
\eq{
\text{Exp-Lift}(\varphi)(\Omega)\coloneqq q^Ay^Bp^{t_oA}\prod_{(n,\ell,m)>0}\left(1-q^ny^\ell p^{t_om}\right)^{c_o(nm,\ell)}\,.\label{eq:explift}
}
Here $(n,\ell,m)>0$ means $n,m\in\mathbb{Z}_{\ge0}$ and $\ell\in\mathbb{Z}$ such that $m>0\vee (m=0\wedge n>0)\vee(n=m=0\wedge \ell<0)$, and
\eq{\label{eq:pabdef}
p= e^{2\pi i \sigma}\,,\qquad A\coloneqq \frac{1}{24}\sum_{\ell\in\mathbb{Z}}c_o(0,\ell)\,,\qquad B\coloneqq\frac{1}{2}\sum_{\ell\in\mathbb{Z}_{>0}}\ell c_o(0,\ell)\,.} 
This defines a meromorphic modular form of weight $k\coloneqq\frac{1}{2}c_o(0,0)$ with respect to the paramodular group $\Gamma^+_{t_o}$ \cite{Gritsenko:1996tm}.
The exponential lift can be split into two factors, one of which can be related to the generating function for wJfs constructed via symmetric product orbifolds
\eq{
\text{Exp-Lift}(\varphi)(\Omega)=q^Ay^Bp^{t_oA}\prod_{(n,\ell)>0}\left(1-q^ny^\ell\right)^{c_o(0,\ell)}\times\prod_{\substack{n\ge0\\m>0\\ \ell\in\mathbb{Z}}}\left(1-q^ny^\ell p^{t_om}\right)^{c_o(nm,\ell)}\,,\label{eq:eldef}
}
where $(n,\ell)>0$ denotes $n>0 \, \vee\, (n=0\, \wedge\,  \ell<0)$. The first factor is a form of weight $k$ and index $t_oA$ that is almost the Hodge factor
\eq{
\phi_{k,t_oA}(\tau,z)\coloneqq q^Ay^B\prod_{(n,\ell)>0}\left(1-q^ny^\ell\right)^{c_o(0,\ell)}\,.
}
When a CFT has a partition function that is a wJf $\varphi$ of index $t_o$, then the generating function of partition functions of the symmetric product orbifold of that theory is 
\eq{\label{eq:symel}
\mathcal{Z}(\tau,z,\sigma)=\sum_{N\in\mathbb{Z}_{\ge0}} p^{t_oN}\varphi\left(\tau,z;\text{Sym}^N(\mathcal{C})\right)=\frac{p^{t_oA}\phi_{k,t_oA}(\tau,z)}{\text{Exp-Lift}(\varphi)(\Omega)}\,.
}
Here $\varphi\left(\tau,z;\text{Sym}^N(\mathcal{C})\right)$ is the wJf connected to the $N$-th symmetric product orbifold, which is related to the Hecke transform of  $\varphi\left(\tau,z;\mathcal{C}\right)$ through the equation
\eq{\label{eq:zsym}
\mathcal{Z}\left(\tau,z,\sigma\right)=\text{exp}\Bigg(\sum_{N\in\mathbb{Z}_{>0}}N^{-1}p^{t_oN}\varphi\left(\tau,z;\mathcal{C}\right)\vert T_-(N)\Bigg)\,.
}

We can show the invariance of $\mathcal{Z}$ under an $S$ transformation using the modular properties of the Hodge factor and the exponential lift. From \eqref{eq:sl2zmatrix} we see that the $S$ transformation is given by
\eq{\label{eq:matrixgenb}
\gamma=\begin{pmatrix}
0 & 0 & 1 & 0\\
0 & 1 & 0 & 0\\
-1 & 0 & 0 & 0\\
0 & 0 & 0 & 1
\end{pmatrix}\,,
}
which maps 
\eq{\label{eq:stransf1}
\tau\mapsto\tauh=-\frac{1}{\tau}\,,\qquad\sigma\mapsto\sh=\frac{\sigma\tau-z^2}{\tau}\,,\qquad  z\mapsto\zh=-\frac{z}{\tau}\,.
}
Using the transformation rules of (Siegel) modular forms we see that
\eqsp{
\text{Exp-Lift}(\varphi)(\Omega)&=(-\tau)^{-k}\text{Exp-Lift}(\varphi)(\hat\Omega)\, ,\\
\phi_{k,t_oA}(\tau,z)&=(-\tau)^{-k}e^{-2\pi i t_oAz^2/\tau}\phi_{k,t_oA}(\tauh,\zh)\,.
}
Therefore, we have
\eq{
\mathcal{Z}(\tau,z,\sigma) =\frac{p^{t_oA}\phi_{k,t_oA}(\tau,z)}{\text{Exp-Lift}(\varphi)(\Omega)} 
 =\mathcal{Z}(\tauh,\zh,\sh)\,.
}

\subsection*{\refstepcounter{subsection}\label{sec:zeros}\thesubsection\quad Zeros and poles of Siegel paramodular forms}

In this section we discuss the zeros and poles of Siegel paramodular forms. We start with the most general case, and then restrict to the forms interesting for this work: Siegel paramodular forms that are exponential lifts of weight zero wJfs that have maximal polarity given by $q^0y^{\pm b_o}$. In other words, we are interested in the zeros and poles of $\text{Exp-Lift}(\varphi)\in M_k\left(\Gamma_{t_o}^+\right)$ where $\varphi$ is consistent with \eqref{eq:most-polar}. Exponential lifts and their zeros and poles are described in Theorem 2.1 of \cite{Gritsenko:1996tm}, which we state here. 

The exponential lift of a wJf has a product expansion as in \eqref{eq:eldef}. Therefore, its divisors are easy to identify. We just need to satisfy $q^ny^\ell p^{t_om}=1$ for one of the factors to diverge or vanish, depending on the sign of $c_o(nm,\ell)$. Because the paramodular forms we consider are invariant under $\Gamma_{t_o}^+$, these divisors must come in orbits under the group. We can package the orbits conveniently by introducing Humbert surfaces.\footnote{We use the notation of Sec.\,1.3 of \cite{Gritsenko:1996tm}.} We take five coprime integers $a,b,c,e,f$, and define their discriminant as
\eq{
D\coloneqq b^2-4t_oef-4t_oac\,.
}
There is an action of $\Gamma_{t_o}^+$ that leaves $D$ invariant, which allows us to package the orbits of the divisors of the exponential lifts. The coprime integers $a,b,c,e$, and $f$ give the divisors of the exponential lift through the following quadratic equation:
\eq{\label{eq:zero1}
t_of(z^2-\tau\sigma)+t_oc\sigma+bz+a\tau+e=0\,.
}
Since we are interested in exponential lifts of weight zero wJfs, we can use their additional symmetries to simplify this expression. First, we use invariance under $\sigma\mapsto\sigma+1/t_o$ to set $a=0$. We can then set $c=0$ using the symmetry $\tau\mapsto\tau+1$. Finally, we can use the symmetry $z\mapsto z+1$ to map $b$ to $b-2t_o$, so that we only need to consider $b$ mod $2t_o$. We are thus left with divisors of the form
\eq{\label{eq:zero2}
t_of(z^2-\tau\sigma)+bz+e=0\,.
}
Ref. \cite{Gritsenko:1996tm} then showed that \textit{all} divisors are given by Humbert surfaces, which encode the orbits of $\Gamma_{t_o}^+$, and are defined as
\eq{\label{eq:humbert}
H_D(b)\coloneqq\pi_{t_o}^+\big(\{\Omega\in\mathbb{H}_2\ : \ \hat{a}\tau+\hat{b}z+t_o\sigma=0\}\big)\,,
}
with $\pi_{t_o}^+$ denoting the images under the paramodular group $\Gamma_{t_o}^+$. To see that \eqref{eq:zero2} is actually in the Humbert surface $H_D(b)$, we use the $S$ transformation \eqref{eq:matrixgenb}, which maps \eqref{eq:zero2} to
\eq{e\tau+bz+f t_o\sigma=0\,,}
so that we can identify
\eq{
\frac{e}{f}=\hat{a}\,,\qquad \frac{b}{f}=-\hat{b}\,.
}
The multiplicity of the divisors \eqref{eq:humbert} is given by
\eq{
m_{D,b}=\sum_{n>0} c_o(n^2\hat{a},n\hat{b})=\sum_{n>0} c_o\left(\frac{n^2e}{f},\frac{nb}{f}\right)\,.
}
For reasons that will become clear later, we investigate $f=-1$. Since $c_o(n,\ell)=0$ when $n<0$ for wJfs, we see that for $f=-1$ only $e\le0$ leads to a divisor with positive multiplicity.
Furthermore, we notice that for the most polar term of the underlying wJf, which is of the form $q^0y^{-b_o}$ so that $b=b_o$ and $e=0$, the multiplicity simplifies to just the ground state degeneracy
\eq{
m_{D_0,b_o}=c_o(0,-b_o)\,.
}

\subsection*{\refstepcounter{subsection}\label{sec:dominantpole}\thesubsection\quad Location of the dominant pole}

Now that we understand the zeros and poles of Siegel paramodular forms, we can find the dominant contribution to \eqref{eq:d-integral}. As described in the main text, we are interested in the regime $\Delta\abs{\dmin}\gg1$ and we assume that the integrand in \eqref{eq:d-integral} is dominated by the explicit exponential $p^{-t}q^{-n}y^{-\ell}$. 

First, we note that $\mathcal{Z}$ is related to the exponential lift in a nontrivial way. We therefore need to understand how the location of the zeroes and poles of $\mathcal{Z}$ are related to the zeros and poles of the exponential lift. To see that the Hodge factor cannot change the location of poles and zeroes we have to look at the $\sigma$ dependence. The Humbert surfaces have nontrivial dependence on $\sigma$, while from \eqref{eq:zero1} it follows that the $\sigma$ dependence of $\mathcal{Z}$ is related in a very simple way to that of the exponential lift. We can thus conclude that the zeroes and poles of $\mathcal{Z}$ are also given by Humbert surfaces.

Now that we know that the dominating pole is given by a Humbert surface we can implement our assumption. This implies that we should look for the pole that maximizes the explicit exponential $p^{-t}q^{-n}y^{-\ell}$. Therefore, we have to maximize
\eq{
g(\lambda)=\tau n+\sigma t+z\ell+\lambda\left(t_of(z^2-\tau\sigma)+bz+e\right)\,,
}
where we have introduced a Lagrange multiplier $\lambda$ that constrains the maximum to coincide with a pole of $\mathcal{Z}$. It can be shown that $g(\lambda)$ is extremized by
\eq{\label{eq:saddlelambda}
\lambda=\pm i\sqrt{\frac{2nt-\ell^2/2}{D}}\,,\qquad \tau=\frac{t}{t_of\lambda}\,,\qquad \sigma=\frac{n}{t_of\lambda}\,,\qquad z=-\frac{1}{2t_of}\left(\frac{\ell}{\lambda}+b\right)\,,
}
such that at the saddle point, we have 
\eq{\label{eq:saddleexp}
\log\big(p^{-t}q^{-n}y^{-\ell}\big)\sim-\frac{\pi}{t_of}\sqrt{D\left(4nt-\ell^2\right)}\,.
}
Because the discriminant $D$ is positive, we have chosen the minus sign for $\lambda$. The leading contribution comes from the zero with maximal discriminant and $f=-1$. Other values of $f$ give exponentially suppressed contributions to $d(m,n,\ell)$. The contributions from the zeros with maximal discriminant and $f\in\mathbb{Z}_{<-1}$ will, once taken into account, reproduce the $c>1$ terms in the Rademacher expansion \eqref{eq:rademacher}.\footnote{To get an exact agreement with the Rademacher expansion, one also needs to carefully take into account exponentially suppressed contributions originating from the deformation of the contour.} Restricting our analysis to just the $f=-1$ contribution is one origin for the nonperturbative error in \eqref{eq:srcomp}. The discriminant reads
\eq{
D=b^2+4t_oe\,,
}
which is maximized by setting $b=b_o$ and $e=0$ (recall that to have a nonzero multiplicity we need $e\le0$).

To summarize, one of the integrals in \eqref{eq:d-integral} can be approximated by the residue of the most dominant pole of $\mathcal{Z}$, which is given by the zero with maximal discriminant $D_0=(b_o)^2$. Furthermore, contributions from other poles are exponentially suppressed.
That is, we can approximate \eqref{eq:d-integral} by the residue of a single pole that is located at
\eq{
t_o(\tau\sigma-z^2)+b_oz=0\,.
}
We can map this pole to the much simpler expression
\eq{
b_o\zh-t_o\sh=0\,,
}
using the $S$ transformation \eqref{eq:matrixgenb}, and the order of the pole is given by the ground state degeneracy of the underlying wJf.

\subsection*{\refstepcounter{subsection}\label{sec:exppole}\thesubsection\quad Expanding around the dominating pole}

We can expand the exponential lift around the dominant pole $b_o\zh-t_o\sh=0$ as follows
\begin{align}
&\text{Exp-Lift}(\varphi)(\hat\Omega)=\phi_{k,t_oA}(\tauh,\zh) \hat p^{t_oA}\prod_{\substack{\ell\in\mathbb{Z}\\n\ge0\\m>0}}\left(1-\hat q^n\hat y^\ell \hat p^{t_om}\right)^{c_o(nm,\ell)}\\
&=\phi_{k,t_oA}(\tauh,\zh) \hat p^{t_oA}\left(1-\hat p^{t_o}\hat y^{-b_o}\right)^{c_o(0,-b_o)}\prod_{\substack{\ell\in\mathbb{Z}\\n\geq0,m>0\\
(m,n,\ell)\neq(1,0,-b_o)}}\left(1-\hat q^n\hat y^\ell \hat p^{t_om}\right)^{c_o(nm,\ell)}\label{eq:d42}\\
&\approx (2\pi i)^{c_o(0,-b_o)}\phi_{k,t_oA}(\tauh,\zh) \hat p^{t_oA}(b_o\zh-t_o\sh)^{c_o(0,-b_o)}\prod_{\substack{\ell\in\mathbb{Z}\\n\geq0,m>0\\
(m,n,\ell)\neq(1,0,-b_o)}}\left(1-\hat q^n\hat y^{\ell+b_om}\right)^{c_o(nm,\ell)}\label{eq:d43}\\
&=(2\pi i)^{c_o(0,-b_o)}\phi_{k,t_oA}(\tauh,\zh) \hat p^{t_oA}(b_o\zh-t_o\sh)^{c_o(0,-b_o)}\prod_{\substack{\tilde\ell\in\mathbb{Z}\\n\geq0\\
(n,\tilde\ell)\neq(0,0)}}\left(1-\hat q^n\hat y^{\tilde \ell}\right)^{f(n,\tilde \ell)}\label{eq:d44}\\
&=(2\pi i)^{c_o(0,-b_o)}\phi_{k,t_oA}(\tauh,\zh) \hat p^{t_oA}(b_o\zh-t_o\sh)^{c_o(0,-b_o)}\varphi_\infty(\tauh,\zh)^{-1}\,.\label{eq:d45}
\end{align}
Here in \eqref{eq:d42} we have isolated the factor responsible for the divisor $b_o\zh-t_o\sh=0$. In \eqref{eq:d43} we then expand the exponentials $\hat p$ and $\hat y$, and keep only the leading term around $b_o\zh-t_o\sh=0$. This explains what we mean with $\approx$. Then in \eqref{eq:d44} and \eqref{eq:d45} we give a definition for the remaining infinite product in terms of
\eq{
\tilde\ell\coloneqq\ell+b_om\,,\qquad f(n,\tilde\ell)\coloneqq\sum_{m=1}^\infty c_o(nm,\tilde \ell-b_om)\,,
}
\eqsp{
\varphi_\infty(\tau,z)\coloneqq&
\prod_{\substack{n\ge0\\\tilde\ell\in\mathbb{Z}\\\left(n,\tilde \ell\right)\neq(0,0)}}\left(1-q^n y^{\tilde\ell}\right)^{-f\left(n,\tilde \ell\right)}\,.
}
The function $\varphi_\infty$ encodes the growth of polar states in the limit $m\rightarrow\infty$, see \cite{Belin:2019jqz,Belin:2019rba} for details.

Note that from the expansion of the exponential lift around the dominant pole, we can trivially find the expansion of $\mathcal{Z}$ around the dominating pole
\eq{
\mathcal{Z}(\tauh,\sh,\sh)\approx(2\pi i)^{-c_o(0,-b_o)}(b_o\zh-t_o\sh)^{-c_o(0,-b_o)}\varphi_\infty(\tauh,\zh)\,.
}

\subsection*{\refstepcounter{subsection}\label{sec:omega}\thesubsection\quad Growth of polar states for symmetric product orbifolds}

Finally, we derive the weight of $\varphi_\infty$ when the modular form has a slow-growing growth of light states. This determines the parameter $\omega$ in \eqref{eq:assumptions1}.

Recall the definition
\eq{
\varphi_\infty(\tau,z)=\prod_{\substack{n\ge0\\\ell\in\mathbb{Z}\\\left(n, \ell\right)\neq(0,0)}}\left(1-q^ny^{\ell}\right)^{-f\left(n, \ell\right)}\eqqcolon\sum_{n',l'}d_{\infty}(n',l') q^{n'} y^{l'}\,,\label{eq:chiinf2}
}
where the function $f$ is defined as
\eq{
f(n,\ell)\coloneqq\sum_{m=1}^\infty c_o(nm,\ell-b_om)\,.
}
This function was first introduced in \cite{Belin:2019jqz}, where it was shown that $f$ extracts the growth of light states in the large central charge limit ($\dmin\rightarrow\infty$), and that $ d_\infty$ corresponds to the regularized density of states of $\varphi\left(\tau,z;\text{Sym}^N(\mathcal{C})\right)$ in the limit $N\rightarrow\infty.$\footnote{Compared to \cite{Belin:2019jqz} the summation range in $f$ is slightly different. The difference is caused by the fact that we consider symmetric product orbifolds, while \cite{Belin:2019jqz} considered exponential lifts. The precise relation is $f^{\text{here}}(n,\ell)=f_R^{\text{there}}(n,\ell)-c_o(0,0)$. In fact $f^\text{here}$ is equal to $\tilde f$ in \cite{Belin:2019rba}.} Ref. \cite{Belin:2019jqz} showed that there are two possibilities for the values of $f$: they grow exponentially when the underlying wJf has fast growth, and they are constants when the underlying wJf is slow growing. Therefore, when the seed wJf is slow growing, $\varphi_\infty$ is a ratio of theta-like functions, of which we can compute the weight. 

In order to compute the weight it is convenient to introduce the function 
\eq{
g(n,\ell)\coloneqq\sum_{m\in\mathbb{Z}}c_o(nm,\ell-b_om)\,,
}
which is related to $f$ according to
\eq{\label{eq:fg}
f(n,\ell)=g(n,\ell)-c_o(0,\ell)-\delta_{n,0}\sum_{m=1}^\infty c_o(0,\ell+b_om)\,.
}
There are a couple of key facts about $g(n,\ell)$ coming from slow-growing forms that we will use to derive the weights \cite{Belin:2019rba,Belin:2019jqz}:
\begin{enumerate}[label=$\roman*$)]
\item $g(n,\ell)$ vanishes unless $n=0$ or $t_on+b_o\ell=0$\,.
\item $g(n,\ell)$ only depends on $n_b\coloneqq n\text{ mod }b_o$ and $\ell_b\coloneqq\ell\text{ mod }b_o$\,.
\item $g(n,\ell)$ is given by
\eq{\notag
g(n,\ell)=\begin{cases}
\sum_{\hat m \in b_o\mathbb{Z}-\ell-n_bt_o/b_o} c_o(-n_b\hat m/b_o-n_b^2t_o/b_o^2,\hat m) ~& \text{if}~t_on+b_o\ell=0~\text{or}~n=0\,,\\
0&\text{otherwise}\,.
\end{cases}
} 
\item $g(n,\ell)=f(n,\ell)$ for $n>0$ and $\abs{\ell}>b_o$.
\end{enumerate}
From $i$) and $ii$) it follows that all nonzero functions $g$ will be of the form
\eq{\label{gmod}
g(b_o\theta+n_b,-(b_o\theta+n_b)\kappa)=g(n_b,n_b\kappa\text{ mod }b_o)\,,
}
where $\kappa\coloneqq t_o/b_o$, and $\theta\in\mathbb{Z}_{\ge0}$. 

Recall that
\eq{
\eta(\tau)\sim\prod_{n=1}^\infty (1-q^n)\,.
}
Therefore, in order to obtain a weighted function we need 
\eq{\label{eq:1.6}
f(\alpha n,\ell)=\beta\quad\text{for all but finitely many } n\in\mathbb{N}\,,
}
for some $\alpha\in\mathbb{N}$ and $\beta\in\mathbb{Z}$ (here $\ell$ can either be zero, or proportional to $n$). Functions $f$ satisfying this property contribute $-\beta/2$ to the weight. Analogously,
\eq{\label{eq:1.62}
f( n,\alpha\ell)=\beta\quad\text{for all but finitely many } \ell\in\mathbb{N}\,
}
also leads to a contribution to the weight of $-\beta/2$ (here $n$ can either be zero, or proportional to $\ell$). Note that 
\eq{
\prod_{n=1}^\infty \left(1-q^{2n-1}\right)^2\sim\frac{\vartheta_{01}(\tau,0)}{\eta(2\tau)}\,,
}
carries no weight;\footnote{We can generalize this statement to any product of the form $\prod_{n=1}^\infty \left(1-q^{\gamma n-\delta}\right)$ with $\gamma$ and $\delta$ coprime positive integers.} only the eta-like factors will lead to a weight.

Our task is thus to identify from \eqref{eq:fg} all possible eta-like functions that can appear. There are two possible sources for eta functions: nonzero values of $g$ or nonzero $c_o(0,0)$. Let us first consider nonzero values of $g$. The claim is then that all eta-like products contribute to $g(0,0)$. Moreover, nonzero $g(0,0)$ sources two different flavors of eta functions sourced by this nonzero value of $g$.
First, there are eta functions sourced by
\eq{
f(\alpha n, \kappa\alpha n)\,.}
To show this we consider generic $\alpha>0$ in \eqref{eq:1.6}, and see that it leads to a nonzero value of $g(0,0)=\beta$. Let $\alpha>0$, and set $n'=b_o n$. 
Then we have
\eq{
g(\alpha n',-\alpha n'\kappa)=g(\alpha b_o n,-\alpha b_o n\kappa)=g(0,0)\,.
}
Second, there are eta functions sourced by
\eq{
f(0, b_o\ell)=g(0,b_o\ell)=g(0,0)\,,
}
for $\ell>1$. Finally, we evaluate $g(0,0)$ to obtain
\eq{
g(0,0)=\sum_{\hat m\in b\mathbb{Z}}c_o(0,\hat m)=c_o(0,0)+2c_o(0,-b_o)\,.
}

Next, we consider nonzero $c_o(0,0)$. Then $f(n,0)=-c_o(0,0)$ for $n>0$. Note that $f(n,\ell)=-c_o(0,\ell)=c_o(0,-\ell)=f(n,-\ell)$ lead to factors of the form $\vartheta_{1,1}(\tau,\ell z)/\eta(2\tau)$, which carry no weight. Therefore, the total weight equals
\eq{\label{eq:weight}
\text{weight}=\frac{2g(0,0)-c_o(0,0)}{2}=2c_o(0,-b_o)+c_o(0,0)/2\,.
}
By a standard Laplace transform on the eta functions we finally obtain the following expression for $\omega$
\eq{
\omega=-c_o(0,-b_o)-c_o(0,0)/4-3/2\,.
}

\bibliographystyle{ytphys}
\bibliography{ref}
\end{document}